\documentclass[bibyear]{aa}  

\usepackage{txfonts}
\usepackage{longtable}
\usepackage{graphicx}
\usepackage{float,lscape}
\usepackage{gensymb}

\begin{document}

   \title{CARMENES input catalogue of M dwarfs}
   \subtitle{II. High-resolution imaging with FastCam}
   \titlerunning{High-resolution imaging of CARMENES M dwarfs with FastCam}
        %

   \author{
        M. Cort\'es-Contreras \inst{1}
          \and
          V.\,J.\,S. B\'ejar \inst{2}
          \and
          J.\,A. Caballero \inst{3,4}
          \and
          B. Gauza \inst{2}
          \and
          D. Montes \inst{1}
          \and \\
          F.\,J. Alonso-Floriano \inst{1}
          \and
          S.\,V. Jeffers \inst{5}
          \and
          J.\,C. Morales \inst{6}
          \and
          A. Reiners \inst{5}
          \and
          I. Ribas \inst{6}
                  \and
          P. Sch\"ofer \inst{5}
          \and
          A. Quirrenbach \inst{4}
          \and \\
          P.\,J. Amado \inst{7}
          \and
          R. Mundt \inst{8}
          \and
          W. Seifert \inst{4}
          }  

        \authorrunning{M.~Cort\'es-Contreras~et~al.}

   \institute{
        Departamento de Astrof\'isica y Ciencias de la Atm\'osfera, Facultad de Ciencias F\'isicas, Universidad Complutense de Madrid, 28040 Madrid, Spain,
        \email{micortes@ucm.es}
         \and
        Instituto de Astrof\'isica de Canarias, V\'ia L\'actea s/n, 38205 La Laguna, Tenerife, Spain, and Departamento de Astrof\'isica, Universidad de La Laguna, 38206 La Laguna, Tenerife, Spain
         \and
        Centro de Astrobiolog\'ia (CSIC-INTA), PO Box 78, 28691 Villanueva de la Ca\~nada, Madrid, Spain
        \and
        Landessternwarte, Zentrum f\"ur Astronomie der Universit\"at Heidelberg, K\"onigstuhl 12, 69117 Heidelberg, Germany
        \and
        Institut f\"ur Astrophysik, Friedrich-Hund-Platz 1, 37077 G\"ottingen, Germany
        \and
        Institut de Ci\`ences de l'Espai (CSIC-IEEC), Campus UAB, c/ de Can Magrans s/n, 08193 Bellaterra, Spain
        \and
        Instituto de Astrof\'isica de Andaluc\'ia (CSIC), Glorieta de la Astronom\'ia s/n, 18008 Granada, Spain
        \and
        Max-Planck-Institut f\"ur Astronomie, K\"onigstuhl 17, 69117 Heidelberg, Germany
        }

   \date{Received 06 Jul 2016; accepted dd Aug 2016}

 
  \abstract
   {
        }
  {We search for low-mass companions of M dwarfs and characterize their multiplicity fraction with the purpose of helping in the selection of the most appropriate targets for the CARMENES exoplanet survey.
   }
  {We obtained high-resolution images in the $I$ band with the lucky imaging instrument FastCam at the 1.5\,m Telescopio Carlos S\'anchez for 490 mid- to late-M dwarfs.
   For all the detected binaries, we measured angular separations, position angles, and magnitude differences in the $I$ band. We also calculated the masses of each individual component and estimated orbital periods, using the available magnitude and colour relations for M dwarfs and our own M$_J$-spectral type and mass-$M_I$ relations.
   To avoid biases in our sample selection, we built a volume-limited sample of M0.0-M5.0 dwarfs that is complete up to 86\,\% within 14\,pc.
   }
  {From the 490 observed stars, we detected 80 companions in 76 systems, of which 30 are new discoveries.
  Another six companion candidates require additional astrometry to confirm physical binding.
  {The multiplicity fraction in our observed sample is 16.7\,$\pm$\,2.0\,\%.
The bias-corrected multiplicity fraction in our volume-limited sample is 19.5\,$\pm$\,2.3\,\% for angular separations of 0.2 to 5.0\,arcsec (1.4--65.6\,au), with a peak in the distribution of the projected physical separations at 2.5--7.5\,au. For M0.0-M3.5\,V primaries, our search is sensitive to mass ratios higher than 0.3 and there is a higher density of pairs with mass ratios over 0.8 compared to those at lower mass ratios. Binaries with projected physical separations shorter than 50\,au also tend to be of equal mass. For 26 of our systems, we estimated orbital periods shorter than 50\,a, 10 of which are presented here for the first time. We measured variations in angular separation and position angle that are due to orbital motions in 17 of these systems.
The contribution of  binaries and multiples with angular separations shorter than 0.2\,arcsec, longer than 5.0\,arcsec, and of spectroscopic binaries identified from previous searches, although not complete, may increase the multiplicity fraction of M dwarfs in our volume-limited sample to at least 36\%.}
   }
{}

   \maketitle

\section{Introduction} \label{intro}

The multiplicity of low-mass stars provides constraints to models of stellar and planet formation and evolution (Goodwin et~al. 2007; Burgasser et~al. 2007; Duch\^ene \& Kraus 2013).
M dwarfs, which have approximate masses of between 0.1 and 0.6\,$M_\odot$, account for two thirds of the stars in the solar neighbourhood and probably the Galaxy.
However, in spite of their abundance and the increasing number of M-dwarf high-resolution imaging surveys in the past decade (Beuzit et~al. 2004; Law et~al. 2008; Bergfors et~al. 2010; Janson et~al. 2012, 2014a; J\'odar et~al. 2013; Bowler et~al. 2015; Ward-Duong et~al. 2015), the multiplicity of M dwarfs is not yet well constrained, at least by comparison with the better determination for Sun-like stars (Duquennoy \& Mayor 1991; Raghavan et~al. 2010; Tokovinin 2011). Published values range between 13.6\,\% and 42\,\%. Thus, the binary fraction of M dwarfs seems intermediate between the one of Sun-like stars and very low mass binaries.
In Table~\ref{table.stellar_multfrac} we summarise the multiplicity fractions and semi-major axis coverage of some of the main multiplicity surveys carried out from F6 to T dwarfs.

\begin{table*}
        \centering
        \caption {Stellar multiplicity fractions.}
        \label{table.stellar_multfrac}
        \begin{tabular}{l c c c c c}
        \hline \hline
        \noalign{\smallskip}
Reference                               &       Investigated            &       $d_{\rm lim}$   &       Multiplicity    &       Projected physical      &               Survey  \\
                                                &       spectral type            &       [pc]    &       fraction        [\%]    &       separation, $s$ [au]                &       method\tablefootmark{a}         \\
        \noalign{\smallskip}
        \hline
        \noalign{\smallskip}
        \noalign{\smallskip}
        
Duquennoy \& Mayor 1991         & F7--G9&       22              &       $\sim$\,65                              &               $\sim$\,0.01--225               & RV, WI  \\ 
        \noalign{\smallskip}
Raghavan et~al. 2010    & $\sim$\,F6--K3                &       25              &       44\,$\pm$\,3            &       $\sim$\,0.005--100\,000                         &       RV, AO, S, WI\\
        \noalign{\smallskip}
Reid \& Gizis 1997              & K2--M6                                        &         8               &       32                              &               $\sim$\,0.1--1800                               & RV, S, WI       \\
        \noalign{\smallskip}            
Leinert et~al. 1997             & M0--M6                                &       5               &       26\,$\pm$\,9                    &       $\sim$\,1--100                                          & S       \\
        \noalign{\smallskip}
Fischer \& Marcy 1992   & M                                     &       20              &       42\,$\pm$\,9            &       0--10\,000      &                               RV, WI\\
        \noalign{\smallskip}
J\'odar et~al. 2013     & K5-M4                                 &       25              &       20.3$_{-5.2}^{+6.9}$    &       $\sim$\,0--80   &                       LI      \\
        \noalign{\smallskip}
Ward-Duong et~al. 2015& K7--M6                  &               15      &               23.5\,$\pm$\,3.2                &       $\sim$\,3--10\,000      &                       AO, WI              \\
        \noalign{\smallskip}
Bergfors et~al. 2010    & M0.0--M6.0            &       52              &       32\,$\pm$\,6                            &       3--180          &                       LI      \\
        \noalign{\smallskip}
Janson et~al. 2012      &               M0.0--M5.0                      &               52      &                       27\,$\pm$\,3                            &       3--227                                  &       LI      \\
        \noalign{\smallskip}
Law et~al. 2008 & M4.5--M6.0                    &       <15.4>          &       13.6$_{-4.0}^{+6.5}$    &       $\sim$\,0--80                                                   & LI      \\
        \noalign{\smallskip}
Siegler et~al. 2005     &       M6.0--M7.5      &       30      &       9$_{-3}^{+4}$   &       $\geqslant$\,3  &       AO      \\
        \noalign{\smallskip}
Janson et~al. 2014a& M5.0--M8.0                         &       36              &       21--27                                          &       $\sim$\,0.5--100                                &       LI      \\
        \noalign{\smallskip}
Close et~al. 2003& M8.0--L0.5           &       33              &       15\,$\pm$\,7                                            &       <15             &               AO      \\
        \noalign{\smallskip}
Bouy et~al. 2003        & M7.0--L8.0                            &       20              &       10--15                                          &       1--8                            &       HST     \\ 
        \noalign{\smallskip}
Reid et~al. 2008        & L                                     &       20              &       12.5$_{-3.0}^{+5.3}$                                    &       <3                              &       HST     \\
        \noalign{\smallskip}
Burgasser et~al. 2003   & T                                     &       <10>            &       9$_{-4}^{+15}$                          &               1--5                    &       HST     \\
        \noalign{\smallskip}
        \hline
        \end{tabular}
\tablefoot{
        \tablefoottext{a}{AO: Adaptive optics; HST: {\it Hubble Space Telescope}; LI: Lucky imaging; RV: Radial velocity; S: Speckle; WI: Wide-field imaging.}
}
\end{table*}

The typical separation of low-mass stars in a binary system tends to decrease with the mass of the primary, which makes the detection of faint companions at resolvable separations more difficult (Jeffries \& Maxted 2005; Burgasser et~al. 2007; Caballero 2007; Bate 2012; Luhman 2012).
In addition, the presence of a stellar companion influences planet formation (Wang et~al. 2014a, 2014b, 2015a, 2015b). The limited number of exoplanet hosts in binary and multiple systems (Mugrauer et~al. 2007; Mugrauer \& Neuh\"auser 2009; Ginski et~al. 2015) and the relatively small number of M dwarfs with known exoplanets detected with radial-velocity and transit methods (Rivera et~al. 2005; Charbonneau et~al. 2009; Bonfils et~al. 2013) prevents a significant statistical analysis of how stellar multiplicity at such low masses affects planet formation.

Because of their low effective temperatures, M dwarfs emit the bulk of their energy in the near-infrared.
It makes them difficult to observe with the required radial-velocity precision with the current spectrographs for exoplanet hunting (e.g. HARPS at the 3.6\,m ESO La Silla Telescope, HARPS-N at the 3.6\,m TNG, and UVES at the 8.2\,m ESO VLT), which operate in the optical.
The prompt development of stable near-infrared spectrographs with wide wavelength coverage and high spectral resolution for radial-velocity surveys of M dwarfs has therefore been identified as critical by numerous decadal panels, funding agencies, and international consortia.
Some noteworthy high-resolution near-infrared spectrographs currently under developement are IRD at 8.2\,m Subaru (Tamura et~al. 2012), HPF at 9.2\,m HET (Mahadevan et~al. 2014), and SPIRou at 3.6\,m CFHT (Donati et~al. 2014).
The high-resolution spectrograph CARMENES (Amado et~al. 2013; Quirrenbach et~al. 2014\footnote{\url{http://carmenes.caha.es}}) at 3.5\,m Calar Alto covers from 520\,nm to 1710\,nm and has started its science survey in January 2016.

CARMENES is the name of the double-channel spectrograph (near-infrared and optical) of the Spanish-German consortium that built it, and of the science project that is being carried out during guaranteed-time observations (GTO).
For at least 600 GTO clear nights in the time frame between 2016 and 2018, CARMENES will spectroscopically monitor about 300 carefully selected M dwarfs with the goal of detecting low-mass planets in their habitable zones.
With a long-term 1\,m\,s$^{-1}$ radial-velocity precision, the consortium aims at being able to detect 2\,M$_\oplus$ planets orbiting in the habitable zone of M5\,V stars and super-Earths around earlier stars (Garc\'ia-Piquer et~al. 2016). 
In addition to the detection of the individual planets themselves, the ensemble of objects will provide sufficient statistics to assess the overall distribution of planets around M dwarfs: frequency, masses, and orbital parameters.

To optimise the observational strategy of the instrument and its scientific return, the consortium has built Carmencita, the CARMENES input catalogue (Caballero et~al. 2013; Quirrenbach et~al. 2015; Alonso-Floriano et~al. 2015a). 
It consists of almost 2200 of the brightest M dwarfs of each spectral subtype observable from Calar Alto, from which we will select the approximately 300 {\em \textup{single}} GTO stars.
By single we mean stars without close visual (physically bound) or optical (unbound) stellar or substellar companions that may induce real or artificial radial-velocity variations and, therefore, contaminate the precise CARMENES measurements (Guenther \& Wuchterl 2003; Ehrenreich et~al. 2010; Guenther \& Tal-Or 2010; Bonfils et~al. 2013). 

As part of our efforts to determine the multiplicity of M dwarfs and to select the best targets for radial-velocity surveys for exoplanets, we performed a high-resolution imaging search of close companions with the lucky imaging instrument FastCam at the Telescopio Carlos S\'anchez, as described in this paper.
Preliminary results of this work were presented as conference proceedings by B\'ejar et~al. (2012) and Cort\'es-Contreras et~al. (2015a, 2015b). 
This paper is the second item of  the series called the CARMENES input catalogue of M dwarfs.
In the first paper, Alonso-Floriano et~al. (2015a) carried out a low-resolution optical spectroscopic analysis of a number of poorly known dwarfs to constrain their spectral types.
Furthermore, this work will soon be complemented with on-going searches of unresolved spectroscopic binaries and triples identified in a large collection of high-resolution optical spectra (Montes et~al. 2015; Jeffers et~al. in prep.) and of wide companions to M dwarfs supported by virtual observatory tools (cf., Cort\'es-Contreras et~al. 2013, 2014; Alonso-Floriano et~al. 2015b).

\begin{figure*}
   \centering
   \includegraphics[width=\textwidth]{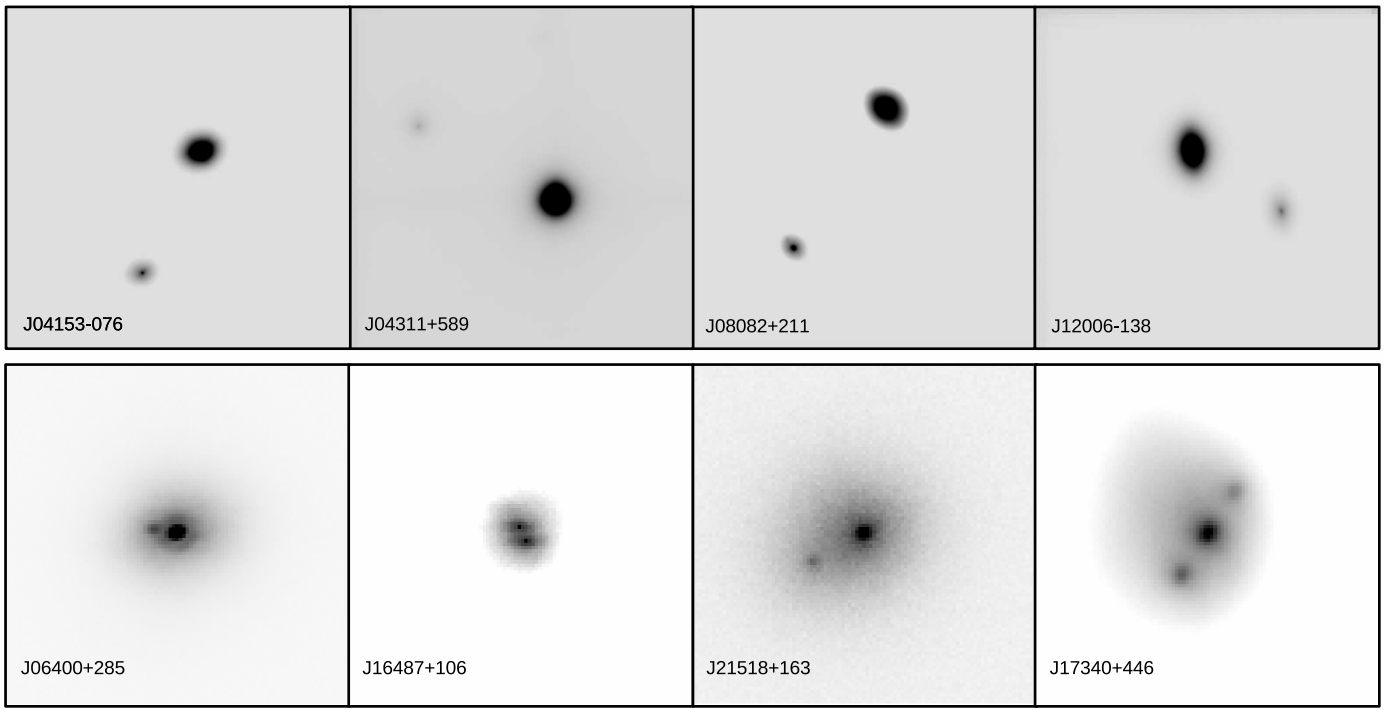}
      \caption{Selection of images of multiple systems identified by us with FastCam. North is up and east is left. The upper row scale is 20\,$\times$\,20\,arcsec$^{2}$, that of the lower row 4\,$\times$\,4\,arcsec$^{2}$. Images at the top were obtained with the shift \& add mode, while the bottom images were obtained with the ``lucky image'' mode.} The bottom right image (J17340+446) is an examlpe of the so-called false triple effect.
         \label{fig.example_FC_bins}
   \end{figure*}


\section{Observations}

Of the almost 2200 M dwarfs currently in Carmencita, we selected 490 Carmencita targets for being observed with the FastCam lucky imager (Oscoz et~al. 2008) at the 1.5\,m Telescopio Carlos S\'anchez at the Observatorio del Teide (Tenerife, Spain).
The high-resolution imager FastCam is equipped with an L3CCD Andor 512\,$\times$\,512 detector with very low electron noise and high readout speed.
It has a field of view of 21.2\,$\times$\,21.2\,arcsec$^{2}$ and an approximate pixel size and orientation of the detector of 0.0425\,arcsec and 91.9\,deg, respectively.
FastCam delivers nearly diffraction-limited images, which at the Telescopio Carlos S\'anchez and in the $I$ band have full-width at half maxima of approximately 0.15\,arcsec.

We carried out the observations during 26 nights in 15 runs from October 2011 to January 2016. 
For each target, we obtained typically ten blocks of 1000 frames each in the Johnson-Cousins $I$ band using the electron multiplication mode. 
Typical frame exposure times were in the 35--50\,ms range.
On average, each star was imaged during 500\,s in total.
The typical Strehl ratio in our observations varies with the percentage of the best-quality frames chosen in the reduction process: from 0.2 for the 100\% to 0.4 for the 1\%.
For astrometric calibration purposes,  we also observed the globular cluster \object{M3} and 18 astrometric standard binary stars from the Aitken Double Star catalogue (ADS -- Aitken 1932; Scardia et~al. 1995) with the same method and on several occasions.

Each frame was bias subtracted and then processed with the FastCam dedicated software developed at the Universidad Polit\'ecnica de Cartagena (see Labadie et~al. 2010; J\'odar et~al. 2013). 
We ran the lucky image (on five blocks) and shift \& add processing modes separately. 
The first allows selecting the fraction of the best-quality frames (we chose 1\,\%, 10\,\%, and 50\,\%), aligns the selected frames using the brightest speckle, and combines them, producing six final lucky images per target. 
The second mode aligns all the block frames and then combines them, resulting in one unique image per target.
Shift \& add produces deeper images than the lucky image mode, but with slightly poorer resolution.
The M3 standard field was reduced only with the shift \& add mode. In Fig.~\ref{fig.example_FC_bins} we show a selection of the processed images at two different spatial scales.

\begin{figure}
        \centering
        \includegraphics[width=\hsize]{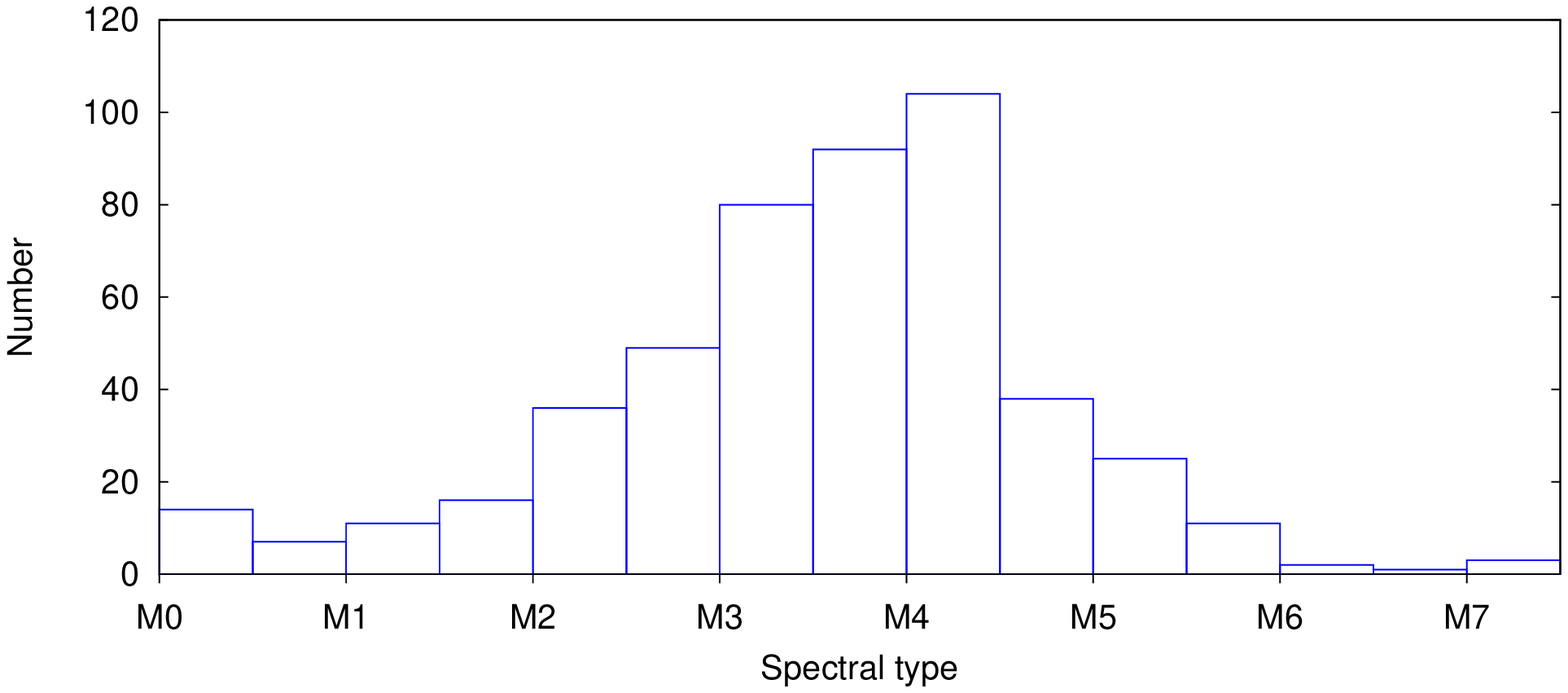}
        \includegraphics[width=\hsize]{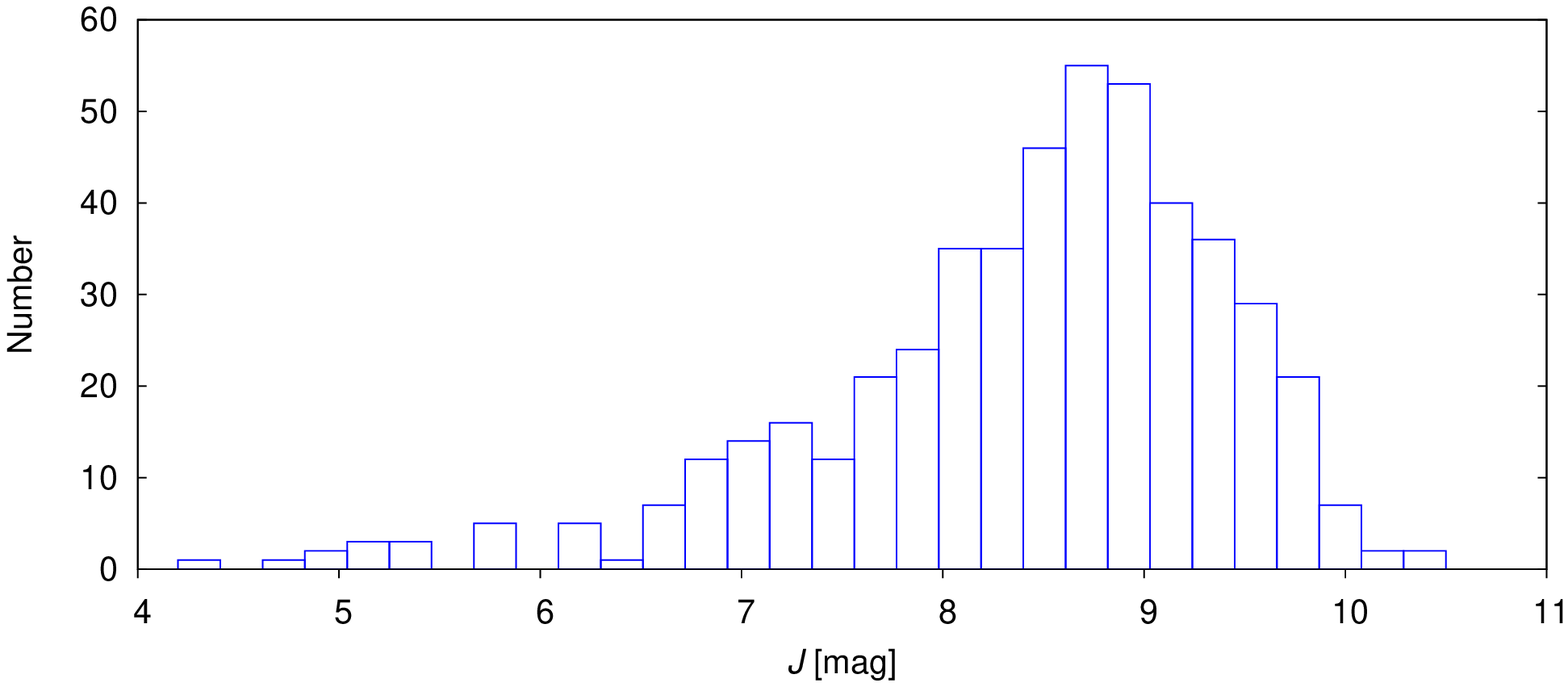}
        \includegraphics[width=\hsize]{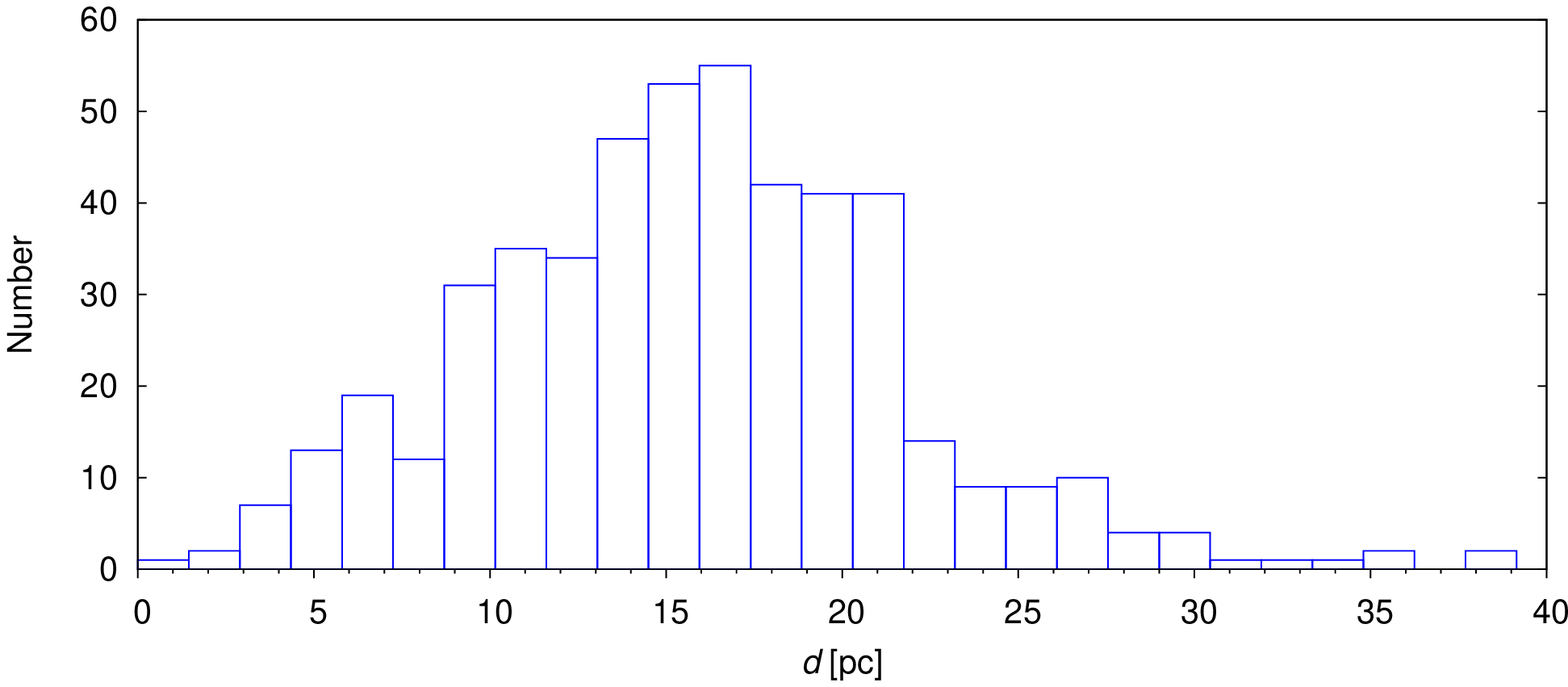}
        \includegraphics[width=\hsize]{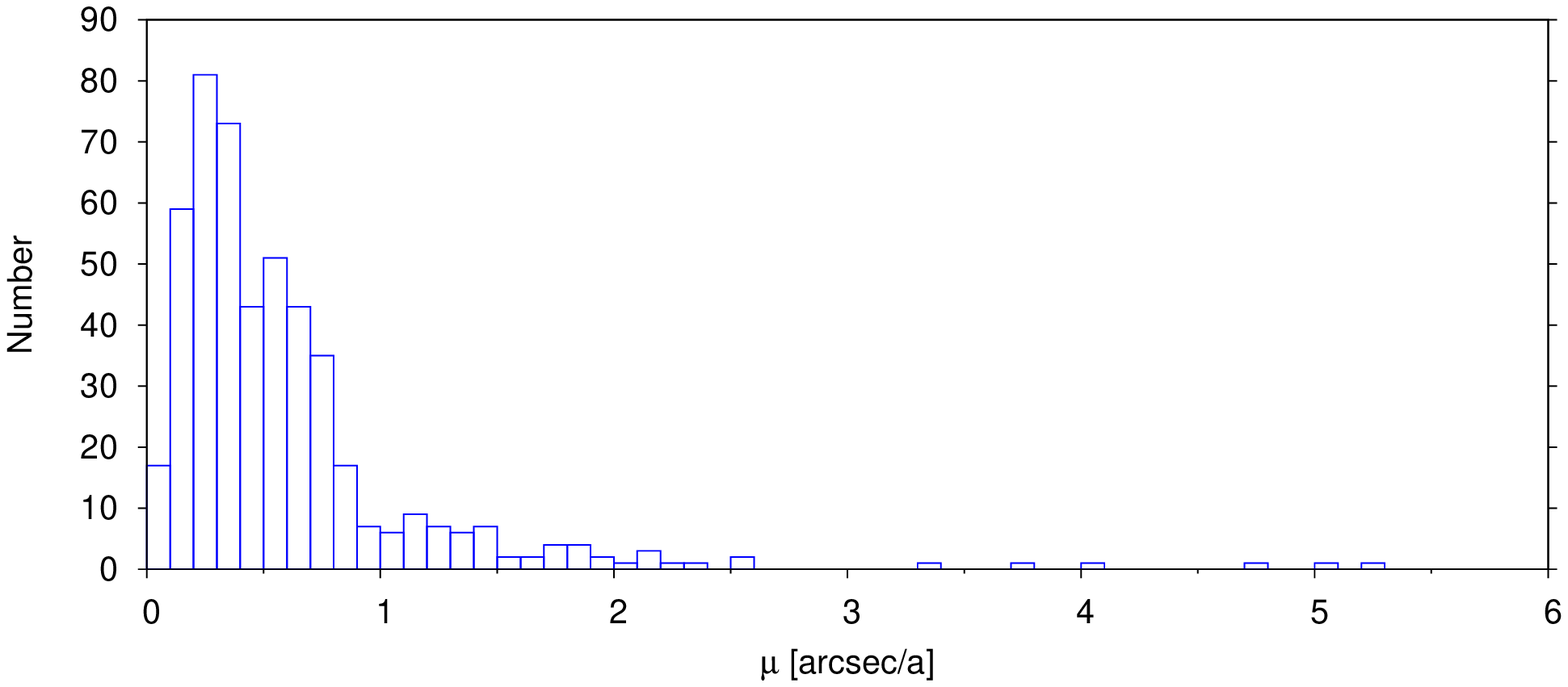}
        \caption{Distributions of spectral type, $J$-band magnitude, distance, and proper motion of the 490 observed M-dwarf targets. 
        The sizes of the bins follow the definitions given by Freedman \& Diaconis (1981). The lowest panel does not display Barnard's star, with $\mu$ = 10.4\,arcsec\,a$^{-1}$.
}
         \label{fig.histograms}
  \end{figure}

In Table~\ref{table.observedstars}, we provide the list of 490 observed M-dwarf targets with the following column information: identification number, our Carmencita identifier (Quirrenbach et~al. 2015; Alonso-Floriano et~al. 2015a), J2000 coordinates and $J$-band magnitude from the Two-Micron All-Sky Survey (Skrutskie et~al. 2006), spectral type and its reference, distance and its reference, and the FastCam observation date and exposure time. 
Figure~\ref{fig.histograms} shows the histograms of spectral types, $J$-band magnitudes, heliocentric distances, and total proper motions of the observed sample. 
Spectral types range from M0.0\,V to M7.0\,V, $J$ from 4.2\,mag to 10.4\,mag, distances from 1.8\,pc to 39.1\,pc, and proper motions from 0.03\,arcsec\,a$^{-1}$ to 10.6\,arcsec\,a$^{-1}$. Because of their closeness, 97\,\% of our targets have total proper motions larger than 100\,mas\,a$^{-1}$.

Our sample of 490 observed Carmencita targets consisted mainly of the brightest stars in the $J$ band for each spectral subtype (see Sect.~2 in Alonso-Floriano et~al. 2015a) that were
        ($i$) not known spectroscopic binaries,
        ($ii$) not resolved systems with visual or optical companions at angular separations smaller than 5\,arcsec, and
        ($iii$) not studied with high-resolution imaging devices before the start of our observations by speckle, adaptive optics, or lucky imaging (Beuzit et~al. 2004; Law et~al. 2008; Bergfors et~al. 2010; J\'odar et~al. 2013; Janson et~al. 2012).
Some high-resolution imaging (Janson et~al. 2014a; Ansdell et~al. 2015; Bowler et~al. 2015; Ward-Duong et~al.~2015) and spectroscopic (Bonfils et~al. 2013; Llamas 2014; Sch\"ofer 2015) surveys have been performed afterwards and have tabulated several objects in common with our target list.
In addition, we also observed
        ($a$) some dubious or poorly investigated close multiple systems (including spectroscopic binary candidates), 
        ($b$) a few stars with possible visual companions at angular separations smaller than 5\,arcsec that needed confirmation or better characterisation, and 
        ($c$) four known binaries with estimated orbital periods shorter than five years that were previously proposed for follow-up by Cort\'es-Contreras et~al. (2013): J05085--181 (\object{GJ~190}), J13317+292 (\object{DG~CVn}), J23174+196 (\object{G~067--053}), and J23455--161 (\object{LP~823--004}).

To confirm the physical binding of pairs (i.e. that the components share a common proper motion), we observed 54 targets more than once, and up to eight times.
Accounting for the 490 M dwarfs, 18 ADS pairs and M3 calibration field, and the different epochs, we acquired 7670 images in total with FastCam.

 \begin{table}
        \centering
        \caption {FastCam adopted plate scale and orientation for each run night.}
        \label{table.M3calibration}
        \begin{tabular}{l c c c c }
        \hline \hline
        \noalign{\smallskip}
Observation     &       \multicolumn{2}{c}{Pixel scale\,[mas/pix]}      &       \multicolumn{2}{c}{Orientation\,[deg]}         \\
date$^{a}$      &       $x$     &       $y$     &       $x$             &       $y$             \\
        \noalign{\smallskip}
        \hline
        \noalign{\smallskip}
        \noalign{\smallskip}
23 Oct 2011*    &       42.25           &       42.56           &       92.08           &       91.60   \\
24 Oct 2011     &       42.25           &       42.56           &       92.08           &       91.60   \\
25 Oct 2011     &       42.25           &       42.56           &       92.08           &       91.60   \\
        \noalign{\smallskip}
30 Jan 2012     &       42.25           &       42.56           &       92.08           &       91.60   \\
31 Jan 2012     &       42.25           &       42.56           &       92.08           &       91.60   \\      
        \noalign{\smallskip}
25 Mar 2012*    &       42.31           &       42.61           &       91.79           &       91.64   \\
26 Mar 2012*    &       42.30           &       42.62           &       91.82           &       91.65   \\
27 Mar 2012     &       42.30           &       42.62           &       91.82           &       91.65   \\                                                                      
        \noalign{\smallskip}
10 Jul 2012*    &       42.48           &       42.61           &       92.11           &       91.91   \\
11 Jul 2012*    &       42.49           &       42.64           &       92.03           &       91.77   \\
12 Jul 2012*    &       42.32           &       42.54           &       91.96           &       91.99   \\
        \noalign{\smallskip}
16 Sep 2012     &       42.32           &       42.54           &       91.96           &       91.99   \\
17 Sep 2012     &       42.32           &       42.54           &       91.96           &       91.99   \\
        \noalign{\smallskip}
13 Jan 2013*    &       42.26           &       42.69           &       91.94           &       91.63   \\
14 Jan 2013*    &       42.21           &       42.59           &       91.85           &       91.63   \\
        \noalign{\smallskip}
28 Feb 2014*    &       42.26           &       42.69           &       91.99           &       91.70   \\
01 Mar 2014     &       42.26           &       42.69           &       91.99           &       91.70   \\
02 Mar 2014     &       42.26           &       42.69           &       91.99           &       91.70   \\
        \noalign{\smallskip}
22 May 2014     &       42.26           &       42.69           &       91.99           &       91.70   \\      
        \noalign{\smallskip}
09 Dec 2014*    &       42.26           &       42.99           &       91.97           &       91.96   \\
        \noalign{\smallskip}
14 Apr 2015*    &       42.28           &       42.37           &       92.18           &       91.65   \\
15 Apr 2015     &       42.28           &       42.37           &       92.18           &       91.65   \\
        \noalign{\smallskip}
09 Jun 2015     &       42.28           &       42.37           &       92.18           &       91.65   \\
        \noalign{\smallskip}
29 Jul 2015     &       42.28           &       42.37           &       92.18           &       91.65   \\
        \noalign{\smallskip}
17 Nov 2015     &       42.28           &       42.37           &       92.18           &       91.65   \\      
        \noalign{\smallskip}    
07 Jan 2016     &       42.28           &       42.37           &       92.18           &       91.65   \\      
        \noalign{\smallskip}    
        \hline
        \end{tabular}
        \tablefoottext{a}{M3 calibration field was observed on nights marked with an asterisk.}                
\end{table}

\section{Analysis}
   
\subsection{Astrometry}

The first step of the analysis was computing the pixel size and detector orientation with common IRAF tasks (Tody 1986). 
To do this, we determined the centroids of the brightest stars in the M3 standard field with {\tt imcentroid}.
Using the celestial coordinates in the ACS Survey of Galactic Globular Clusters (Sarajedini et~al. 2007) and the pixel coordinates in our images, we then determined the transformation equations with {\tt ccmap} by fitting to a general transformation of order two. 
Table~\ref{table.M3calibration} lists the pixel scales and orientations of the detector for each night. 
For nights without M3 images, we used the calibration of the closest night with computed plate solution. 
Pixel scale and rotation angle in the centre of the detector in the {\it x} and {\it y} axes are similar within the different campaigns with almost negligible variations from night to night. 
Their mean values are 42.31\,$\pm$\,0.09\,mas/pixel and 42.63\,$\pm$\,0.15\,mas/pixel in pixel scale and 91.98\,$\pm$\,0.12\,deg and 91.74\,$\pm$\,0.15\,deg in orientations of the detector in the {\it x} and {\it y} axes, respectively. 
The uncertainties are the standard deviations of the measurements.

To double-check that our astrometric solutions were correct, we calculated angular separations ($\rho$) and position angles ($\theta$) for each ADS binary.
To do that, we measured the $x$ and $y$ positions of each star with {\tt imcentroid}, and transformed them into equatorial coordinates using the astrometric solution of the corresponding night with {\tt cctran}.
Table~\ref{table.ads} shows the previously published values of $\rho$, $\theta$, the epochs of observation and references, and our measured values in different epochs. 
Our errors in $\rho$ and $\theta$ were derived from the standard deviation of the measurements in all images within the same night and the determined astrometric solutions on different nights.
In general, the measured values of $\rho$ and $\theta$ of the same pair on different nights were consistent within 3$\sigma$ between them and with tabulated values from recent works. 
In some cases, the quality of our measurements surpassed previous publications.

We carried out a visual inspection for companions to our 490 Carmencita targets and found 137 additional sources in 116 systems, for which we measured the relative positions and position angles following the same procedure as described above for the ADS binaries.
In some epochs of nine stars with companions very close to the resolution limit of our images, we were unable to measure the photocentroid of both components with {\tt imcentroid }and, hence, we used the brightest pixel to measure their positions. In these cases, the uncertainties in the determination of $\rho$ and $\theta$ were larger and we adopted a typical error bar of one pixel.

We classified the 137 sources into three groups:
($i$) 51 optical companions (i.e. unbound, Table~\ref{table.visuals}),
($ii$) 80 physical companions (i.e. bound, Table~\ref{table.rhotheta}), and
($iii$) six unconfirmed companions (bottom of Table~\ref{table.rhotheta}).
For the classification, we used old photographic plate digitisations and all-sky surveys provided by the Aladin sky atlas (Bonnarel et~al. 2000), previous astrometry tabulated by the Washington Double Star Catalogue (WDS, Mason et~al. 2001) and/or our own multi-epoch astrometric measurements together with the target proper motions (mostly from van~Leeuwen 2007 and Roeser et~al. 2010).

Most of the 51 optical companions are null-proper-motion point-like sources in photographic plates of the first National Geographic Society -- Palomar Observatory Sky Survey in the mid-1950s. For the rest of the companions, we performed a multi-epoch analysis of their relative positions.
We considered as optical (unbound) companions those that show $\rho$ and $\theta$ values in different epochs consistent within 3$\sigma$ with null proper motion and inconsistent by more than 3$\sigma$ with the proper motion of the M dwarfs. Otherwise, we considered them as physically bound. For five of the six unconfirmed binaries, we only had one epoch, and for the other (J07349+147), the $\rho$ and $\theta$ values at different epochs did not allow us to distinguish between null or common proper motion.

\begin{figure}
   \centering
   \includegraphics[width=1\hsize]{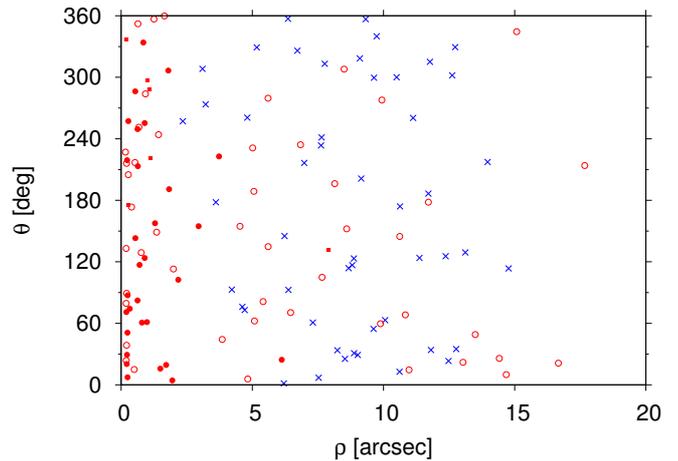}
      \caption{Diagram of $\theta$ vs. $\rho$ for all the 137 measured pairs. 
       Filled red circles are new physically bound pairs, 
       small filled red squares are unconfirmed related pairs,
      open red circles are known physically bound pairs,        
      and blue crosses are optically unrelated pairs.}
         \label{fig.rhothetameasured}
   \end{figure}

Figure~\ref{fig.rhothetameasured} displays the measured $\rho$ and $\theta$ values of all the detected pairs. It shows a homogeneous distribution of the position angle of the companions, which discards possible false detections associated with, for example, optical ghosts.


\subsection{Photometry}

\begin{figure}
        \centering
        \includegraphics[width=1\hsize]{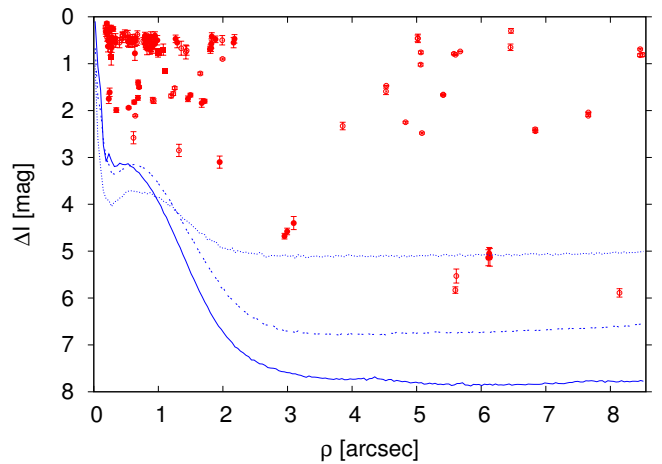}
         \caption{Diagram of $\Delta I$ vs. $\rho$ up to 8.5\,arcsec for the physical pairs.
Colour and symbol code is as in Fig.~\ref{fig.rhothetameasured}.
Solid, dashed, and dotted lines indicate the 3$\sigma$ detection limits for primaries in the magnitude ranges $I <$\,10\,mag, 10\,mag $\leq I <$\,11\,mag, and $I \geq$\,11\,mag, respectively.
}
         \label{fig.curvadedetectabilidad}
\end{figure}

In Table~\ref{table.rhotheta} we list magnitude differences in the $I$ band for the 80 physical and six likely physical pairs.
To measure the magnitude difference  of the binaries, we performed aperture and point spread function (PSF) photometry using the {\tt phot}, {\tt psf}, and {\tt allstar} routines in the {\tt daophot} package of IRAF.

For wide enough pairs, we used the primary star PSF as a reference for the secondary. In these cases, magnitude differences from aperture photometry and PSF fitting did not differ significantly. 
Since the PSF varies depending on the focus and sky position, for close pairs we chose the most appropriate single star observed during the same night as a reference to compute the PSF.
For five pairs, we were unable to measure the $\Delta I$ between components using PSF photometry, and we estimated it from the peak flux ratio of the PSF subtracted image, and for J23455--161, we perceived the companion and could not measure the magnitude difference.

A few close pairs showed a so-called false triple effect associated with the reduction process by the FastCam software, based on the selection of the brightest pixel.
When both components are of similar brightness, this software may not distinguish between the primary and secondary and, in the process of aligning, selects the brightest pixel in one or another star, resulting in an apparent triple system. For equal brightness binaries, this may lead to a degeneracy in the determination of the position angle of 180\,deg.
The option {\tt 2stars} in the FastCam reduction software, which takes this ambiguity into account, solved this effect in most cases. For the rest, we determined the real flux ratio of the pair by following the procedure described by Law~(2006):

\begin{equation}
        F_R = \frac{2I_{13}}{I_{12}I_{13} + \sqrt{I_{12}^2I_{13}^2-4I_{12}I_{13}}},
\end{equation}

\noindent where $I_{12} = F_1/F_2$ and $I_{13} = F_1/F_3$, and $F_1$, $F_2$ and $F_3$ are the fluxes of the images in the positions of the true primary, true secondary, and spurious tertiary, respectively.

In Fig.~\ref{fig.curvadedetectabilidad} we plot the measured magnitude differences in the $I$ band and angular separations of the companions. Most of them are of similar brightness ($\Delta I \approx$ 0.0--1.0\,mag) and are located at angular separations smaller than 2.5\,arcsec. Figure~\ref{fig.curvadedetectabilidad} also shows the contrast curves of our survey as a function of angular separation. The maximum magnitude difference in each stacked image depends on the brightness of the primary star. For this reason, we considered three different groups in our sample according to their $I$ magnitude, from which we selected four single stars covering different spectral types to obtain a representative mean contrast curve. For each star, we estimated the detection limit as a function of the angular separation as three times the standard deviation of the number of counts in ten-pixel-wide annuli centred on the target. This detection limit was converted into $\Delta I$ using the peak flux value of the star.
The maximum magnitude difference in the detection of possible companions at angular separations between 0.2 and 1.0\,arcsec varies from 3 to 4\,mag and from 5 to 7\,mag at separations larger than 2\,arcsec, depending on the brightness of the primary star.
The limiting magnitude of our survey is about $I \approx$ 17\,mag, and we were able to detect all sources brighter than this limit at angular separations greater than 3\,arcsec. This implies that at separations larger than 3\,arcsec, the detection of companions earlier than M8 dwarfs is complete up to 40\,pc, which corresponds to the entire sample, and the detection of companions earlier than M9 dwarfs is complete up to 25\,pc, which is in most of our sample.


        \section{Results and discussion}
        
        \subsection{Detected binaries}
        
        Of the 490 observed stars, we confirmed with our data 80 companions in 76 systems, of which 30 are presented here for the first time. In addition, there are also six unconfirmed binaries
that need additional epochs to confirm the physical binding. 
        The majority of the optical components of the survey were easily identified using previous available data, and most of the remaining ones were confirmed as physically bound companions using our own measurements at different epochs. Therefore, we considered the six unconfirmed binaries as very probably physically bound rather than unbound pairs. We took into account the six binaries for the determination of the multiplicity fraction.

        The 86 pairs are listed in Table~\ref{table.rhotheta}. In the last column of the table, we include the multiplicity flag from version 1.2 of the Guide Star Catalog (Morrison et~al. 2001), which is ``False'' for 18 of the 30 new confirmed binaries,``True'' for 10 of them and has no entry for the close companions of J08082+211 and J15191--127. For the 30 new binaries, and to our knowledge, there are no other references to binarity. 

Of the 80 physical companions, 48 are tabulated by WDS (second column in Table~\ref{table.rhotheta}), of which two were previously suggested by Behall \& Harrington (1976; J05333+448) and Bowler et~al. (2015; J15496+348) and confirmed here. Another two were recently presented by Ward-Duong et~al. (2015; J05034+531) and Ansdell et~al. (2015; J06212+442), and one of the new companions resolved here is most likely associated with a spectroscopic binary identified by Bonfils et~al. (2013; J15191--127). The remaining 29 are pairs with no previous binarity references to our knowledge.

Some of the measured companions were not detected in all epochs because of the relative motion of the components and the crossing of the companion behind or in front of the primary star (J05078+179 and J05333+448), presence of the companion near the diffraction limit (J13317+292, J15496+348, J16487+106, and J21012+332), and a focus problem (J06400+285).


        \subsection{Multiplicity fraction}\label{sec.multfrac}

\begin{figure}
   \centering
   \includegraphics[width=1\hsize]{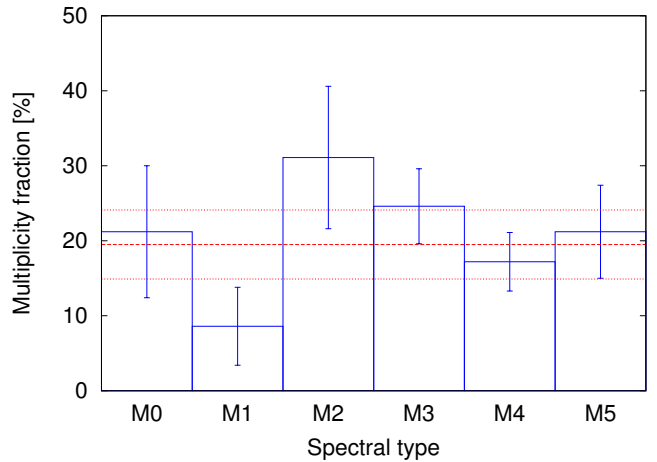}
      \caption{Multiplicity fraction as a function of spectral type from M0.0 V to M5.0 V in the volume-limited sample with separations from 0.2 to 5.0\,arcsec.
      Error bars are Poissonian.
Horizontal dashed and dotted lines are the global multiplicity fraction and the $\pm$\,2\,$\sigma$ values.
      }
         \label{fig.multfrac_spt}
\end{figure}

Of the 490 observed M dwarfs, 408 are single and 82 (76+6) are in binary or multiple systems within the FastCam field of view. This gives a close multiplicity fraction of 16.7\,$\pm$\,2.0\,\%, by assuming a Poissonian distribution of the errors. Nevertheless, it must not be taken as a real M-dwarf multiplicity fraction because of the selection bias of the observed sample: we did not include many stars that were previously observed in similar studies or that had known visual companions at less than 5\,arcsec.

For statistical purposes, we grouped all our Carmencita (Sect.~\ref{intro}) and FastCam targets in a combined sample.
Of the 2176 Carmencita stars, 1141 M dwarfs have been surveyed with FastCam or with high-resolution imagers with similar capabilities (Beuzit et~al. 2004; Lafreni\`ere et~al. 2007; Law et~al. 2008; Bergfors et~al. 2010; Janson et~al. 2012, 2014a; J\'odar et~al. 2013; Bowler et~al. 2015; Ward-Duong et~al. 2015). For completeness, we considered a range of angular separations from 0.2 to 5\,arcsec to our targets. The lower limit was given by the FastCam spatial resolution and the upper limit by the maximum separation at which we could detect companions to at least 90\,\% of the observed stars. 
Of the 1141 surveyed M dwarfs, 219 have physical companions in this interval of angular separations (55 from this work and 164 from other publications), which gives a close multiplicity fraction of 19.2\,$\pm$\,1.4\,\%.

To avoid any selection bias and give a more reliable multiplicity fraction, we proceeded by building a volume-limited sample with a maximum distance of 14\,pc and a completeness of 86\,\%. This completeness was estimated by assuming that all M0--M5 dwarfs are known within 7\,pc and that their density in the solar vicinity is constant.
This third sample is composed of 425 dwarfs with spectral types between M0.0 V and M5.0 V, of which 83 have companions (either from FastCam and other works) in the range from 0.2 to 5.0\,arcsec. This translates into a close multiplicity fraction of 19.5\,$\pm$\,2.3\,\%, which is consistent within error bars with the 13.6\,\%--27\,\% fractions obtained for M dwarfs in most surveys (Table~\ref{table.stellar_multfrac}), although some authors provided higher multiplicity fractions (Fischer \& Marcy 1992; Bergfors et~al. 2010). In Sects.~\ref{undetectedbins} and \ref{sec.wide} we estimate the contribution to the multiplicity fraction of pairs separated by less than 0.2\,arcsec and more than 5\,arcsec.


  \begin{figure}
   \centering
   \includegraphics[width=1\hsize]{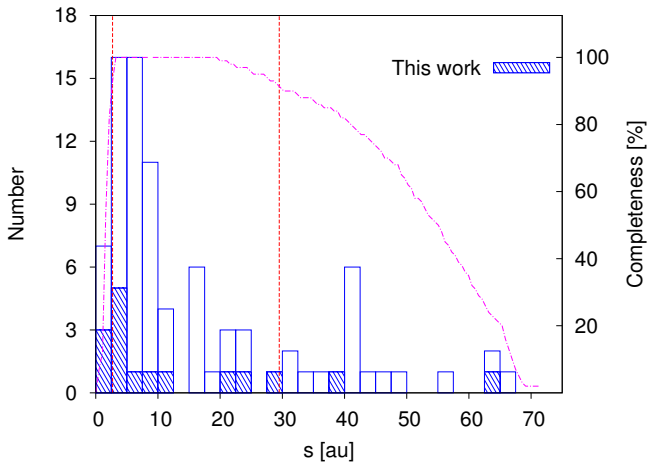}
      \caption{Projected physical separation distribution of the binaries in the volume-limited sample. Dashed bars represent our binaries. Vertical dashed lines mark the 90\,\% completeness limits, and the dash-dotted curve represents the completeness as a function of the projected physical separation.}
         \label{fig.semimajor_dist}
   \end{figure}

        \subsection{Dependence of multiplicity on spectral type}\label{subsection.sptdist}

To estimate the spectral types of the individual components of the binaries, we used the $I-J$ colours and $M_I$ absolute magnitudes as a function of spectral type for M dwarfs from Table~3 in Kirkpatrick et~al. (1994), together with the 2MASS photometry and spectral type of the pair.

For pairs resolved by 2MASS, we used these relations and values to obtain the $I$ magnitude of the primary, and obtained the $I$ magnitude of the secondary from our measured $\Delta I$. We derived the absolute $M_I$ magnitudes through the distance modulus and inferred the spectral types of the secondaries with the $M_I$-spectral type relation of Kirkpatrick et~al. (1994).

For pairs not resolved by 2MASS, the $J$-band magnitude involves the contribution of all the components in the system. In these cases, we obtained the $I$ magnitude of the system from the $I-J$ colours and the global spectral types of the pairs from the literature.
Using the $I$ magnitude and our measured $\Delta I$, we computed the individual $I$ magnitudes. We calculated the individual $M_I$ absolute magnitudes by applying the distance modulus, and estimated individual spectral types from the $M_I$-spectral type relation.

The distances in our sample come mostly from literature parallax determinations (see references in Table~\ref{table.observedstars}). For stars without parallactic distance, we calculated spectro-photometric distances from our own $M_J$-spectral type relation. This relation was obtained from a polynomial fit using single stars with well-determined spectral types between M0\,V and M6\,V, parallactic distances, and 2MASS $J$-band photometry from the Carmencita sample, and has the form:

\begin{equation}
        M_J = a~\rm{SpT}^2 + b~\rm{SpT} + c,
\end{equation}

\noindent where $a$~=~0.078\,$\pm$\,0.007\,mag, $b$~=~0.265\,$\pm$\,0.038\,mag and $c$~=~5.895\,$\pm$\,0.044\,mag, and SpT indicates the numerical spectral subtype within the M range.

For very close binaries, spectro-photometric distances are not reliable since the 2MASS photometry and the spectral type determination do not provide the contribution of the two components separately.
In these cases, in an iterative way, we estimated new individual spectro-photometric distances for the two components in the system from spectral type estimations based on the global spectral type, the $M_I$-spectral type relation, the individual $I$ magnitudes, and the distance modulus. These updated distances are given in Table~\ref{table.observedstars}.
 Given the low number of close binaries not resolved in our survey ($\sim$\,10\,\%, see Sect.~\ref{undetectedbins}), we do not expect many additional unresolved components.

The individual spectral types are listed in Table~\ref{table.rhotheta_magmasses}. \rm{SpT} column indicates the combined spectral type of the system from which individual spectral types were derived. In the \rm{SpT$_1$} and \rm{SpT$_2$} columns, the spectral types indicated with capital ``M'' come from the literature, and with lower case ``m'' refer to our estimated spectral types.

In Fig.~\ref{fig.multfrac_spt} we show the dependence of the multiplicity fraction of M dwarfs on the spectral type in our volume-limited sample.
The multiplicity fractions for different spectral subtypes are consistent within the error bars among them, except for M1 stars, for which it is lower.
We compared this distribution with the global multiplicity fraction obtained in the previous section and performed a $\chi^2$ test. Without the M1 contribution, the distribution is consistent with a flat distribution with a confident level of 96\,\%.

In addition, our determined multiplicity fraction has intermediate values between Sun-like (44\,\%--65\,\%) and very low mass stars and brown dwarfs (9\,\%--15\,\%). This agrees with the generally accepted decreasing trend of the multiplicity fraction with decreasing mass of the primaries (Table~\ref{table.stellar_multfrac}).


\subsection{Projected physical separation distribution}\label{projphys_dist}

To study the distribution of the binaries in the volume-limited sample, we converted angular separations ($\rho$) into projected physical separations ($s$) by using the small-angle approximation $\tan{\rho} \approx \rho$. Hence, $s\approx \rho d$. The distances $d$ come from parallax or photometry as in Sect.~\ref{subsection.sptdist}.

In Fig.~\ref{fig.semimajor_dist} we show the projected physical separation distribution of the binaries in the volume-limited sample. We also represent the completeness of the volume-limited sample as a function of projected physical separation, and draw the completeness limits with a confidence level of 90\,\%, which correspond to the $s$ interval between 2.6 and 29.5\,au. We estimated these values as those separations at which we are able to detect companions in 90\,\% of the sample.

The projected physical separations of the observed pairs in the volume-limited sample range from 1.4 to 65.6\,au and their distribution peaks at 2.5--7.5\,au. This is consistent with the values of 5--10\,au found by J\'odar et~al. (2013) for M0--M4 dwarfs, and of $\sim$6\,au found by Janson et~al. (2014) for M3--M8 dwarfs and Ward-Duong et~al. (2015) for M0--M6 dwarfs. However, these values are lower than those found for Sun-like stars (Duquennoy \& Mayor 1991; Raghavan et~al. 2010) and more similar to those found for ultracool dwarfs (4--6\, au for M8.0--L0.5, Close et~al. 2003; 2--4\,au for M7.0--L8.0, Bouy et~al. 2003; <3\,au for L dwarfs, Reid et~al. 2008).

Within the physical separation completeness range from 2.6 to 29.5\,au, there are 61 M dwarfs with low-mass companions in our volume-limited sample. This translates into a multiplicity fraction of 14.4\,$\pm$\,2.0\%, which is lower than the fraction derived in Sect.~\ref{sec.multfrac} as a result of the missing systems at larger separations (see Fig.~\ref{fig.semimajor_dist}).

        
                   \begin{figure}
   \centering
   \includegraphics[width=1\hsize]{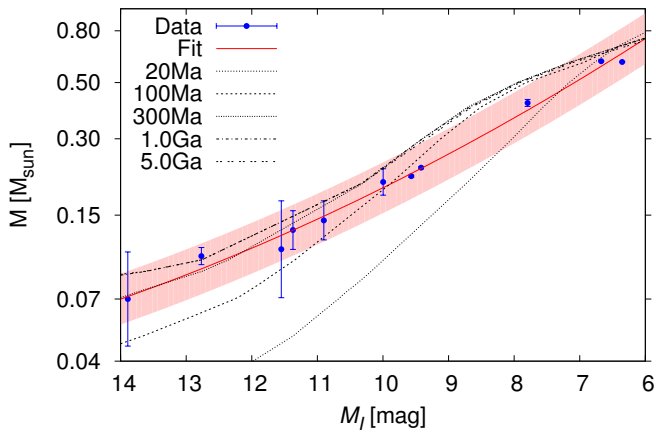}
      \caption{Mass $\mathcal{M}$ vs. Absolute magnitude $M_I$. Blue points represent the dynamical masses and absolute magnitudes taken from the literature. The red solid line and shadowed area represent the best-fit $\pm$\,3\,$\sigma$.
      Different dashed lines display the BT-Settl evolutionary models at 20\,Ma, 100\,Ma, 300\,Ma, 1\,Ga, and 5\,Ga.
      }
         \label{fig.mimass}
   \end{figure}

        \subsection{Masses}

We derived masses from our own mass-luminosity relation in the Johnson-Cousins $I$ band. To our knowledge, there is no published mass-luminosity relation employing this band.
We collected dynamical masses and $I$ -band magnitudes of eleven low-mass stars from different works (Delfosse et~al. 2000; Henry 2004; Reid et~al. 2004; Tokovinin 2008) and obtained an $M_I$-$\mathcal{M}$ relation using a parabolic fit of the form:

\begin{equation} \label{MI_mass}
        \log {\mathcal{M}} = a~M_I^2+b~M_I + c,
\end{equation}

\noindent where $\mathcal{M}$ is the mass, $M_I$ is the absolute $I$ -band magnitude, $a$~=~0.005\,$\pm$\,0.002\,mag$^{-2}$, $b$~=~--0.222\,$\pm$\,0.037\,mag$^{-1}$, and $c$~=~1.035\,$\pm$\,0.180.
This relation is valid for main-sequence stars in the $M_I$ interval between 6 and 14\,mag, which corresponds to $\sim$\,M0--M8 spectral types. 
Figure~\ref{fig.mimass} shows the data taken from the literature, the corresponding best fit, and the comparison with BT-Settl evolutionary models from the Lyon group (Baraffe et~al. 2015).

In some of our detected pairs, one or both components are also spectroscopic binaries (see Table~\ref{table.sbs}). For these we estimated individual masses assuming equally bright components.

For main-sequence stars, the luminosity and effective temperatures are unambiguosly related to the mass, and thus, the relation in Eq.~\ref{MI_mass} is only valid for stars older than $\sim$\,300\,Ma, as inferred from Fig.~\ref{fig.mimass}. For stars younger than 300\,Ma, the mass-luminosity relation strongly depends on the age.
We searched for young stars in our sample by collecting radial velocities from the literature (Caballero et~al. in prep.) and computing UVW Galactocentric space velocities as in Montes et~al. (2001) for 452 of the 490 observed stars (there are 38 stars without radial velocities). Of these, 155 have U and V velocity components inside or near the boundaries that delineate the young-disc population (Montes et~al. 2016).
In total, 42 stars of our 82 multiple systems are candidate members in young stellar kinematic groups. We checked the literature and found that 26 of the 42 are relatively old interloper stars that do not show any youth feature or have been poorly investigated. The remaining 16 stars are confirmed members of stellar kinematic groups or the young-disc population. Their associations and ages are listed in Table~\ref{table.mov_groups}.

Since $I-J$ colours of young stars and field stars do not show significant differences (see Bihain et~al. 2010; Pe\~na-Ram\'irez et~al. 2016), we applied the colour-spectral-type relation from Kirkpatrick et~al. (1994) to derive the individual $I$ magnitudes of these stars as explained in Sect.~\ref{subsection.sptdist}.
We considered Castor, Ursa Majoris, and young-disc members old enough to be main-sequence stars, and thus, to apply our mass-$M_I$ relation with confidence.
For these calculations, we assumed the ages given in Table~\ref{table.mov_groups}.
These stars appear in italics in Table~\ref{table.rhotheta_magmasses}.
The candidate pair to IC 2391 does not have a parallactic distance. Hence, we estimated its mass from the $I-J$ colours and the BT-Settl evolutionary models from the Lyon group (Baraffe et~al. 2015).
We also applied these models to derive masses from the individual $I$ magnitudes and parallactic distances for $\beta$ Pic, Columba/Carina and Local Association members.

\begin{table*}
        \centering
        \caption {Target members of stellar kinematic groups.}
        \label{table.mov_groups}
        \begin{tabular}{l l l l c l}
        \hline \hline
        \noalign{\smallskip}
Karmn                   &       Name            &       Moving  &       Ref.\tablefootmark{a}   &       Assumed         &       Ref.\tablefootmark{b}           \\
                                &                               &       group                   &                       &       age [Ma]    &               \\
        \noalign{\smallskip}
        \hline
        \noalign{\smallskip}
        \noalign{\smallskip}    
J01221+221      &       G 034--023      &       Young disc      &       Abe14   &       $\geq$\,300             &       This work\\
        \noalign{\smallskip}
J04153--076     &       $o$\textsuperscript{02} Eri C   &       $\beta$ Pic     &       AF15    &       $\sim$\,20      &       Bell15  \\
                \noalign{\smallskip}
J05019+099      &       LP 476--207     &       $\beta$ Pic     &       AF15    &       $\sim$\,20              &       Bell15\\
        \noalign{\smallskip}
J05068--215E    &       BD--21 1074 A   &       $\beta$ Pic     &       AF15    &       $\sim$\,20      &               Bell15\\        
        \noalign{\smallskip}
J05068--215W    &       BD--21 1074 BC  &       $\beta$ Pic     &       AF15    &       $\sim$\,20      &               Bell15\\
        \noalign{\smallskip}
J05103+488      &       G 096--021 AB   &       IC 2391?        &       This work&   $\sim$\,50      &       Barr04\\
        \noalign{\smallskip}
J10028+484      &       G 195--055      &       Local Association?      &       This work&   $\sim$\,100     & Bas96 \\
        \noalign{\smallskip}
J10196+198      &       BD+20 2465      &       Castor  &       Cab10   &       $\geq$\,300     &       Barr98, Mam13 \\        
        \noalign{\smallskip}
J12123+544S     &       BD+55 1519 A    &       UMa     &       Mon01   &       $\geq$\,300     &       Gia79,SM93\\    
        \noalign{\smallskip}
J12123+544N     &       BD+55 1519 B    &       UMa     &       Mon01   &       $\geq$\,300     &       Gia79, SM93\\
        \noalign{\smallskip}
J13317+292      &       DG CVn AB       &       Columba/Carina  &       Ried14  &       $\sim$\,40      &       Bell15  \\
        \noalign{\smallskip}
J18548+109      &       V 1436 Aql B    &       Castor  &       Cab10   &       $\geq$\,300     &       Barr98, Mam13 \\
        \noalign{\smallskip}
J23293+414S     &       G 190--027      &       Local Association       &       Klu14   &$\sim$\,100    & Bas96   \\
        \noalign{\smallskip}
J23293+414N     &       G 190--028      &       Local Association       &       Klu14   &$\sim$\,100    & Bas96   \\
        \noalign{\smallskip}
J23318+199 E    &       EQ Peg Aab      &       Castor  &       Cab10   &       $\geq$\,300     &       Barr98, Mam13  \\
        \noalign{\smallskip}
J23318+199 W    &       EQ Peg Bab      &       Castor  &       Cab10   &       $\geq$\,300     &       Barr98, Mam13 \\
        \noalign{\smallskip}
        \noalign{\smallskip}
        \hline
        \end{tabular}
\tablefoot{
        \tablefoottext{a}{Abe14: Aberasturi et~al. 2014; AF15: Alonso-Floriano et~al. 2015b; Cab10: Caballero 2010; Klu14: Klutsch et~al. 2014; Mon01: Montes et~al. 2001; Ried14: Riedel et~al. 2014.}
        \tablefoottext{b}{Barr04: Barrado y Navascu\'es et~al. 2004; Barr98: Barrado y Navascu\'es 1998; Bas96: Basri et~al. 1996; Bell15: Bell et~al. 2015; Gia79: Giannuzzi 1979; Mam13: Mamajek et~al. 2013; SM93: Soderblom \& Mayor 1993.}
}
\end{table*}

Table~\ref{table.rhotheta_magmasses} lists the inferred $I$ magnitudes (Sect.~\ref{subsection.sptdist}) and mass values of the components of 76 of our systems.
All of the detected companions have absolute magnitudes brighter than 14\,mag, the lowest limit of our empirical mass-magnitude relation, which corresponds to masses close to the hydrogen-burning limit ($\sim$\,0.07\,M$_\odot$).
The only exception is the unconfirmed companion of J04352--161, which has an absolute magnitude fainter than 14\,mag, and we were unable to determine its mass with the method explained before.


        \begin{figure}
   \centering
   \includegraphics[width=\hsize]{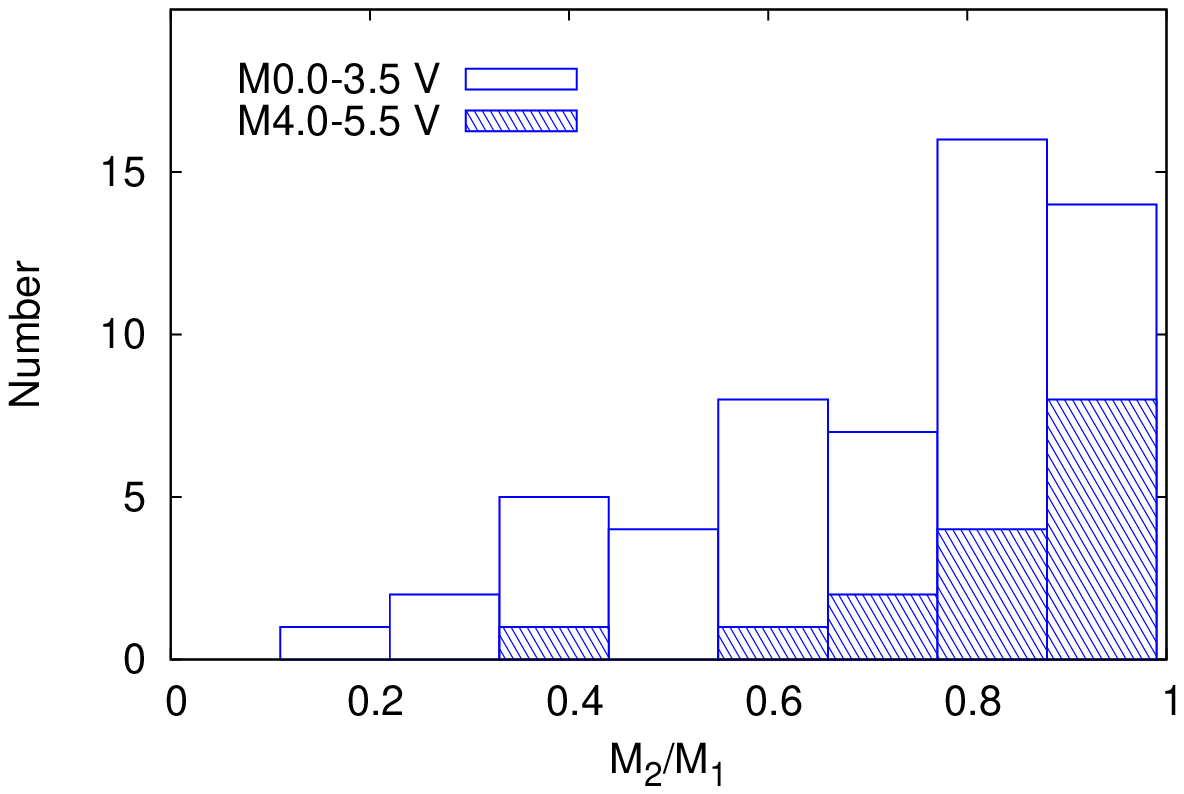}
   \includegraphics[width=1\hsize]{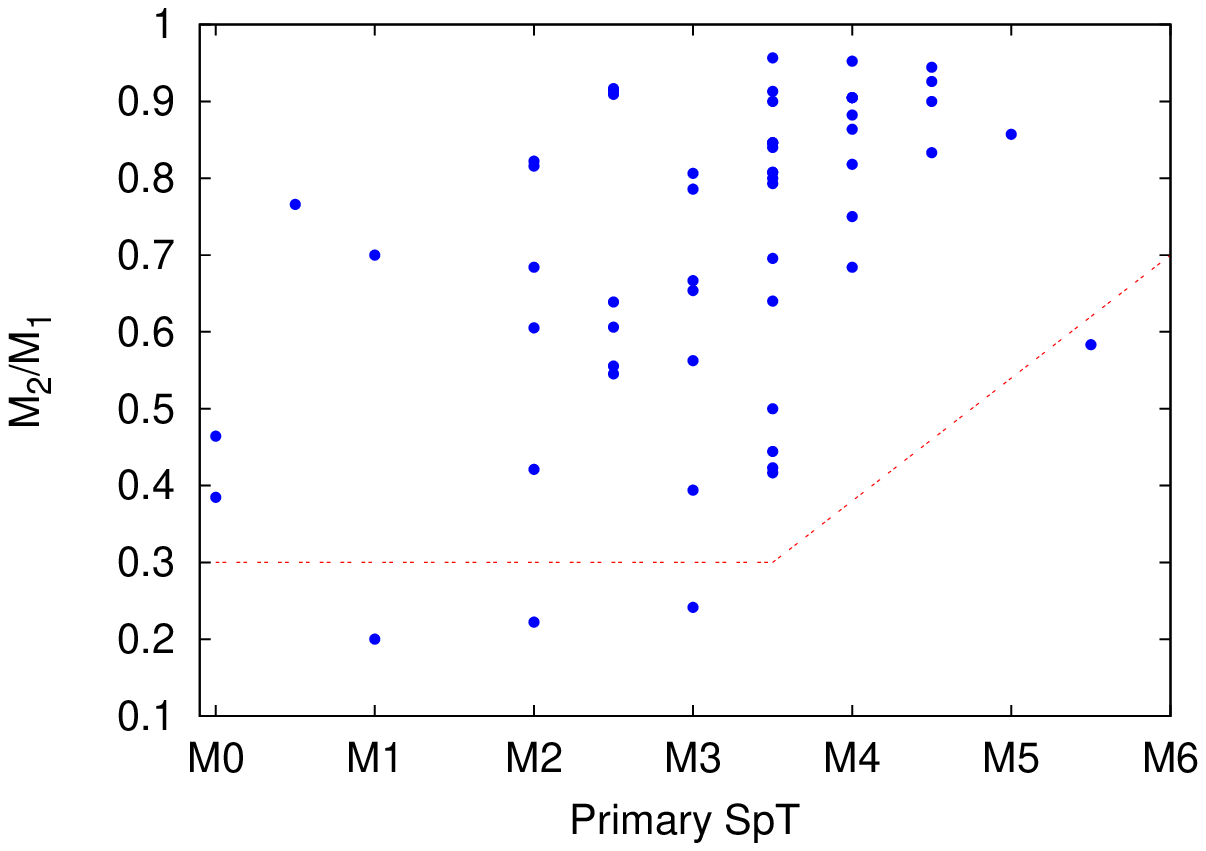}  
      \caption{{\it Top panel}: mass ratio distribution of our binaries. Empty and dashed bars separate the mass ratio distribution of M0.0--M3.5 and M4.0--M5.5 dwarfs.
      {\it Bottom panel}: mass ratio of the pairs vs. spectral type of the primary. The red dashed line represents the mass ratio completeness limits. The standard error of the mean mass ratio is 0.03 and the error bar is $\pm$\,0.5 in spectral type.}
         \label{fig.massratio_dist}
   \end{figure}

                  \begin{figure}
   \centering
   \includegraphics[width=1\hsize]{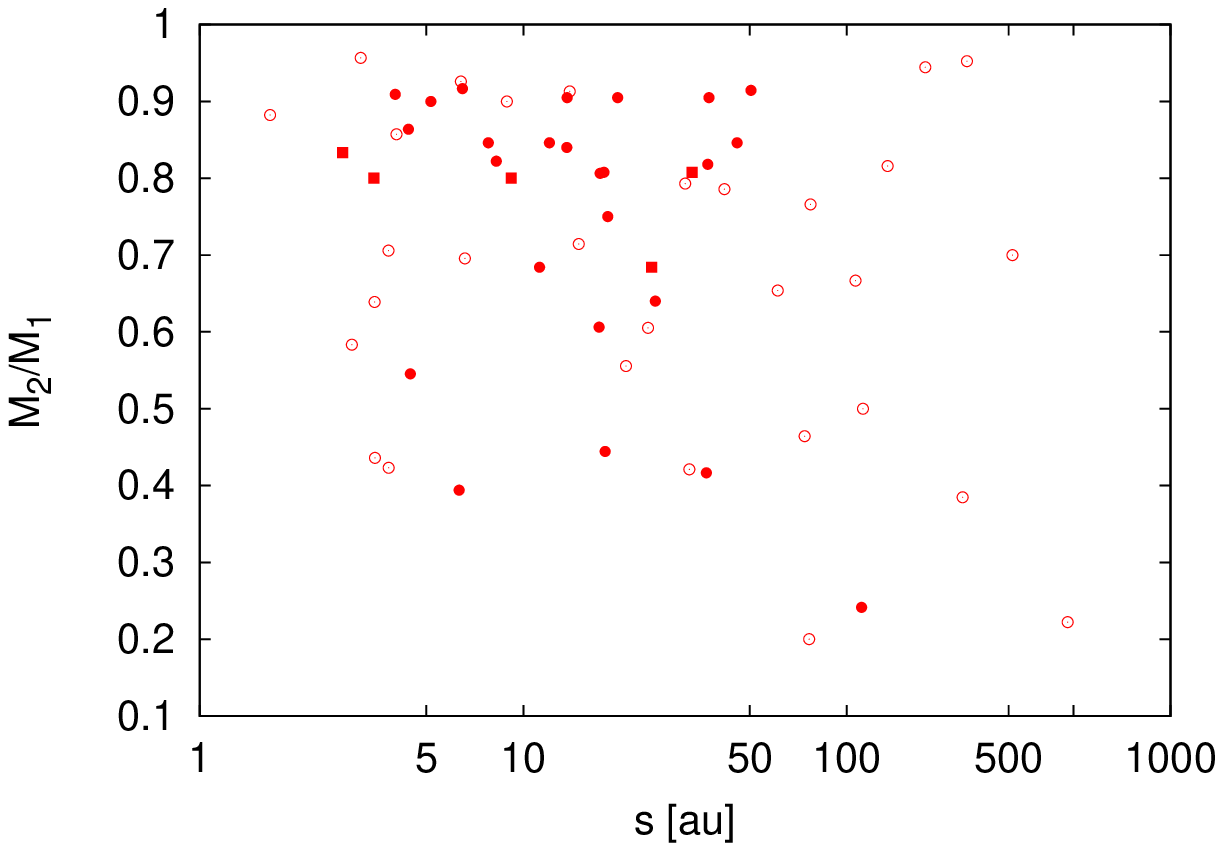}
      \caption{Mass ratio vs. projected physical separation. Colour and symbol code is as in Fig.~\ref{fig.rhothetameasured}.}
         \label{fig.s_massratio}
   \end{figure}

        \subsection{Mass ratios}
        
        The upper panel in Fig.~\ref{fig.massratio_dist} shows the mass ratio ($\mathcal{M}_2/\mathcal{M}_1$) histogram of our binaries in Table~\ref{table.rhotheta_magmasses}.
This global distribution slightly increases towards higher mass ratios and has its maximum above 0.8.
The slightly lower number of equal-mass pairs with mass ratios near unity is not significant and could be related to the effect of the reduction process using the brightest pixel, which artificially sharpens the PSF of the primary with respect to the PSF of the secondary, and may produce a lower flux ratio than expected.
This distribution is also affected by our sensitivity limit. While in spectral types earlier than M3.5 (i.e. more massive stars) our search of companions is complete for mass ratios greater than 0.3, in later spectral types (i.e. less massive stars) the search is complete for mass ratios greater than 0.35--0.60.

Empty and dashed bars represent the mass ratio distributions of M0.0-M3.5 and M4.0-M5.5 primaries, respectively. The distribution of the former shows the same trend as the global distribution, with a peak around 0.8--0.9. For the latter, the distribution increases towards higher ratios. As explained before, this might be due to our observational bias.

The high occurrence of binaries with mass ratios above 0.8 can also be seen in the lower panel in Fig.~\ref{fig.massratio_dist}, which represents the spectral type of the primary versus the mass ratio.
For later spectral types, our detected binaries also tend to have similar masses. This may be due to the lack of sensitivity to lower mass ratios at later spectral types. The distribution differs with the more homogeneous mass ratio distributions observed by Janson et~al. (2012, 2014).
The number of binaries with mass ratios closer to unity (i.e. similar masses) for M0.0-M3.5 contrasts with the relatively low numbers presented in Bergfors et~al. (2010) in this range, but is more similar to their distribution for later M4.0-M5.5 spectral types.

        Figure~\ref{fig.s_massratio} displays the occurrence of mass ratios with physical separations. Pairs with separations shorter than 50\,au tend to have mass ratios over 0.8, while pairs at larger separations present a more homogeneous distribution.
        
        Similar studies also show this observed trend in the relation between separation of the components and mass ratio: near equal-mass pairs (mass ratios $\geq$\,0.8) are found at smaller separations. Moverover, the lower the mass of the primary, the higher the mass ratio and the closer the semi-major axis at which companions are found (J\'odar et~al. 2013: Janson et~al. 2012, 2014).
        The closer distance to the Sun of our sample compared to the samples of Bergfors et~al. (2010) and Janson et~al. (2014), who investigated the mass ratio at larger separations, may explain the difference with our results in the mass ratio distribution.
However, Monte Carlo simulations of Sun-like stars and M-dwarf surveys from Duquennoy \& Mayor (1991) and Raghavan et~al. (2010), and Fischer \& Marcy (1992) and Janson et~al. (2012), respectively, suggest that the mass ratio distributions could be independent of the separation and dynamical evolution (Reggiani \& Meyer 2011, 2013).

  
        \subsection{Periods and orbital motion}
        
We derived periods for 70 systems with Kepler's third law, the masses of the components, and the maximum projected physical separations (Sect.~\ref{projphys_dist}). Since these measures are a lower limit estimate to the semi-major axis, the periods given in Table~\ref{table.rhotheta_magmasses} should be also considered as a lower limit.
        
In total, 26 systems have periods shorter than 50\,a, of which 13 are known bound systems, 10 are newly discovered binaries, and three are the unconfirmed pairs J01221+221, J07349+147, and J10028+484.

Of the 26 systems, we consider four triple systems here: J05078+179, J08082+211, and J16554--083S, which are formed by a spectroscopic binary plus a third resolved component, and J23293+414S, for which we resolved the three components of the system.  In addition, the ``triples'' J08082+211 and J16654--083S belong to a hierarchical quadruple and quintuple system, respectively, with the fourth and fifth components outside the field of view of FastCam (Sect.~\ref{sec.wide}).
 
Several systems were observed repeatedly during the programme, which allowed us to perform a multi-epoch analysis. 
Some of them showed appreciable variation of angular separation and position angle in different epochs of our data. When these variations were larger than 3\,$\sigma$ with respect to constant values of $\rho$ and $\theta$ and were consistent with an orbital trajectory, we considered that the orbital motion of the pair was detected.
Because of the large uncertainties, the variations of $\rho$ and $\theta$ of the pairs J05333+448, J08066+558, and J20407+199 lie within 3\,$\sigma$ and therefore they do not fulfil our criterion, but they show appreciable variations that are probably related to the orbital motion.
However, the time baseline is not long enough to provide a precise estimate of the orbital parameters of the systems.

\begin{figure}[]
   \centering
   \includegraphics[width=1\hsize]{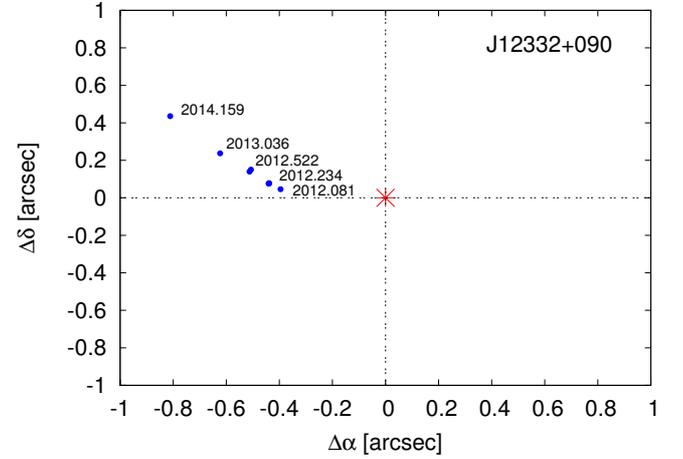}
      \caption{Orbital variation of the pair J12332+090 from our FastCam data. The asterisk marks the position of the primary. Five of the eight epochs are labelled.}
         \label{fig.orbitalmotion}
   \end{figure}

Table~\ref{table.orbital} lists these 16 systems, of which 13 are new. We tabulate the WDS discoverer code of the previously known pairs, the number of used epochs, the time interval between the first and last measured epoch, and the estimated periods. 
We show an example of one of these binaries (J12332+090) in Fig.~\ref{fig.orbitalmotion}.

        \begin{table}[]
        \centering
        \caption{Systems with measurable orbital motion.}
        \label{table.orbital}
        \begin{tabular}{lcccc}
        \hline\hline
        \noalign{\smallskip}
Karmn   &       WDS     &Epochs &       $\Delta t$      &       $P$\\
        &               &       &       [a]     &       [a]\\
        \noalign{\smallskip}

        \hline
        \noalign{\smallskip}
J02518+294      &       ...     &       3       &       4.2     &       130\\ 
J05068--215W&   DON\,93 &       3       &       1.2     &       62\\ 
J05078+179      &       ...     &       2       &       1.1     &       50\\ 
J05333+448      &       BH\,76  &       6       &       2.3     &       8.3\\
J06400+285      &       ...     &       3       &       1.0     &       20\\
J08066+558      &       ...     &       4       &       3.2     &       26\\
J08082+211      &       ...     &       2       &       3.8     &       33\\
J08595+537      &       ...     &       2       &       1.1     &       19\\
J11355+389      &       ...     &       5       &       3.0     &       31\\ 
J11521+039      &       ...     &       2       &       1.1     &       15\\
J12332+090      &       REU\,1  &       8       &       2.1     &       16\\
J13180+022      &       ...     &       4       &       3.1     &       73\\
J14210+275      &       ...     &       2       &       3.0     &       97\\
J16487+106      &       ...     &       3       &       1.1     &       10\\
J17530+169      &       ...     &       5       &       3.0     &       110\\
J20407+199      &       RAO\,23 &       2       &       2.8     &       8.4     \\
J21518+136      &       ...     &       3       &       3.0     &       66\\ 

        \noalign{\smallskip}
        \hline
        \end{tabular}
        \end{table}


        \subsection{Known close and spectroscopic binaries (not detected in our search)}\label{undetectedbins}

In the observed sample there were also previously known pairs
that we were unable to resolve because of the small separation of the components ($\rho \lesssim$ 0.2\,arcsec) and/or the faintness of the companion. These pairs are listed in Table~\ref{table.notresolvedbins}. In addition, there were also previously known spectroscopic binaries, taken into account for the period estimation of our detected binaries in Table~\ref{table.rhotheta_magmasses}. They are listed in Table~\ref{table.sbs}. 

The close multiplicity fraction of 19.5\,$\pm$\,2.3\,\% given in Sect.~\ref{sec.multfrac} is a lower limit of the total multiplicity fraction of M dwarfs, since it only includes physical companions in the interval of angular separations between 0.2 and 5.0\,arcsec. Although studies of spectroscopic binaries and very close binaries ($\rho <$ 0.2\,arcsec) are not complete, we know from the literature that we are missing 47 very close additional binaries in this range in our volume-limited sample (e.g. Delfosse et~al. 2013; Sch\"ofer et~al. 2015; Tokovinin et~al. 2015). This number is consistent with the fractional incidence of eclipsing binaries obtained from surveys like {\it Kepler} (Shan et~al. 2015), and increases the given binary fraction by 11\,\%. Hence the multiplicity fraction at separations smaller than 5\,arcsec would be at least $\sim$\,30\,\%.

        \begin{table*}
        \centering
        \caption{Astrometric properties of previously known imaging companions at $\rho < 5$\,arcsec not resolved or detected in our data.}
        \label{table.notresolvedbins}
        \begin{tabular}{lccccccc}
        \hline\hline
        \noalign{\smallskip}
Karmn   &       WDS     &       Discoverer      &       $\rho$          &       $\theta$        &        Epoch   &       Ref.~\tablefootmark{a}  &       $\Delta$mag (band)              \\
        &                       &               code    &       [arcsec]        &       [deg]           &               [a]     &                       &       [mag]                           \\
        \noalign{\smallskip}

        \hline
        \noalign{\smallskip}
J00088+208      &       00089+2050      &       BEU\,1  &       0.133  &      271.9   &       2012.02         & Jan14a        &       1.59 ($i'$)          \\      
J05085--181     &       05086--1810     &       WSI\,72 &       0.07  &      44.4    &       2011.04 &       WD15    &       0.1 ($K_{s}$)         \\ 
J04311+589      &       ...             &       ...             & 0.07    &               &       1965.702        &       Str77   &       0.5 ($V$)\\
J06523--051     &       06523--0510     &       WSI\,125  &     0.18    &       149.6   &       2010.068        &        Mas01&  0.5     (o)                     \\

J07307+481      &       ...     &       ...     &       0.054   &       ...     &       1960.60 &       Harr81  &       ...     \\      
J09177+462      &       09177+4612      &       JNN\,68 &       0.204   &       37.5    &       2011.073&       Bow15           &       0.102 ($K_{s}$)               \\      
J10513+361      &       10513+3607      &       BWL\,26 &       0.206   &       119.60          & 2012.357        &       Bow15   &       3.3 ($H$)               \\      
J12290+417      &       12290+4144      &       BWL\,31 &       0.0503  &      255.5   &       2011.469        &       Bow15   &       0.647 ($H$)                   \\      
J16241+483      &       16240+4822      &       HEN\,1  &       0.1387  &       295.4   & 2006.62 & Mar07 &       2.781 ($H_{\rm cont}$)          \\      
J16354+350~\tablefootmark {b}   &       16355+3501              &       BWL\,44         &       0.092   &       25.62   &       2011.469        &       Bow15   &       0.406 ($H$)   \\
J17177+116~\tablefootmark {c}   &       ...     &...    &       ...     & ...     &       1977    & Chr78 &       ...     \\
J18387--144     &       18387--1429     &       HDS\,2641 &     0.107   &       358     &       1991    &       DN00    &       0.04 ($Hp$)          \\      
J19122+028      &       19121+0254      &       AST\,1  &       0.16    &       319.7   &       2007.36 &       WD15    &       0.80 ($H$)           \\      
J20298+096~\tablefootmark {d}   &       20298+0941      &       AST\,2  &       0.160   &       89.1    &       2012.66 & Jan14a  &       2.72 ($z'$)     \\      
J20433+553      &       20433+5521      &       LLO\,1  &       0.854   &       20.2    &       2007.66 &       Ire08   &       5.06 ({\it H})       \\      
J21013+332      &       21013+3314      &       JNN\,288        &       0.142   &       34.0            &       2012.01 & Jan14a  &       1.07 ($i'$)     \\      
J21160+298E     &       21161+2951      &       BWL\,56 &       0.0543  &      354.6   &       2011.47 & Bow15 &       0.37 ({\it H})         \\      
J21313--097     &       21313--0947     &       BLA\,9  &       0.16    &       128.2   &       2005.33 &       WD15            &       1.12 ($H$)   \\      
J23174+196      &       23175+1937      &       BEU\,23 &       0.145   &       220.2   &       2012.65 &       Jan14b  &       1.17 ($J$)      \\      
        \noalign{\smallskip}
        \hline
        \end{tabular}
        \tablefoot{
        \tablefoottext{a}{Bow15: Bowler et~al. 2015; Chr78: Christy 1978; DN00: Dommanget \& Nys 2000; Harr81: Harrington et~al. 1981; Ire08: Ireland et~al. 2008; Jan14a: Janson et~al. 2014a; Jan14b: Janson et~al. 2014b; Mar07: Martinache et~al. 2007; Mas01: Mason et~al. 2001; Str77: Strand 1977; Tok15: Tokovinin et~al. 2015; WD15: Ward-Duong et~al. 2015}             
                \tablefoottext{b}{The BWL\,44 companion at 2.2\,arcsec is optical.}
                \tablefoottext{c}{Astrometric perturbation with a 10\,a period estimation in Chr78.} 
                \tablefoottext{d}{Spectroscopic binary identified by Benedict et~al. 2000 and resolved by Janson et~al. 2014a for the first time.}
        }
        \end{table*}

        \begin{table}[ht]
        \centering
        \caption{Known spectroscopic binaries in the observed sample.}
        \label{table.sbs}
        \begin{tabular}{lcc}
        \hline\hline
        \noalign{\smallskip}
Karmn   &       Spectroscopic   &       Ref.~\tablefootmark {a} \\
        &                binarity       &               \\
        \noalign{\smallskip}

        \hline
        \noalign{\smallskip}
J03346--048     &       SB3     & Llam14                \\      
J03526+170      &       SB2     & Bon13                 \\     
J04252+080S     &       SB2     & Llam14                \\      
J04352--161     &       SB2     &       RB09    \\
J04488+100      &       SB2     & Jeff16                \\
J05019+099      &       SB2     &       Del99   \\      
J05032+213      &       SB2     & Jeff16        \\
J05078+179      &       SB1     & Jeff16        \\
J05342+103S~\tablefootmark {b}  &       SB      &Rein12 \\
J05466+441      &       SB2     & Jeff16                \\
J07418+050      &       SB2     & Llam14                \\                           
J08082+211      &       SB2     & Shk10  \\
J09011+019      &       SB2     & Jeff16        \\
J09120+279      &       SB2     & Jeff16                \\
J09143+526      &       SB1     & Jeff16                        \\
J11036+136      &       SB1     & Jeff16                \\
J12142+006      &       SB2     & Bon13         \\     
J12191+318      &       SB2     & Jeff16                \\
J12290+417      &       SB2     & Jeff16                \\
J14171+088      &       SB2     & Jeff16                \\
J14368+583      &       SB2     & Jeff16                \\
J15191--127     &       SB      & Bon13  \\
J16255+260      &       SB2     & Jeff16                \\
J16487+106      &       SB2     & Jeff16                \\
J16554-083S     &       SB      & Pett84        \\
J18411+247S     &       SB2     & GR96  \\
J19354+377      &       SB1     & Jeff16                                \\
J20433+553      &       SB2     &       Ire08           \\
J20445+089N~\tablefootmark {c}& SB1     & Jeff16                        \\
J23096--019     &       SB2     & Jeff16                \\
J23174+382      &       SB2     & Jeff16                \\
J23318+199E&    SB1     & Del99                 \\
J23318+199W&    SB1     & Del99                 \\
J23573--129W    &       SB2     & Jeff16                \\
        \noalign{\smallskip}
        \hline
        \end{tabular}
\tablefoot{
\tablefoottext{a}{Bon13: Bonfils et~al. 2013; Del99: Delfosse et~al. 1999; GR96: Gizis \& Reid 1996; Ire08: Ireland et~al. 2008; Jeff16: Jeffers et~al. in prep; Llam14: Llamas 2014; Pett84: Pettersen et~al. 1984; RB09: Reiners \& Basri 2009; Rein12: Reiners et~al. 2012; Shk10: Shkolnik et~al. 2010}
\tablefoottext{b}{From the spectral types and magnitude differences of the components, we infer that the spectroscopic binary is the B companion.}
\tablefoottext{c}{Equal-brightness close binary previously suggested by Cort\'es-Contreras et~al. 2014.}
}
        \end{table}


        \subsection{Known companions at separations larger than 5\,arcsec}\label{sec.wide}

Many of our FastCam stars have stellar or substellar companions outside the field of view of the instrument or at angular separations larger than the 5.0\,arcsec cut-off defined for statistical purposes.
We compiled the multiplicity information of all of them using our observations and the WDS catalogue. Of the wide binaries present in the WDS catalogue, nearly 60\,\% come from the Luyten Double Star Catalogue (Luyten 1997) and the Lowell Proper Motion Survey (Giclas et~al. 1971).
In Table~\ref{table.knownwide}, we list for each wide system the WDS discoverer code, names, spectral types, and angular separation.
    
        As a summary, of the 490 observed stars,  50 are M-dwarf primaries with M-type wide companions, four with white dwarf companions, one with an L-dwarf, and three with a T-dwarf secondary. 
In addition, 11 M secondaries have F\,(2), G\,(3), K\,(4) or white dwarf\,(3) primaries.
Five tertiary M dwarfs are in triple systems involving K+M\,(1), G+K\,(1), K+DA\,(1) or K+K\,(2) primaries.

In our volume-limited sample are 25 M dwarf primaries with wide M, L, or T dwarf secondaries at separations larger than 5\,arcsec.
Although our search at wide separations is not complete, since we carried out a compilation from different studies in the literature, we estimated an increment in the multiplicity fraction of 6\,\% (25 systems out of 425 M dwarfs in our volume-limited sample), which added to the percentage estimated for pairs at separations closer than 0.2\,arcsec, and spectroscopic binaries would translate into a minimum multiplicity fraction at all separations of $\sim$\,36\,\%.


\section{Summary}

We obtained high-resolution images in the $I$ band of 490 M dwarfs of the CARMENES input catalogue (Carmencita) with the lucky imaging instrument FastCam at the 1.5\,m Telescopio Carlos S\'anchez.

Among the 490 observed M dwarfs, we identified 80 physically bound companions in 76 systems, of which 30 are presented here for the first time, plus six unconfirmed companions. 
For all of them, we measured angular separations, position angles, and $I$-band magnitude differences.
From the $\Delta I$ differences, together with 2MASS photometry, spectral type, and colour-magnitude relations for field M dwarfs, we estimated individual $I$-band magnitudes and spectral types of each component.
We also derived individual masses $\mathcal M$ and estimated orbital periods for these pairs from our own $\mathcal{M}$-$M_I$ relation. For these calculations, we used parallactic distances. When not available, we derived spectro-photometric distances from our determined $M_J$-spectral type relation.

For our observed sample, we determined a multiplicity fraction of 16.7\,$\pm$\,2.0\,\%. However, our sample has a strong selection bias because we discarded M stars with previously known companions at separations smaller than 5\,arcsec. To obtain an unbiased multiplicity fraction, we built a volume-limited sample of Carmencita stars observed with FastCam and similar high-resolution imagers. It contains 425 M0--5 dwarfs and is complete up to 86\,\% within 14\,pc. For this sample, we derived a multiplicity fraction of 19.5\,$\pm$\,2.3\,\% in the completeness range of angular separations between 0.2 and 5.0\,arcsec, which agrees with previously reported values (Leinert et~al. 1997; Janson et~al. 2012, 2014; J\'odar et~al. 2013; Ward-Duong et~al. 2015).
The multiplicity fraction is consistent with a flat distribution from M0\,V to M5\,V within Poissonian error bars, and has intermediate values between solar-type stars and very low mass stars and brown dwarfs in accordance with the decreasing tendency observed towards lower masses.

The distribution of the number of pairs as a function of projected physical separation has a maximum between 2.5 and 7.5\,au and decreases at wider separations. 
The pairs with projected physical separations smaller than 50\,au tend to have mass ratios higher than 0.8, while for larger separations this distribution is more uniform.

We estimated that 26 of our systems have orbital periods shorter than 50\,a, of which 10 are newly discovered systems. 
In 17 of them, we were able to detect orbital variations within our own multi-epoch measurements. These systems are especially interesting for future astrometric follow-up for determining their orbital solutions and measuring dynamical masses.

For our volume-limited sample, we also collected from the literature the physically bound companions at separations closer than 0.2\,arcsec and larger than 5\,arcsec, and unresolved spectroscopic binaries.
The addition of these systems may increase the multiplicity fraction derived in this work to at least 36\,\%, a value consistent with the 42\,$\pm$\,9\,\% obtained by Fischer \& Marcy (1992).
Nevertheless, the sample is not complete at separations beyond the completeness limit of our survey (0.2--5.0\,arcsec) and, hence, this value must only be considered as a rough estimation.

Finally, we provided a complete sample of multiple M dwarfs useful for studying the effect of low-mass stellar multiplicity on planet formation with the help of CARMENES and other near-infrared high-resolution spectrographs.


\begin{acknowledgements}

We thank A. P\'erez-Garrido for the provision and support of the FastCam reduction software and X. Bonfils for the supply of radial velocity measurements from the ESO HARPS GTO Program ID 072.C-0488.
MCC thanks L. Peralta de Arriba, V. Pereira and H.\,M. Tabernero for their assistance and valuable conversations.
This article is based on observations made with the Telescopio Carlos S\'anchez operated on the island of Tenerife jointly by the Instituto de Astrof\'isica de Canarias and the Universidad de La Laguna in the Spanish Observatorio del Teide.
This research made use of SIMBAD, operated at Centre de Donn\'ees astronomiques de Strasbourg (France), the NASA's Astrophysics Data System, the Washington Double Star catalogue (WDS) maintained at the U.S. Naval Observatory, and the Image Reduction and Analysis Facility (IRAF), distributed by the National Optical Astronomy Observatory and operated by the Association of Universities for Research in Astronomy (AURA) under a cooperative agreement with the National Science Foundation.
CARMENES is funded by the German Max-Planck-Gesellshaft (MPG), the Spanish Consejo Superior de Investigaciones Cient\'ificas (CSIC), the European Union through FEDER/ERF funds, and the members of the CARMENES Consortium (Max-Planck Institut f\"ur Astronomie, Instituto de Astrof\'isica de Andaluc\'ia, Landessternwarte K\"onigstuhl, Institut de Ci\`ences de l'Espai, Institut f\"ur Astrophysik G\"ottingen, Universidad Complutense de Madrid, Th\"uringer Landessternwarte Tautenburg, Instituto de Astrof\'isica de Canarias, Hamburger Sternwarte, Centro de Astrobiolog\'ia, and the Centro Astron\'omico Hispano-Alem\'an), with additional contributions by the Spanish Ministry of Economy, the state of Niedersachsen, the German Science Foundation (DFG), and by the Junta de Andaluc\'ia. 
Financial support was also provided by the Junta de Andaluc\'ia, and the Spanish Ministries of Science and Innovation and of Economy and Competitiveness, under grants 2011-FQM-7363,
AP2009-0187, AYA2014-54348-C3-01/02/03-R, AYA2015-69350-C3-2-P, ESP2013-48391-C4-1-R, and ESP2014-57495-C2-2-R.

 \end{acknowledgements}



\newpage

\begin{appendix}

\onecolumn

\section{Long tables}

\centering
	\begin{landscape}
	\setlength\LTleft{0pt}
	\setlength\LTright{0pt}
\begin{longtable}{cllcccllcllc}    
	\label{table.observedstars}\\
	\caption{Log of observed stars.}\\
	\hline \hline
	\noalign{\smallskip}

No.	&	Karmn	&	Name	&	$\alpha$	&	$\delta$	&	{\it J}	&	SpT	&	Ref. \tablefootmark {a}	& {\it d}	& Ref. \tablefootmark {b}		&	Observation	&	N $\times$ $t_{\rm exp}$	\\
 	&    		& 		&	(J2000)		& 	 (J2000)	&	[mag]	& 		&		& [pc]		&	&	date		&	[s]		\\
	\noalign{\smallskip}

	\hline
	\noalign{\smallskip}
	\endfirsthead
	\caption{Log of observed stars (continued).}\\
	\hline\hline	
	\noalign{\smallskip}

No.	&	Karmn	&	Name	&	$\alpha$	&	$\delta$	&	{\it J}	&	SpT	&	Ref.\tablefootmark {a}		& {\it d}	& Ref. \tablefootmark {b}	&	Observation	&	N $\times$ $t_{\rm exp}$	\\
 	&    		& 		&	(J2000)		& 	 (J2000)	&	[mag]	& 		&		& [pc]		&	&	date		&	[s]		\\
	\noalign{\smallskip}
	\hline
	\noalign{\smallskip}

	\endhead
	\hline
	\noalign{\smallskip}

	\endfoot
															
1	& J00067-075	& GJ 1002		& 00:06:43.26	& --07:32:14.7	& 8.323	& M5.5 V	& PMSU	& 4.82	& Wein16	& 23 Oct 2011	& 10 $\times$ 50	\\
2	& J00088+208	& LP 404--033		& 00:08:53.92	& +20:50:25.2	& 8.870	& M4.5 V	& PMSU	& 14.8	& Dit14	& 24 Oct 2011	& 10 $\times$ 50	\\
3	& J00154-161	& GJ 1005 AB		& 00:15:27.99	& --16:08:00.9	& 7.215	& M4.0 V	& PMSU	& 6.0	& HT98	& 10 Jul 2012   & 10 $\times$ 50	\\
4	& J00158+135	& GJ 12			& 00:15:49.20	& +13:33:21.9	& 8.619	& M3.0 V	& PMSU	& 10.0	& Dit14	& 16 Sep 2012	& 10 $\times$ 50	\\	
5	& J00169+051	& GJ 1007		& 00:16:56.29	& +05:07:26.1	& 9.398	& M4.5 V	& PMSU	& 17.57	& vAl95	& 25 Oct 2011	& 10 $\times$ 50	\\	
6	& J00169+200	& G 131--047		& 00:16:56.78	& +20:03:55.1	& 9.681	& M3.5 V	& PMSU	& 24.5	& This work& 14 Jan 2013	& 10 $\times$ 50	\\	 
	&		&			&		&		&	&		&	&	&	& 17 Nov 2015	& 10 $\times$ 50	\\	
7	& J00182+102	& GJ 16			& 00:18:16.59	& +10:12:10.1	& 7.564	& M1.5 V	& PMSU	& 16.34	& HIP2	& 17 Nov 2015	& 10 $\times$ 50	\\
\end{longtable}
\tablefoot{
\tablefoottext{a}{AF15: Alonso-Floriano et~al. 2015a; Cru03: Cruz et~al. 2003; Dea12: Deacon et~al. 2012; Gig10: Gigoyan et~al. 2010; Gra03: Gray et~al. 2003; Gra06: Gray et~al. 2006; Klutsch et~al. priv. comm.; Koe10: Koen et~al. 2010; Lep13: L\'epine et~al. 2013; Mon01: Montes et~al. 2001; New14: Newton et~al. 2014; PMSU: (Palomar/Michigan State University survey catalogue of nearby stars) Reid et~al. 1995., Hawley et~al. 1996, Gizis et~al. 2002; Ria06: Riaz et~al. 2006; Sch05: Scholz et~al. 2005; Sim15: Simon-D\'iaz et~al. 2015; ZS04: Zuckerman \& Song 2014}
\tablefoottext{b}{Ben00: Benedict et~al. 2000; Cru03: Cruz et~al. 2003; Daw05: Dawson et~al. 2005; Dit14: Dittmann et~al. 2014; GC09: Gatewood \& Coban 2009; HD80: Harrington \& Dahn 1980; Hen06: Henry et~al. 2006; HT98: Hershey \& Taff 1998; HIP2: van Leeuwen 2007; Ire08: Ireland et~al. 2008; Jen52: Jenkins 1952; Jen63: Jenkins 1963; Jen09: Jenkins et~al. 2009; Lep13: L\'epine et~al. 2013; New14: Newton et~al. 2014; PMSU: (Palomar/Michigan State University survey catalogue of nearby stars) Reid et~al. 1995., Hawley et~al. 1996, Gizis et~al. 2002; Rei02: Reid et~al. 2002; Ried10: Ria06: Riaz et~al. 2006; Reidel et~al. 2010; Ried14: Riedel et~al. 2014; Sub09: Subasavage et~al. 2009; vAl95: van Altena et~al. 1995; Wein16: Weinberger et~al. 2016. An ``a'' or ``b'' preceding the reference indicates that no measure for this component was found but we used instead the measure of the A or B companion, respectively.}
}

	\end{landscape}

\centering
	\begin{longtable}{l c c c c c c c}
	\label{table.ads}\\
	\caption{ADS standard stars.}\\
	\hline \hline
	\noalign{\smallskip}
	
	ADS 	& \multicolumn{2}{c}{Literature}		& Epoch	&	Ref.~\tablefootmark{a}	&	\multicolumn{2}{c}{This work} 	&	Epoch		\\
		&	$\rho$\,[arcsec]	&	$\theta$\,[deg]	&	&			&	$\rho$\,[arcsec]	&	$\theta$\,[deg]	&	\\

	\noalign{\smallskip}

	\hline
	\noalign{\smallskip}
	\endfirsthead
	\caption{ADS standard stars (continued).}\\
	\hline\hline	
	\noalign{\smallskip}

	ADS 	& \multicolumn{2}{c}{Literature}		& Epoch	&	Ref.~\tablefootmark{a}	&	\multicolumn{2}{c}{This work} 	&	Epoch		\\
		&	$\rho$\,[arcsec]	&	$\theta$\,[deg]	&	&			&	$\rho$\,[arcsec]	&	$\theta$\,[deg]	&	\\

	\noalign{\smallskip}
	\hline
	\noalign{\smallskip}
	\endhead
	\hline
	\noalign{\smallskip}

	\endfoot
	
	2999		&	3.55		&	222.2			&	J2011.662		&	Mas12			&		3.634 $\pm$ 0.016		&	222.56 $\pm$ 0.22	&	J2012.078	\\
			&			&				&				&				&	3.621 $\pm$ 0.010		&	222.43 $\pm$ 0.16	&	J2012.081	\\	
			&			&				&				&				&	3.611 $\pm$ 0.010		&	222.19 $\pm$ 0.17	&	J2013.034	\\

	\noalign{\smallskip}
	3297		&	2.93		&	276.8		&	J2011.064	&	Mas12		&3.073 $\pm$ 0.012		&	276.98 $\pm$ 0.23	&	J2012.078	\\
			&			&				&				&				&	3.070 $\pm$ 0.009		&	276.92 $\pm$ 0.18	&	J2012.081	\\	
			&			&				&				&				&	3.067 $\pm$ 0.013		&	277.05 $\pm$ 0.17	&	J2012.229	\\
			&			&				&				&				&	3.069 $\pm$ 0.008		&	277.03 $\pm$ 0.16	&	J2012.231	\\
				&			&				&				&				&	3.073 $\pm$ 0.010		&	276.74 $\pm$ 0.16	&	J2013.034	\\
	\noalign{\smallskip}	
	3853	&	3.041			&	74.2			&	J2008.867	&	Har11		&3.072 $\pm$ 0.012		&	74.80 $\pm$ 0.30	&	J2012.078	\\		
    			&			&					&				&				&	3.085 $\pm$ 0.013		&	74.85 $\pm$ 0.13	&	J2012.081	\\		
			&			&					&				&				&	3.070 $\pm$ 0.009		&	74.92 $\pm$ 0.15	&	J2012.231	\\
			&			&					&				&				&	3.079 $\pm$ 0.023		&	74.66 $\pm$ 0.12	&	J2013.034	\\

\noalign{\smallskip}	
	4241 (A--B)~\tablefootmark{b}	&	0.2526 $\pm$ 0.0010	&	80.1 $\pm$ 0.4	&	J2013.710	&	Sim15	&	0.243 $\pm$ 0.008	&	85.89 $\pm$ 2.46		&		J2012.234	\\
	4241 (AB--C)	&	11.24	&	239.2		&	J2013.168		&	Schl13	&	11.439 $\pm$ 0.071	&	238.25 $\pm$ 0.24		&		J2012.234	\\
	\noalign{\smallskip}	
	7878	&	3.743 $\pm$ 0.021	&	161.7 $\pm$ 0.3		&	J2008.036	&	Pru09		&	3.768 $\pm$ 0.011		&	161.81 $\pm$ 0.13	&	J2012.234	\\
			&						&						&				&				&	3.769 $\pm$ 0.010		&	161.74 $\pm$ 0.18	&	J2014.936	\\

\noalign{\smallskip}	
	8105	&	3.678 $\pm$ 0.035	&	96.8 $\pm$ 0.3		&	J2010.392	&	Pru12		&	3.665 $\pm$ 0.012		&	97.64 $\pm$ 0.13	&	J2012.234	\\
     			&						&						&				&				&	3.673 $\pm$ 0.012		&	97.54 $\pm$ 0.15 	&	J2014.936	\\
\noalign{\smallskip}	
	8220	&	3.495			&	208.14			&	J2013.314	&	Ben14		&	3.506 $\pm$ 0.009		&	207.84 $\pm$ 0.14	&	J2012.234	\\
       			&					&					&				&				&	3.510 $\pm$ 0.010		&	207.91 $\pm$ 0.18	&	J2014.936	\\
            
	\noalign{\smallskip}	
	9168	&	2.199 $\pm$ 0.013	&	254.9 $\pm$ 0.3		&	J2011.491	&	Sca13		&	2.200 $\pm$ 0.006		&	255.52 $\pm$ 0.19	&	J2012.231	\\	
 			&						&						&				&				&	2.203 $\pm$ 0.008		&	255.19 $\pm$ 0.16	&	J2012.527	\\	   
 			&						&						&				&				&	2.209 $\pm$ 0.011		&	255.31 $\pm$ 0.13	&	J2013.034	\\   
  			&						&						&				&				&	2.197 $\pm$ 0.011		&	255.42 $\pm$ 0.25	&	J2013.036	\\          
			&						&						&				&				&	2.182 $\pm$ 0.008		&	255.62 $\pm$ 0.20	&	J2014.159	\\
    		&						&						&				&				&   2.188 $\pm$ 0.010		&	255.52 $\pm$ 0.23	&	J2014.162	\\	 
 	\noalign{\smallskip}		
	9312		&	3.029 $\pm$ 0.016			&	37.6 $\pm$ 0.3		&	J2011.496	&	Sca13		&	3.049 $\pm$ 0.010		&	37.95 $\pm$ 0.19	&	J2012.229	\\
 			&						&						&				&				&	3.046 $\pm$ 0.007		&	37.88 $\pm$ 0.19	&	J2012.231	\\	
			&						&						&				&				&	3.052 $\pm$ 0.011		&	37.81 $\pm$ 0.13	&	J2012.234	\\	
 			&						&						&				&				&	3.048 $\pm$ 0.010		&	37.96 $\pm$ 0.19	&	J2012.527	\\	
 			&						&						&				&				&	3.047 $\pm$ 0.009		&	37.83 $\pm$ 0.18	&	J2013.036	\\	
 			&						&						&				&				&	3.042 $\pm$ 0.012		&	37.54 $\pm$ 0.18	&	J2014.159	\\
			&						&						&				&				&	3.052 $\pm$ 0.016		&	37.55 $\pm$ 0.19	&	J2014.162	\\			
			&						&						&				&				&	3.035 $\pm$ 0.018		&	37.57 $\pm$ 0.14	&	J2014.164	\\

\noalign{\smallskip}
	9461		&	4.130 $\pm$ 0.015			&	276.95 $\pm$ 0.16	&	J2008.608	&	Des11		&	4.111 $\pm$ 0.012		&	277.05 $\pm$ 0.16	&	J2012.229	\\ 
			&						&						&				&				&	4.118 $\pm$ 0.017		&	277.01 $\pm$ 0.16	&	J2012.231	\\			
 			&						&						&				&				&	4.104 $\pm$ 0.013		&	276.76 $\pm$ 0.16	&	J2012.527	\\	
 			&						&						&				&				&	4.092 $\pm$ 0.014		&	276.76 $\pm$ 0.14	&	J2013.036	\\			
  			&						&						&				&				&	4.084 $\pm$ 0.012		&	276.95 $\pm$ 0.16	&	J2014.159	\\
			&						&						&				&				&	4.094 $\pm$ 0.012		&	277.10 $\pm$ 0.22	&	J2014.162	\\			
			&						&						&				&				&	4.094 $\pm$ 0.012		&	277.06 $\pm$ 0.13	&	J2014.164	\\

\noalign{\smallskip}	
	14708	&	2.48	&	28.6	&	J2010.693	&	Thor11	&	2.453 $\pm$ 0.015	&	28.15 $\pm$ 0.16	&	J2011.812	\\	
   			&			&			&				&			&	2.454 $\pm$ 0.014	&	28.45 $\pm$ 0.13	&	J2012.708	\\	
			&			&			&				&			&	2.460 $\pm$ 0.014	&	28.57 $\pm$ 0.16	&	J2012.710	\\
			&			&			&				&			&	2.451 $\pm$ 0.011	&	28.06 $\pm$ 0.12	&	J2014.386	\\

\noalign{\smallskip}	
	14733	&	2.601 $\pm$ 0.021	&	123.9 $\pm$ 0.4	&	J2011.876		&	Sca13	&	2.662 $\pm$ 0.012	&	125.11 $\pm$ 0.13	&	J2011.812	\\	
   			&						&					&					&			&	2.631 $\pm$ 0.013	&	125.19 $\pm$ 0.16	&	J2012.708	\\
 			&						&					&					&			&	2.647 $\pm$ 0.015	&	125.54 $\pm$ 0.16	&	J2012.710	\\
			&						&					&					&			&	2.638 $\pm$ 0.012	&	125.60 $\pm$ 0.14	&	J2014.386	\\

\noalign{\smallskip}	
	14878	&	6.85				&	113.9			&	J2012.595		&	Mas13	&	6.897 $\pm$ 0.012	&	113.60 $\pm$ 0.12	&	J2011.807	\\
  			&						&					&					&			&	6.842 $\pm$ 0.015	&	113.57 $\pm$ 0.14	&	J2012.708	\\
			&						&					&					&			&	6.864 $\pm$ 0.013	&	113.98 $\pm$ 0.13	&	J2012.710	\\
			&						&					&					&			&	6.842 $\pm$ 0.012	&	113.96 $\pm$ 0.13	&	J2014.386	\\

\noalign{\smallskip}	
	15935	&	3.593	&	224.19	&	J1990.809	&	Ha4	&	3.899 $\pm$  0.014	& 	225.11 $\pm$ 0.16	&	J2011.807	\\
		&		&		&			&		&		3.906 $\pm$ 0.011	&	225.29 $\pm$ 0.14	&	J2012.710	\\

\noalign{\smallskip}	
	16389 		&	3.893 $\pm$ 0.019	&	13.9 $\pm$ 0.3		&	J2009.957	&	Sca11	&	3.970 $\pm$ 0.012		&	13.85 $\pm$ 0.15	&	J2011.810	\\
    			&				&				&			&		&	3.999 $\pm$ 0.011		&	14.20 $\pm$ 0.17	&	J2012.522	\\ 
				&				&				&			&		&	3.973 $\pm$ 0.011		&	14.55 $\pm$ 0.18	&	J2012.524	\\			
				&				&				&			&		&	3.983 $\pm$ 0.013		&	14.47 $\pm$ 0.16	&	J2012.527	\\			
				&				&				&			&		&	3.982 $\pm$ 0.013		&	14.45 $\pm$ 0.14	&	J2012.708	\\	

\noalign{\smallskip}	
	16496 	&	2.664 $\pm$ 0.027	&	164.9 $\pm$ 0.5		&	J2011.860	&	Sca13		&	2.702 $\pm$ 0.010		&	165.07 $\pm$ 0.20	&	J2012.522	\\
		 	&					&					&				&				&	2.681 $\pm$ 0.012		&	165.59 $\pm$ 0.13	&	J2012.708	\\

	\noalign{\smallskip}	
	16982	&	2.590 $\pm$ 0.028 &	210.8 $\pm$ 0.7	&	J2011.874	&	Sca13	&	2.626 $\pm$ 0.015	&	211.37 $\pm$ 0.14	&	J2011.810	\\
		 	&					&					&				&			&	2.625 $\pm$ 0.014	&	211.81 $\pm$ 0.13	&	J2012.708	\\
		 	&					&					&				&			&	2.637 $\pm$ 0.014	&	211.96 $\pm$ 0.13	&	J2012.710	\\

\noalign{\smallskip}	
	17140 	&	3.094 $\pm$ 0.030		&	325.4 $\pm$ 0.5		&	J2011.874	&	Sca13		&	3.197 $\pm$ 0.012		&	325.81 $\pm$ 0.18	&	J2012.522	\\
			&				&				&			&			&	3.155 $\pm$ 0.007		&	326.24 $\pm$ 0.14	&	J2012.524	\\
			&				&				&			&			&	3.149 $\pm$ 0.015		&	326.08 $\pm$ 0.12	&	J2012.708	\\
\noalign{\smallskip}

\end{longtable}

\tablefoot{
\tablefoottext{a}{Ben14: Benavides 2014; Des11: Desidera et~al.2011; Har11: Hartkopf et~al. 2011; Mas12: Mason et~al. 2012; Mas13: Mason et~al. 2013; Sca11: Scardia et~al. 2011 ; Sca13: Scardia et~al. 2013;  Schl13: Schlimmer 2013; Sim15: Simon--D\'iaz et~al. 2015; Thor11: Thorel et~al. 2011.}
\tablefoottext{b}{$\Delta I$ = 0.35\,mag for A-B and $\Delta I$ = 5.39\,mag for AB-C (this work).}
	}

\centering    
\begin{longtable}{llccl}
\label{table.visuals}\\
\caption{List of observed stars with confirmed visual (unbound) companions.}\\
\hline \hline
\noalign{\smallskip}

Karmn	&	No.&	$\rho$		&	$\theta$	&	Observation	\\
		&		&	[arcsec]	&	[deg]		&	date		\\
\noalign{\smallskip}
\hline
\noalign{\smallskip}
\endfirsthead
\caption{List of observed stars with confirmed visual (unbound) companions (continued).}\\
\hline\hline	
\noalign{\smallskip}

Karmn	&	No.	&	$\rho$		&	$\theta$	&	Observation	\\
		&		&	[arcsec]	&	[deg]		&	date		\\
\noalign{\smallskip}
\hline
\noalign{\smallskip}
\endhead
\hline
\noalign{\smallskip}

\endfoot

J00234+243	&\#1	&	2.4			&	257			&		J2011.812	\\	
J00413+558	&\#1	&	13.1			&	129			&		J2011.812	\\	      
J01033+623	&\#1	&	10.6			&	13			&		J2011.807	\\	
J01593+585	&	\#1	&	7.5			&	7			&		J2011.807	\\    
		& \#1	&	7.8			&	5			&		J2013.034	\\	
		& \#2	&	4.2			&	93			&		J2013.034	\\	
J02565+554W&	 \#1	&	7.6			&	241			&		J2012.710	\\	
		&	 \#1&	7.7			&	242			&		J2013.034	\\	
J03102+059	&\#1	&	10.6			&	174			&		J2012.710	\\	
J05106+297	&\#1	&	3.1			&	308			&		J2014.162	\\	
J06000+027	&	 \#1&	7.8			&	313			&		J2011.807	\\	
		&	 \#1&	7.9			&	315			&		J2012.231	\\	
		&	 \#1&	8.5			&	311			&		J2014.936	\\	
J06361+116	&	 \#1&	8.7			&	114			&		J2011.807	\\	
	&		 \#1&	8.7			&	112			&		J2012.231	\\	
J06422+035	&\#1	&	8.2			&	34			&		J2011.807	\\	
J06490+371	&	\#1&	12.6			&	302  		&		J2011.810	\\		
J07033+346	&\#1	&	9.3			&	357			&		J2011.810	\\	
J07227+306	&\#1	&	9.0			&	29			&		J2012.234	\\	
J07518+055	&\#1	&	9.7			&	340			&		J2012.081	\\	
J07581+072	&	\#1	&	11.1			&	260			&		J2012.081	\\
			& \#2	&	11.8			&	315			&		J2012.081	\\
J08105-138	&	 \#1&	9.6			&	55			&		J2012.234	\\	
		&	 \#1&	9.74			&	55			&		J2013.036	\\	      
J08126-215	&\#1	&	12.7			&	329			&		J2012.234	\\	
J08428+095	&	 \#1&	6.2			&	145			&		J2012.081	\\	
	&		 \#1&	5.62			&	144			&		J2013.034	\\	
J11289+101	&\#1	&	14.0			&	217			&		J2012.229	\\	
J11420+147	&\#1	&	6.4			&	93		&		J2012.234	\\		
J13165+278	&\#1	&	6.2			&	1			&		J2012.231	\\	
J16462+164	&	 \#1&	12.8			&	35			&		J2012.229	\\	
		&	 \#1&	13.0			&	36			&		J2012.524	\\	
		&	 \#1&	13.2			&	35			&		J2012.710	\\	
		&	 \#1&	14.6			&	36			&		J2015.284	\\	
J16509+224	&\#1	&	9.1			&	318			&		J2012.524	\\	
J17177-118	&\#1	&	5.3			&	149			&		J2012.231	\\	
J17321+504	&\#1	&	12.5			&	23			&		J2012.234	\\	
J17425-166	&\#1	&	8.9			&	123			&		J2012.234	\\	
			&\#2	&	3.2			&	276			&		J2012.234	\\
J17460+246	&\#1	&	6.4			&	357			&		J2012.524	\\	
J17530+169	&	 \#1&	8.9			&	31			&		J2012.522	\\	      
		&	 \#1&	8.9			&	31			&		J2012.527	\\	      
		&	 \#1&	9.0			&	31			&		J2012.708	\\	
J18240+016&	\#1	&	7.0			&	216			&		J2012.524	\\	
J18264+113&	\#1	&	4.2			&	10			&		J2012.522	\\	
J18387-144	& \#1	&	4.8			&	261			&		J2012.522	\\	
		&	 \#2	&	9.6			&	300			&		J2012.522	\\
		& \#3	&	7.3			&	61			&		J2012.522	\\	
J18427+139	&		 \#1&	3.6			&	178			&		J2012.524	\\    
		&	 \#1&	3.7			&	177			&		J2012.710	\\	
J18480-145	&	\#1&	7.6			&	234			&		J2012.522	\\	
J19098+176	&	\#1&	6.7			&	326			&		J2012.522	\\	
J19124+355	&	\#1&	12.4			&	125			&		J2012.527	\\	
J19220+070	& \#1	&	4.7			& 	73			& 		J2015.284	\\	
J19463+320	&\#1	&	11.8			&	186			&		J2012.524	\\	
		& \#1	&	11.7			&	186			&		J2012.710	\\	
		& \#1	&	11.0			&	194			&		J2015.284	\\	
		&\#2	&	8.5			&	25			&		J2012.524	\\
		& \#2	&	8.7			&	27			&		J2012.710	\\	
		&\#3	&	8.8			&	117			&		J2012.524	\\
		& \#3	&	9.0			&	117			&		J2012.710	\\	
J19539+444W&		 \#1&	14.9			&	116			&		J2012.527	\\	
		&	 \#1&	14.9			&	112			&		J2012.710	\\	
J20298+096	&	\#1&	9.2			&	201			&		J2012.522	\\	
J20347+033	&	\#1&	10.5			&	300			&		J2012.527	\\	
J21013+332	&	\#1&	10.1			&	63			&		J2012.524	\\	
J21518+136	&	\#1&	11.4			&	124			&		J2012.524	\\	
J22252+594	&\#1	&	11.8			&	34			&		J2012.708	\\	
J23577+233	&	\#1&	4.6			&	76			&		J2012.524	\\	

\noalign{\smallskip}
\end{longtable}

	\centering
	\begin{longtable}{l c c c c c c}
	\label{table.rhotheta}\\
	\caption{Astrometric properties of the physically and likely bound binaries in the sample.}\\
	\hline \hline
	\noalign{\smallskip}
	
Karmn	&	WDS		&	$\rho$		&	$\theta$	&	$\Delta I$	&	Epoch	&	GSC 1.2 \tablefootmark{a}	\\
	& 			&	[arcsec]	&	[deg]		& 	 [mag]		&		& 		\\
	\noalign{\smallskip}

	\hline
	\noalign{\smallskip}
	\multicolumn{7}{c}{Physically bound systems}	\\
		\noalign{\smallskip}
	\hline
		\noalign{\smallskip}
	\endfirsthead

	\caption{Astrometric properties of the physically and likely bound binaries in the sample (continued).}\\
	\hline\hline	
	\noalign{\smallskip}
Karmn	&	WDS	&	$\rho$		&	$\theta$	&	$\Delta I$	&	Epoch	&	GSC 1.2	\\
	& 		&	[arcsec]	&	[deg]		& 	 [mag]		&		& 		\\
	\noalign{\smallskip}
	\hline
	\noalign{\smallskip}

	\endhead
	\hline
	\noalign{\smallskip}

	\endfoot

J00154-161	&	00155-1608 HEI299	&	0.275 $\pm$ 0.013	&	204.96 $\pm$ 3.10	&	0.58 $\pm$ 0.08		&	J2012.522	&	F	\\			
J00413+558	&	00415+5550 GIC13	&	10.833 $\pm$ 0.047	&	68.25 $\pm$ 0.28	&	2.92 $\pm$ 0.13		&	J2011.812	&	F	\\			
J02518+294	&	New			&	0.626 $\pm$ 0.008	&	249.44 $\pm$ 1.29	&	0.53 $\pm$ 0.10		&	J2011.810	&	F	\\			
		&				& 	0.798 $\pm$ 0.011	&	247.05 $\pm$ 0.11	& 	0.50 $\pm$ 0.07		&	J2015.875	&		\\
		&				& 	0.803 $\pm$ 0.015	&	246.30 $\pm$ 1.54	& 	...			&	J2016.015	&		\\             
J02565+554W	&	02565+5526 LDS5401	&	16.664 $\pm$ 0.051	&	21.15 $\pm$ 0.17	&	0.97 $\pm$ 0.04		&	J2012.710	&	F	\\			
J02591+366 \tablefootmark{b}	&	New	&	1.949 $\pm$ 0.010	&	4.35 $\pm$ 0.28  	&	3.10 $\pm$ 0.13		&	J2013.036	&	F	\\	
		&				&	1.900 $\pm$ 0.054	&	358.58 $\pm$ 0.83	&	...			&	J2016.094	&		\\   
J03574-011	&	03575-0110 BU543	&	10.972 $\pm$ 0.027	&	14.61 $\pm$ 0.16	&	2.29 $\pm$ 0.01		&	J2013.034	&	F	\\			     	  
J04153-076	&	04153-0739 STF518	&	8.598 $\pm$ 0.030	&	152.13 $\pm$ 0.14	&	1.62 $\pm$ 0.04		&	J2011.812	&	...	\\			
J04311+589	&	04312+5858 STI2051	&	9.885 $\pm$ 0.039	&	59.64 $\pm$ 0.23	&	3.64 $\pm$ 0.05		&	J2011.812	&	...	\\						  
J05019+099	&	05020+0959 HDS654	&	1.354 $\pm$ 0.012	&	148.90 $\pm$ 0.34	&	0.67 $\pm$ 0.15		&	J2014.159	&	F	\\			
J05034+531	&	Ward-Duong et~al. 2015	&	5.600 $\pm$ 0.021	&	279.58 $\pm$ 0.19	&	5.83 $\pm$ 0.07		&	J2014.162	&	F	\\			
		&				&	5.614 $\pm$ 0.006	&	279.27 $\pm$ 0.14	&	5.53 $\pm$ 0.15		&	J2015.875	&		\\
J05068-215E(A)	&	05069-2135 DON93	&	8.499 $\pm$ 0.053	&	307.94 $\pm$ 0.33	&	0.81 $\pm$ 0.04		&	J2011.812	&	T	\\			
		&				&	8.459 $\pm$ 0.024	&	308.08 $\pm$ 0.17	&	0.82 $\pm$ 0.03 	&	J2012.710	&		\\			
		&				&	8.462 $\pm$ 0.019	&	308.04 $\pm$ 0.17	&	0.69 $\pm$ 0.02		&	J2013.036	&		\\			
J05068-215W(BC)	&	05069-2135 DON93	&	0.767 $\pm$ 0.004	&	128.97 $\pm$ 1.58	&	0.51 $\pm$ 0.07		&	J2011.812	&	T	\\
		&				&	0.803 $\pm$ 0.029	&	126.86 $\pm$ 1.52	&	0.61 $\pm$ 0.04	    	&	J2012.710	&		\\
		&				&	0.811 $\pm$ 0.010	&	125.54 $\pm$ 0.60	&	0.64 $\pm$ 0.02		&	J2013.036	&		\\
J05078+179	&	New			&	0.540 $\pm$ 0.003	&	286.30 $\pm$ 0.66	&	1.94 $\pm$ 0.02		&	J2013.034	&	T	\\	
		&				&	0.348 $\pm$ 0.010	&	288.28 $\pm$ 1.34	&	1.99 $\pm$ 0.05		&	J2014.164	&		\\	
		&				&	Unresolved		&	...			&	...			&	J2015.284	&		\\	
J05103+488	&	05104+4850 HEI321	&	1.993 $\pm$ 0.051	&	113.00 $\pm$ 0.87	&	0.90 $\pm$ 0.02		&	J2013.034	&	F	\\	
		&				&	1.989 $\pm$ 0.009	&	112.78 $\pm$ 0.17	&	0.50 $\pm$ 0.10		&	J2015.284	&		\\	
J05333+448~\tablefootmark{c}	&05333+449 BH76	&	0.205 $\pm$ 0.008	&	38.55 $\pm$ 2.80	&	0.24 $\pm$ 0.05		&	J2011.810	&	 T	\\	 
		&				&	0.192 $\pm$ 0.011	&	36.47 $\pm$ 0.19	&	...			&	J2012.229	&		\\	
		&				&	0.183 $\pm$ 0.004	&	35.75 $\pm$ 1.18	&	0.22 $\pm$ 0.04		&	J2012.231	&		\\	
		&				&	0.164 $\pm$ 0.066	&	33.95 $\pm$ 5.65	&	...			&	J2012.710	&		\\	
		&				&	0.144 $\pm$ 0.060	&	35.24 $\pm$ 5.29	&	...			&	J2013.034	&		\\	
		&				&	0.140 $\pm$ 0.070	&	34.13 $\pm$ 5.10	&	...			&	J2014.159	&		\\	
J05342+103N	&	05342+1019 LDS6189	&	5.061 $\pm$ 0.012	&	188.72 $\pm$ 0.14	&	1.02 $\pm$ 0.03		&	J2013.034	&	F	\\	
J05466+441	& New				&	3.729 $\pm$ 0.016	&	222.80 $\pm$ 0.20	&	5.71 $\pm$ 0.04		&	J2011.810	&	F	\\	
		&	 			&	3.712 $\pm$ 0.022	&	222.70 $\pm$ 0.41	&	6.21 $\pm$ 0.08		&	J2014.162	&		\\	
J06212+442	&	Ansdell et~al. 2015	&	1.222 $\pm$ 0.060	&	204.6 $\pm$ 3.69	&	...			&	J2013.036	&	F	\\	
		&				&	1.319 $\pm$ 0.019	&	203.67 $\pm$ 0.22	&	2.85 $\pm$ 0.13		&	J2015.875	&		\\
J06400+285	&	New			&	0.258 $\pm$ 0.004	&	80.35 $\pm$ 1.76	&	...			&	J2012.081	&	F	\\	 
		&				&	0.251 $\pm$ 0.001	&	87.30 $\pm$ 0.18	&	0.30 $\pm$ 0.06		&	J2012.710	&		\\	
		&				&	0.269 $\pm$ 0.003	&	91.35 $\pm$ 1.65	&	0.23 $\pm$ 0.05		&	J2013.034	&		\\	 
		&				&	Unresolved		&	...			&	...			&	J2014.159	&		\\	
		&				&	Unresolved		&	...			&	...			&	J2015.284	&		\\	
J07395+334	&	07397+3328 LDS3755	&	13.675 $\pm$ 0.066	&	48.97 $\pm$  0.32	&	4.81 $\pm$ 0.08		&	J2012.081	&	F	\\	
J08066+558	&	New			&	0.274 $\pm$ 0.004	&	257.27 $\pm$ 1.23	&	0.53 $\pm$ 0.11		&	J2012.081	&	T	\\	
		&				&	0.248 $\pm$ 0.004	&	253.75 $\pm$ 0.88	&	0.55 $\pm$ 0.08		&	J2013.034	&		\\	
		&				&	0.202 $\pm$ 0.011	&	245.61 $\pm$ 3.40	&	0.54 $\pm$ 0.03		&	J2014.159	&		\\	
		&				&	0.174 $\pm$ 0.060	&	236.52 $\pm$ 8.48	&	...			&	J2015.284	&		\\	
J08082+211~\tablefootmark{d}	&08082+2106 COU91&	10.617 $\pm$ 0.030	&	144.74 $\pm$ 0.18	&	1.03 $\pm$ 0.03		&	J2012.081	&	F	\\	
        	&				&	10.633 $\pm$ 0.095	&	144.10 $\pm$ 1.52	&	0.88 $\pm$ 0.01		&	J2015.875	&		\\	
  		&	New			&	0.329 $\pm$ 0.020	& 	74.26 $\pm$ 0.74	&	2.31 $\pm$ 0.05		&	J2012.081	&	...	\\	
     		&				&	0.580 $\pm$ 0.045	&	36.48 $\pm$ 1.57	&	3.22  $\pm$0.05 	&	J2015.875	&		\\	
J08105-138	&	08107-1348 JOD4		&	0.927 $\pm$ 0.006	&	283.74 $\pm$ 0.34	&	1.80 $\pm$ 0.05		&	J2012.234	&	T	\\	
		&				&	0.913 $\pm$ 0.009	&	283.11 $\pm$ 0.87	&	1.77 $\pm$ 0.05		&	J2013.036	&		\\	
J08595+537	&	New			&	0.229 $\pm$ 0.011	&	219.10 $\pm$ 4.19	&	0.29 $\pm$ 0.06		&	J2014.162	&	F	\\	
		&				&	0.301 $\pm$ 0.012	&	223.00 $\pm$ 1.49	&	0.26 $\pm$ 0.05		&	J2015.281	&		\\	
J09011+019	&	New			&	2.954 $\pm$ 0.014	&	154.77 $\pm$ 0.17	&	4.68 $\pm$ 0.06		&	J2012.229	&	F	\\	
		&       			&	2.994 $\pm$ 0.016	&	155.08 $\pm$ 0.23	&	4.58 $\pm$ 0.07		&	J2013.034	&		\\	
		&				&	3.096 $\pm$ 0.065	&	156.22	$\pm$ 0.40	&	4.40 $\pm$ 0.14		&	J2015.875	&		\\
J10151+314	&	New			&	1.799 $\pm$ 0.005	&	306.58 $\pm$ 0.20	&	0.69 $\pm$ 0.05		&	J2012.229	&	F	\\	
		&				&	1.806 $\pm$ 0.007	&	306.43 $\pm$ 0.26	&	0.69 $\pm$ 0.09		&	J2013.034	&		\\	
		&				&	1.822 $\pm$ 0.012	&	306.41 $\pm$ 0.58	&	0.57 $\pm$ 0.10		&	J2014.159	&		\\	
		&				&	1.815 $\pm$ 0.010	&	306.40 $\pm$ 0.23	&	0.66 $\pm$ 0.08		&	J2015.281	&		\\	
J10196+198	&	10200+1950 BAG32	&	0.195 $\pm$ 0.061	&	23.81 $\pm$ 3.68	&	2.00 $\pm$ 0.2			&	J2012.078	&	T	\\	
J10260+504W	&	10261+5029 LDS1241	&	14.413 $\pm$ 0.038	&	25.83 $\pm$ 0.18	&	0.24 $\pm$ 0.04		&	J2012.229	&	T	\\	
		&				&	14.410 $\pm$ 0.045	&	25.58 $\pm$ 0.25	&	0.24 $\pm$ 0.09		&	J2014.164	&		\\	
J10379+127	&	New			&	0.848 $\pm$ 0.010	&	333.88 $\pm$ 1.08	&	0.47 $\pm$ 0.09		&	J2014.162	&	F	\\	
		&				&	0.792 $\pm$ 0.013	&	336.27 $\pm$ 0.94	&	0.49 $\pm$ 0.10		&	J2015.281	&		\\	
		&				&	0.806 $\pm$ 0.020	&	336.05 $\pm$ 0.83	&	0.58 $\pm$ 0.11		&	J2015.875	&		\\
J10448+324	&	New			&	1.292 $\pm$ 0.025	&	157.62 $\pm$ 0.81	&	0.55 $\pm$ 0.11		&	J2013.036	&	T	\\	 
		&				&	1.267 $\pm$ 0.014	&	159.63 $\pm$ 0.57	&	0.47 $\pm$ 0.08		&	J2015.284	&		\\	
J10546-073	&	New			&	0.793 $\pm$ 0.028	&	60.71 $\pm$ 1.66	&	0.41 $\pm$ 0.08		&	J2014.164	&	F	\\	
		&				&	0.813 $\pm$ 0.039	&	64.75 $\pm$ 2.66	&	0.45 $\pm$ 0.09		&	J2015.281	&		\\	
J11355+389	&	New			&	0.244 $\pm$ 0.020	&	50.84 $\pm$ 1.90	&	0.51 $\pm$ 0.10		&	J2012.231	&	F	\\	
		&				&	0.219 $\pm$ 0.005	&	46.41 $\pm$ 2.14	&	0.63 $\pm$ 0.12		&	J2012.524	&		\\	
		&				&	0.278 $\pm$ 0.004	&	55.49 $\pm$ 2.22	&	0.64 $\pm$ 0.12		&	J2013.034	&		\\	
		&				&	0.281 $\pm$ 0.007	&	61.37 $\pm$ 1.43	&	0.44 $\pm$ 0.08		&	J2014.159	&		\\	
		&				&	0.339 $\pm$ 0.017	&	65.67 $\pm$ 1.96	&	0.55 $\pm$ 0.10		&	J2015.281	&		\\	
J11521+039	&	New			&	0.224 $\pm$ 0.014	&	29.25 $\pm$ 3.75	&	0.51 $\pm$ 0.11		&	J2014.162	&	F	\\	
		&				&	0.337 $\pm$ 0.008	&	26.06 $\pm$ 2.08	&	0.46 $\pm$ 0.10	 	&	J2015.281	&		\\	
J12006-138	&	12007-1348 LDS4166	&	6.839 $\pm$ 0.021	&	234.14 $\pm$ 0.17	&	2.44 $\pm$ 0.03		&	J2012.081	&	F	\\	
		&				&	6.836 $\pm$ 0.023	&	234.05 $\pm$ 0.16	&	2.40 $\pm$ 0.03		&	J2013.034	&		\\	
J12016-122	&	New			&	6.123 $\pm$ 0.013	&	24.42 $\pm$ 0.13	&	5.05 $\pm$ 0.13		&	J2012.234	&	T	\\	
		&				&	6.134 $\pm$	0.012	&	24.41 $\pm$ 0.19	&	5.13 $\pm$ 0.19		&	J2013.034	&		\\	
		&				&	6.109 $\pm$ 0.018	&	24.58 $\pm$ 0.16	&	5.14 $\pm$ 0.17		&	J2015.284	&		\\	
J12123+544S	&	12123+5429 VYS5		&	14.677 $\pm$ 0.044	&	9.91 $\pm$ 0.15 	&	2.86 $\pm$ 0.04		&	J2012.231	&	F	\\	
J12162+508	&	New			&	1.884 $\pm$ 0.014	&	188.92 $\pm$ 0.84	&	0.48 $\pm$ 0.07		&	J2014.162	&	F	\\	
		&				&	...			&	...			&	...			&	J2014.386	&		\\	
		&				&	1.828 $\pm$ 0.010	&	190.85 $\pm$ 0.33	&	0.42 $\pm$ 0.08		&	J2015.281	&		\\	
J12277-032	&	New			&	1.496 $\pm$ 0.018	&	15.84 $\pm$ 0.45	&	1.67 $\pm$ 0.04		&	J2014.162	&	T	\\	
		&				&	1.459 $\pm$ 0.007	&	15.73 $\pm$ 0.50	&	1.75 $\pm$ 0.06		&	J2015.281	&		\\	
J12332+090	&	12335+0901 REU1		&	0.398 $\pm$ 0.005	&	173.37 $\pm$ 0.76	&	0.55 $\pm$ 0.11		&	J2012.081	&	...	\\	
		&				&	0.447 $\pm$ 0.003	&	170.25 $\pm$ 0.46	&	0.46 $\pm$ 0.09		&	J2012.299	&		\\	
		&				&	0.445 $\pm$ 0.003	&	170.043 $\pm$ 0.49	&	0.41 $\pm$ 0.07		&	J2012.231	&		\\	
		&				&	0.446 $\pm$ 0.004	&	169.92 $\pm$ 0.75	&	0.40 $\pm$ 0.08		&	J2012.234	&		\\	
		&				&	0.529 $\pm$ 0.003	&	163.46 $\pm$ 0.35	&	0.44 $\pm$ 0.04		&	J2012.522	&		\\	
		&				&	0.531 $\pm$ 0.005	&	164.74 $\pm$ 0.64	&	0.52 $\pm$ 0.10		&	J2012.524	&		\\	
		&				&	0.667 $\pm$ 0.007	&	159.16 $\pm$ 0.31	&	0.55 $\pm$ 0.06		&	J2013.036	&		\\	
		&				&	0.921 $\pm$ 0.011	&	151.76 $\pm$ 0.37	&	0.46 $\pm$ 0.08		&	J2014.159	&		\\	
J13168+170	&	13169+1701 BU800	&	7.658 $\pm$ 0.024	&	104.89 $\pm$ 0.18	&	2.11 $\pm$ 0.02		&	J2014.159	&	F	\\	
		&				&	7.659 $\pm$ 0.024	&	104.50 $\pm$ 0.18	&	2.04 $\pm$ 0.01		&	J2015.284	&		\\	
J13180+022	&	New			&	0.637 $\pm$ 0.007	&	213.25 $\pm$ 0.51	&	0.78 $\pm$ 0.15		&	J2012.229	&	T	\\	
		&				&	0.646 $\pm$ 0.012	&	217.48 $\pm$ 0.69	&	0.45 $\pm$ 0.08		&	J2013.034	&		\\	
		&				&	0.592 $\pm$ 0.004	&	222.91 $\pm$ 0.37	&	0.55 $\pm$ 0.10		&	J2014.159	&		\\	
		&				&	0.530 $\pm$ 0.055	&	225.78 $\pm$ 2.51	&	0.53 $\pm$ 0.10		&	J2015.281	&		\\	
J13317+292	&	13318+2917 BEU17	&	0.193 $\pm$ 0.066	&	79.42 $\pm$ 11.91	&	1.5 $\pm$ 0.2			&	J2012.522	&	F	\\	
		&				&	0.190 $\pm$ 0.095	&	85.41 $\pm$  3.58 	&	...			&	J2012.527	&		\\	
		&				&	Unresolved		&	...			&	...			&	J2015.281	&		\\	
J13417+582	&	13418+5815 JNN94	&	0.675 $\pm$ 0.020	&	251.26 $\pm$ 1.42	&	0.38 $\pm$ 0.08		&	J2014.164	&	F	\\	
		&				&	0.699 $\pm$ 0.022	&	251.33 $\pm$ 0.56	&	...			&	J2015.281	&		\\	
J13526+144	&	13526+1425 JOD7		&	1.252 $\pm$ 0.008	&	356.75 $\pm$ 0.30	&	1.52 $\pm$ 0.03		&	J2012.078	&	T	\\	
		&				&	1.231 $\pm$ 0.015	&	357.72 $\pm$ 0.34	&	1.64 $\pm$ 0.04		&	J2013.034	&		\\	
		&				&	1.197 $\pm$ 0.016	&	359.55 $\pm$ 0.79	&	1.69 $\pm$ 0.04		&	J2015.284	&		\\	
J14157+594	&	14157+5928 LDS2707	&	5.014 $\pm$ 0.054	&	231.04 $\pm$ 0.59	&	0.46 $\pm$ 0.07		&	J2014.164	&	F	\\	
		&				&	5.021 $\pm$ 0.075	&	230.87 $\pm$ 0.54	&	0.46 $\pm$ 0.09		&	J2014.386	&		\\	
		&				&	5.064 $\pm$ 0.020	&	230.90 $\pm$ 0.19	&	0.76 $\pm$ 0.03		&	J2015.284	&		\\	
J14210+275	&	New			&	0.626 $\pm$ 0.008	&	82.25 $\pm$ 0.93	&	1.82 $\pm$ 0.04		&	J2012.231	&	F	\\	
		&				&	0.683 $\pm$ 0.009	&	84.96 $\pm$ 0.39	&	1.73 $\pm$ 0.05		&	J2015.281	&		\\	
J14279-003S	&	14279-0032 GIC20	&	13.032 $\pm$ 0.007	&	21.96 $\pm$ 0.28	&	0.19 $\pm$ 0.03		&	J2012.229	&	F	\\	
J14331+610	&	New			&	0.899 $\pm$ 0.020	&	255.34 $\pm$ 1.15	&	0.65 $\pm$ 0.13		&	J2014.162	&	F	\\	
		&				&	0.944 $\pm$ 0.005	&	252.63 $\pm$ 0.27	&	0.58 $\pm$ 0.11		&	J2015.281	&		\\	
J15081+623	&	New			&	0.983 $\pm$ 0.022	&	61.15 $\pm$ 1.62	&	0.50 $\pm$ 0.10		&	J2014.162	&	T	\\	
		&				&	0.947 $\pm$ 0.032	&	68.06 $\pm$ 3.30	&	0.40 $\pm$ 0.08		&	J2014.386	&		\\	
		&				&	0.884 $\pm$ 0.024	&	67.63 $\pm$ 0.37	&	0.42 $\pm$ 0.10		&	J2015.284	&		\\	
J15126+457	&	15126+4544 MCT8		&	0.527 $\pm$ 0.080	&	216.82 $\pm$ 2.36	&	0.37 $\pm$ 0.08		&	J2012.231	&	F	\\	
		&				&	0.481 $\pm$ 0.021	&	219.63 $\pm$ 1.32	&	0.36 $\pm$ 0.08		&	J2015.284	&		\\	
J15191-127~\tablefootmark{e}	&	New	&	0.305 $\pm$ 0.064		&	246.58 $\pm$ 	2.22	&	3.4 $\pm$ 0.15		&	J2012.234	&	...	\\
J15400+434N	&	15400+4330 VBS25	&	4.529 $\pm$ 0.028	&	154.66 $\pm$ 0.21	&	1.47 $\pm$ 0.01		&	J2014.162	&	...	\\	
		&				&	4.525 $\pm$ 0.016	&	154.78 $\pm$ 0.12	&	1.59 $\pm$ 0.07		&	J2014.386	&		\\	
J15496+348~\tablefootmark{f}	&15496+3449 BWL41&	0.211 $\pm$ 0.002	&	89.25 $\pm$ 0.42	&	3.00 $\pm$ 0.10		&	J2012.234	&	...	\\	
		&				&	Unresolved		&	...			&	...			&	J2012.522	&		\\	
		&				&	0.226 $\pm$ 0.003	&	91.31 $\pm$ 1.54	&	...			&	J2012.524	&		\\	
		&				&	0.208 $\pm$ 0.013	&	98.68 $\pm$ 1.40	&	...			&	J2012.708	&		\\	
		&				&	Unresolved		&	...			&	...			&	J2014.159	&		\\	
J16487+106	&	New			&	0.212 $\pm$ 0.010	&	20.25 $\pm$ 1.62	&	0.39 $\pm$ 0.08		&	J2014.162	&	F	\\	
		&				&	Unresolved		&	...			&	...			&	J2014.386	&		\\	
		&				&	0.235 $\pm$ 0.012	&	31.97 $\pm$ 2.08	&	0.30 $\pm$ 0.06		&	J2015.281	&		\\	
J16554-083S	&	16555-0820 KUI75	&	0.206 $\pm$ 0.004	&	216.03 $\pm$ 1.02	&	0.43 $\pm$ 0.09		&	J2012.231	&	...	\\	
  J16578+473	&	16579+4722 A1874	&	5.082 $\pm$ 0.020	&	62.27 $\pm$ 0.25	&	2.48 $\pm$ 0.02		&	J2014.386	&	F	\\	
J17340+446	&	New			&	0.544 $\pm$ 0.018	&	143.11 $\pm$ 2.15	&	0.54 $\pm$ 0.11		&	J2014.386	&	T	\\	
		&				&	0.596 $\pm$ 0.009	&	144.97 $\pm$ 0.71	&	0.47 $\pm$ 0.09		&	J2015.281	&		\\	
J17530+169	&	New			&	0.896 $\pm$ 0.006	&	123.69 $\pm$ 0.49	&	0.60 $\pm$ 0.06		&	J2012.231	&	T	\\	
		&				&	0.887 $\pm$ 0.003	&	122.86 $\pm$ 0.24	&	0.66 $\pm$ 0.09		&	J2012.522	&		\\	
		&				&	0.893 $\pm$ 0.004	&	122.56 $\pm$ 0.21	&	0.61 $\pm$ 0.12		&	J2012.527	&		\\	
		&				&	0.871 $\pm$ 0.010	&	122.20 $\pm$ 0.41	&	0.69 $\pm$ 0.11		&	J2012.708	&		\\	
		&				&	0.823 $\pm$ 0.016	&	114.50 $\pm$ 1.37	&	0.68 $\pm$ 0.13		&	J2015.284	&		\\	
J18180+387E	&	18180+3846 GIC151	&	9.940 $\pm$ 0.031	&	277.79 $\pm$ 0.13	&	1.50 $\pm$ 0.03		&	J2012.524	&	...	\\	
J18264+113	&	18264+1121 NI38		&	8.141 $\pm$ 0.030	&	196.27 $\pm$ 0.33	&	5.89 $\pm$ 0.09		&	J2012.522	&	F	\\	
J18411+247N	&	18411+2247 LDS6330	&	4.828 $\pm$ 0.015	&	5.67 $\pm$ 0.15 	&	2.25 $\pm$ 0.02		&	J2012.527	&	F	\\	
J18427+596N	&	18428+5938 STF2398	&	11.712 $\pm$ 0.034	&	178.19 $\pm$ 0.17	&	0.88 $\pm$ 0.03		&	J2012.527	&	...	\\	
J18548+109	&	18550+1058 VYS8		&	3.854 $\pm$ 0.017	&	44.30 $\pm$ 0.24	&	2.33 $\pm$ 0.08		&	J2014.386	&	F	\\	
J19463+320	&	19464+3201 KAM3		&	5.603 $\pm$ 0.040	&	134.79 $\pm$ 0.43	&	0.81 $\pm$ 0.01		&	J2012.524	&	F	\\	
		&				&	5.576 $\pm$ 0.016	&	134.66 $\pm$ 0.15	&	0.79 $\pm$ 0.02		&	J2012.710	&		\\	
		&				&	5.675 $\pm$ 0.017	&	134.85 $\pm$ 0.19	&	0.74 $\pm$ 0.01		&	J2015.284	&		\\	
J19539+444E	&	19539+4425 GIC59	&	6.461 $\pm$ 0.024	&	70.41 $\pm$ 0.16	&	0.30 $\pm$ 0.05		&	J2012.527	&	F	\\	
		&				&	6.454 $\pm$ 0.027	&	70.28 $\pm$ 0.20	&	0.65 $\pm$ 0.07		&	J2012.710	&	 	\\	
		&	19539+4425 MCY3		&	0.642 $\pm$ 0.016	&	352.19 $\pm$ 1.49	&	2.11 $\pm$ 0.02		&	J2012.527	&		\\	
		&				&	0.615 $\pm$ 0.048	&	349.91 $\pm$ 1.67	&	2.58 $\pm$ 0.13		&	J2012.710	&		\\	
J20407+199	&	20408+19656 RAO23	&	0.166 $\pm$ 0.081	&	227.02 $\pm$ 6.45	&	1.4 $\pm$ 0.2			&	J2012.527	&	F	\\	
		&				&	0.290 $\pm$ 0.080	&	245.37	 $\pm$ 3.69	&	2.1 $\pm$ 0.3			&	J2015.281	&		\\	
J20445+089S	&	20446+0854 LDS1046	&	15.075 $\pm$ 0.064	&	344.47 $\pm$ 0.14	&	1.06 $\pm$ 0.04		&	J2012.527	&	F	\\	
		&				&	15.173 $\pm$ 0.049	&	344.42 $\pm$ 0.14	&	0.97 $\pm$ 0.03		&	J2012.710	&		\\	
J20488+197	&	20488+1943 JNN286	&	0.191 $\pm$ 0.006	&	132.95 $\pm$ 1.68	&	0.33 $\pm$ 0.09		&	J2012.524	&	F	\\	
		&				&	0.184 $\pm$ 0.011	&	134.71 $\pm$ 1.71	&	0.31 $\pm$ 0.06		&	J2012.710	&		\\	
J21012+332	&	New			&	0.246 $\pm$ 0.022	&	7.38 $\pm$ 1.70 	&	1.61 $\pm$ 0.09		&	J2012.524	&	F	\\	
		&				&	0.232 $\pm$ 0.016	&	6.31 $\pm$ 2.67		&	1.75 $\pm$ 0.10		&	J2012.710	&		\\	
		&				&	Unresolved		&	...			&	...			&	J2015.281	&		\\	
		&				&	Unresolved		&	...			&	...			&	J2015.284	&		\\	
J21323+245	&	21324+2434 MCT12	&	1.426 $\pm$ 0.005	&	243.98 $\pm$ 0.14	&	0.75 $\pm$ 0.15		&	J2012.524	&	T	\\	
		&				&	1.438 $\pm$ 0.006	&	243.66 $\pm$ 0.27	&	0.72 $\pm$ 0.10		&	J2012.710	&		\\	
J21518+136	&	New			&	0.701 $\pm$ 0.010	&	116.96 $\pm$ 1.25	&	1.50 $\pm$ 0.03		&	J2012.524	&	F	\\	
		&				&	0.686 $\pm$ 0.022	&	118.08 $\pm$ 1.40	&	1.40 $\pm$ 0.05		&	J2012.710	&		\\	
		&				&	0.674 $\pm$ 0.011	&	130.80 $\pm$ 1.04 	&	...			&	J2015.572	&		\\	
J22279+576	&	22280+5742 KR60 	&	1.649 $\pm$ 0.010	&	359.82 $\pm$ 0.16	&	1.21 $\pm$ 0.03		&	J2011.807	&	T	\\	
J23096-019~\tablefootmark{g}	&	New	&	1.715 $\pm$ 0.006	&	19.37 $\pm$ 0.23	&	1.80 $\pm$ 0.04		&	J2012.708	&	T	\\	
		&				&	1.672 $\pm$ 0.019	&	19.82 $\pm$ 0.37	&	1.84 $\pm$ 0.09		&	J2013.036	&		\\	
J23293+414S(Bab)&	23294+4128 BWL59	&	0.257 $\pm$ 0.027	&	209.09 $\pm$ 2.31	&	0.73 $\pm$ 0.09		&	J2012.524	&	F	\\	
		&				&	Unresolved		&	...			&	...			&	J2015.572	&		\\	
J23293+414N(AB)	&	23294+4128 GIC193	&	17.670 $\pm$ 0.069	&	213.92 $\pm$ 0.18	&	0.38 $\pm$ 0.01		&	J2012.524	&	F	\\	
		&				&	17.676 $\pm$0.024	&	214.40 $\pm$0.15	&	0.38 $\pm$ 0.01		&	J2015.572	&		\\	
J23318+199E	&	23317+1956 WIR1		&	5.412 $\pm$ 0.019	&	81.14 $\pm$ 0.14	&	1.67 $\pm$ 0.02		&	J2011.807	&	F	\\	
		&				&	5.411 $\pm$ 0.018	&	81.07 $\pm$ 0.14	&	1.66 $\pm$ 0.01		&	J2011.812	&		\\	
J23455-161~\tablefootmark{h}	&	23455-1610 MTG5	&	$\sim$0.50		&	$\sim$15		&	...			&	J2012.522	&	...	\\	
J23517+069	&	New			&	2.178 $\pm$ 0.015	&	102.48 $\pm$ 0.40	&	0.47 $\pm$ 0.09		&	J2012.708	&	F	\\	
		&				&	2.165 $\pm$ 0.013	&	102.51 $\pm$ 0.35	&	0.55 $\pm$ 0.11		&	J2013.036	&		\\	
		&				&	2.181 $\pm$ 0.013	&	103.23 $\pm$ 0.13	&	...			&	J2015.572	&		\\	
	\noalign{\smallskip}
	\hline 
	\noalign{\smallskip}
	
\multicolumn{7}{c}{Likely bound systems}	\\
		\noalign{\smallskip}
	\hline 
	\noalign{\smallskip}

J00169+200	&	New			&	1.076 $\pm$ 0.023	&	288.33 $\pm$ 0.54	&	0.70 $\pm$ 0.12		&	J2013.036	&	F	\\		
J01221+221	&	New			&	0.271 $\pm$ 0.014	&	175.04 $\pm$ 2.00	&	0.86 $\pm$ 0.17		&	J2015.875	&	F	\\		
J04352-161	&	New			&	7.887 $\pm$ 0.043	&	131.61 $\pm$ 0.56	&	4.60 $\pm$ 0.38		&	J2014.162	&		\\		
J06277+093	&	New			&	1.103 $\pm$ 0.090	&	221.11 $\pm$ 1.26	&	1.16 $\pm$ 0.04		& 	J2015.875	&	F	\\
J07349+147	&	New			&	1.008 $\pm$ 0.019	&	297.13 $\pm$ 0.64	&	0.77 $\pm$ 0.04		&	J2014.164	&	T	\\		
		&				&	0.988 $\pm$ 0.014	&	293.70 $\pm$ 0.32	&	0.74 $\pm$ 0.03		&	J2015.281	&		\\		
		&				&	0.995 $\pm$ 0.010	&	292.83 $\pm$ 0.21	&	0.78 $\pm$ 0.08		&	J2015.281	&		\\
J10028+484~\tablefootmark{i}	&	New	&	0.208 $\pm$ 0.014	&	336.81 $\pm$ 3.05	&	0.31 $\pm$ 0.06		&	J2012.231	&	F	\\		      

	\noalign{\smallskip}

	\end{longtable}
	\tablefoot{
	\tablefoottext{a}{Guide Star Catalog multiplicity flag. ``F'' is False, ``T'' is true, and ``...'' indicates no data.}
	\tablefoottext{b}{$\rho$ and $\theta$ measures in the second epoch correspond to observations with the CARMENES acquisition and guiding camera, which were carried out only for confirming physical association.}
	\tablefoottext{c}{Similar periods between Behall \& Harrington 1976 and this work suggest that we are resolving the same pair.}
    \tablefoottext{d}{The B component is a double lined spectroscopic binary with a period shorter than 215.0\,d (Shkolnik et~al. 2010) and is not the third component resolved here.}
    \tablefoottext{e}{We considered that this resolved binary is the spectroscopic binary identified by Bonflis et~al. 2013, according to the radial-velocity amplitude and period estimation of the pair. Nevertheless, we can not affirm whether the system is double or triple.}
	\tablefoottext{f}{Close visual binary in Bowler et~al. 2015 in one epoch. Our multi-epoch images confirm physical binding.}
	\tablefoottext{g}{It is also a spectroscopic binary.} 
	\tablefoottext{h}{Faint pair for which we could not measure $\rho$ and $\theta$ with precision. The companion, also detected by Tokovinin et~al. 2015, seems to be physically related and consistent over time with the companion identified by Montagnier et~al. 2006 at 0.068\,arcsec.}
	\tablefoottext{i}{Also observed and identified as single in Law et~al. 2008, probably due to the crossing of the companion behind the primary at the observing epoch, in agreement with the 13\,a period estimated in this work.}
	}

\begin{landscape}
\centering
	\begin{longtable}{lcccccccccc}
	\label{table.rhotheta_magmasses}\\
	\caption{Derived parameters for confirmed physical pairs.}\\
	\hline \hline
	\noalign{\smallskip}

Karmn\tablefootmark{a}	&	Component	&	$\Delta  I$	&	$I_{1}$	&	$I_{2}$	&	$\rm{SpT}$	&	$\rm{SpT}_{1}$	&	$\rm{SpT}_{2}$	&	$\mathcal{M}_{1}$	&	$\mathcal{M}_{2}$	&	$P$\tablefootmark{b}	\\
	& 			&	[mag]		& 	[mag]	&	[mag]	&		&			&		& 	[M$_{\odot}$]	&	[M$_{\odot}$]	&	[a]		\\
	\noalign{\smallskip}

	\hline
	\noalign{\smallskip}
	\endfirsthead
	\caption{continued.}\\
	\hline\hline	
	\noalign{\smallskip}
Karmn	&	Component	&	$\Delta  I$	&	$I_{1}$	&	$I_{2}$	&	$\rm{SpT}$	&	$\rm{SpT}_{1}$	&	$\rm{SpT}_{2}$	&	$\mathcal{M}_{1}$	&	$\mathcal{M}_{2}$	&	$P$		\\
	& 			&	[mag]		& 	[mag]	&	[mag]	&		&			&		& 	[M$_{\odot}$]	&	[M$_{\odot}$]	&	[a]		\\
	\noalign{\smallskip}
	\hline
	\noalign{\smallskip}

	\endhead
	\hline
	\noalign{\smallskip}

	\endfoot

J00154-161	& A, B		&	0.58 $\pm$ 0.08 	& 9.24	& 9.82	& M4.0 V	& m4.0  & m5.0  & 0.17 $\pm$ 0.08	& 0.15 $\pm$ 0.07	& 3.7		\\ 
J00413+558	& A, B		&	2.92 $\pm$ 0.13 	& 11.36& ...	& ...		& M4.0  & DC	& 0.22 $\pm$ 0.10	& ...			& ... 		\\ 
J02518+294	& A, B		&	0.52 $\pm$ 0.10 	& 11.56& 12.08& M4.0 V	& m4.0  & m5.0  & 0.20$\pm$ 0.09	& 0.15 $\pm$ 0.08    	&	130	\\
J02565+554W	& A, B		&	0.97 $\pm$ 0.04 	& 8.56	& 9.53	& ...		& M1.0  & m2.5  & 0.50 $\pm$ 0.16	& 0.35 $\pm$ 0.13 	& 6400 		\\ 
J02591+366	& A, B		&	3.10 $\pm$ 0.13 	& 10.50& 13.60& M3.5 V	& m3.5  & m6.0  & 0.24 $\pm$ 0.13	& 0.10 $\pm$ 0.07  	& 380		\\
J03574-011	& A, B		&	2.29 $\pm$ 0.01 	& ...	& 9.10	& ...		& K4	& M2.5	& ...			& 0.34 $\pm$ 0.12  	& ...		\\ 
{ \it J04153-076}	& AB, C		&	1.62 $\pm$ 0.04 	& ...	& 8.39	& ...		& K0.5+DA& M4.5  & ...			& 0.33 $\pm$ 0.05	&... 		\\ 
J04311+589	& A, B		&	3.64 $\pm$ 0.05 	& 8.14	& ...	& ...		& M4.0	& DC	& 0.23 $\pm$ 0.10	& ...		    	&... 		\\ 
{\it J05019+099}& Aab, B	&	0.67 $\pm$ 0.15 & 9.20	& 9.87	& M4.0 V	& ... 	& ...	& 	...	& 0.33 $\pm$0.05 					&	200			\\
{\it J05019+099}& Aa, Ab	&	... 			& 9.95	& 9.95	& ...		& ...	& ...	& 	0.31 $\pm$0.05	& 0.31 $\pm$0.05	&			\\
J05034+531	& A, B		&	5.68 $\pm$ 0.07 	& 8.04	& 13.72& M0.5 V	& m1.0 & m7.0	& 0.45 $\pm$ 0.16	& 0.09 $\pm$ 0.05   	& 910 		\\ 
{\it J05068-215E}	& A, BC	&	0.77 $\pm$ 0.08 & 8.19	& 8.96	& ...		& M1.5	&	...	&	0.62 $\pm$ 0.05	&	...	&	1600	\\
{\it J05068-215W}	& B, C	&	0.59 $\pm$ 0.08 & 8.67	& 9.26	& M3.5		& ...	&	...	&	0.49 $\pm$ 0.05	& 0.35 $\pm$ 0.05		&	62	\\
J05078+179	& Aab, B		&	1.96 $\pm$ 0.05 	& 9.51	& 11.47& M3.0 V	& m2.0  & m4.0  &...			& 0.23 $\pm$ 0.08  	&...		\\ 
J05078+179	& Aa, Ab	&	...			& 10.26& 10.26& m2.0		& m2.5  & m2.5  & 0.33 $\pm$ 0.05	& 0.33 $\pm$ 0.05   	& 50		\\ 
{\it J05103+488}& A, B	&	0.70 $\pm$ 0.06 	& 9.62	& 10.32& M2.5 V	& ...	& ... 	  & 0.45 $\pm$ 0.05	 	& 0.31 $\pm$ 0.05	 & ...	\\
J05333+448	& A, B		&	0.23 $\pm$ 0.05 	& 10.31& 10.54& M3.5 V	& m3.5  & m3.5  & 0.23 $\pm$ 0.10	& 0.22 $\pm$ 0.09  	& 8.3 		\\ 
J05342+103N	& A, Bab	&	1.02 $\pm$ 0.03 	& 9.88	& 10.90& ...		& M3.0  & m4.0  & ...			& 0.30 $\pm$ 0.10  	& 990 		\\ 
J05342+103N	& Ba, Bb	&	... 			& 11.65& 11.65& M4.5		& m4.5  & m4.5  & 0.18 $\pm$ 0.08	& 0.18 $\pm$ 0.08  	& ... 		\\ 
J05466+441	& Aab, B		&	5.76 $\pm$ 0.09		& 9.98	&15.74	& ...		& M4.0	& m8.0	&	...		& 0.07 $\pm$ 0.03 	& 920		\\
J05466+441	& Aa, Ab	&	...			&10.73	&10.73	& M4		& m3.5	& m3.5	& 0.25 $\pm$ 0.09	& 0.25 $\pm$ 0.09 	&	...	\\
J06212+442	& A, B		&	2.85 $\pm$ 0.13		& 9.99	& 12.84	& M2.0 V	& m2.0	& m5.0	& 0.38 $\pm$ 0.11	& 0.16 $\pm$ 0.06  	& 250		\\
J06400+285	& A, B		&	0.24 $\pm$ 0.10 	& 10.09& 10.33& M2.0 V	& m2.5  & m2.5  & 0.36 $\pm$ 0.15	& 0.33 $\pm$ 0.14  	& 20		\\ 
J07395+334	& A, B		&	4.81 $\pm$ 0.08 	& 9.62	& 14.43& ...		& M2.0  & m6.5  & 0.54 $\pm$ 0.19	& 0.12 $\pm$ 0.07   	& 13000		\\ 
J08066+558	& A, B		&	0.54 $\pm$ 0.07 	& 9.76	& 10.30& M2.0 V	& m2.0  & m2.5  & 0.45 $\pm$ 0.16	& 0.37 $\pm$ 0.14   	& 26 		\\ 
J08082+211	& Bab, C	&	2.76 $\pm$ 0.07		& 8.78	& 11.54& M3.0 V	& m2.5	& m5.0	& ...			& 0.17 $\pm$ 0.09   	& 33		\\ 
J08082+211	& Ba, Bb	&	...			& 9.53	& 9.53& m2.5		& m3.0  & m3.0  & 0.32$\pm$ 0.10	& 0.32 $\pm$ 0.10   	& ... 		\\ 
J08105-138	& B, C		&	1.78 $\pm$ 0.05 	& 9.80	& 11.58& M2.5 V	& m2.5  & m4.5  & 0.36 $\pm$ 0.13	& 0.20 $\pm$ 0.09  	& 130 		\\ 
J08595+537	& A, B		&	0.28 $\pm$ 0.06 	& 11.11& 11.39& M3.5 V	& m3.5	& m3.5	& 0.20 $\pm$ 0.09	& 0.18 $\pm$ 0.08   	& 19 		\\ 
J09011+019	& Aab, B	&	4.55 $\pm$ 0.17		& 9.27	& 13.82& M3.0 V	& m3.0	& m6.5	& ...			& 0.09 $\pm$ 0.04   	& 460 		 	\\
J09011+019	& Aa, Ab	&	...			& 10.02& 10.02 & m3.0 V	& m3.5	& m3.5	& 0.26 $\pm$ 0.10	& 0.26 $\pm$ 0.10   	& ... 		\\
J10151+314	& A, B		&	0.65 $\pm$ 0.10 	& 11.32& 11.97& M4.0 V	& m4.0  & m4.5  & 0.22 $\pm$ 0.11	& 0.18 $\pm$ 0.10    	& 360 		\\ 
{\it J10196+198}	&	A, B	&	2.00 $\pm$ 0.20		& 6.93	& 8.93	& M3.0 V	& m3.0	& m5.0	& 0.32 $\pm$ 0.12 	&	0.18 $\pm$ 0.11		& 1.5			\\
J10260+504W	& A, B		&	0.24 $\pm$ 0.06 	& 10.79& 11.03& ...		& M4.0  & m4.5	& 0.21 $\pm$ 0.09	& 0.20 $\pm$ 0.08   	& 5600 				\\ 
J10379+127	& A, B		&	0.51 $\pm$ 0.12 	& 10.73& 11.24& M3.0 V	& m3.5  & m4.0  & 0.26 $\pm$ 0.10	& 0.22 $\pm$ 0.09   	& 100		\\ 
J10448+324	& Aa, Ab	&	0.51 $\pm$ 0.11 	& 11.07& 11.58& M3.0 V	& m2.5  & m3.5  & 0.35 $\pm$ 0.13	& 0.32 $\pm$ 0.12   	& 440 				\\ 
J10546-073	& A, B		&	0.43 $\pm$ 0.09 	& 11.08& 11.51& M4.0 V	& m4.0  & m4.5  & 0.21 $\pm$ 0.09	& 0.19 $\pm$ 0.08   	& 80 		\\ 
J11355+389	& A, B		&	0.55 $\pm$ 0.13 	& 11.02& 11.57& M3.5 V	& m3.5  & m4.0  & 0.26 $\pm$ 0.10	& 0.22 $\pm$ 0.09   	& 31		\\ 
J11521+039	& A, B		&	0.48 $\pm$ 0.11 	& 10.44& 10.92& M4.0 V	& m4.0  & m4.5  & 0.22 $\pm$ 0.09	& 0.19 $\pm$ 0.08   	& 15 			\\ 
J12006-138	& A, B		&	2.42 $\pm$ 0.04 	& 10.32& 12.74& ...		& M3.5  & m5.5  & 0.24 $\pm$ 0.13	& 0.12 $\pm$ 0.08  	& 2000			\\ 
J12016-122	& A, B		&	5.11 $\pm$ 0.10		& 10.00& 15.11& ...		& M3.0	& m8.0	& 0.29 $\pm$ 0.12	& 0.07 $\pm$ 0.05   	& 2000 			\\
{\it J12123+544S}	& A, B		&	2.86 $\pm$ 0.04 	& 7.90	& 10.76& ...		& M0.0  & m4.0	& 0.52 $\pm$ 0.17	& 0.20 $\pm$ 0.09   	& 4100 		\\ 
J12162+508	& A, B		&	0.45 $\pm$ 0.11 	& 11.36& 11.81& M4.0 V	& m4.0  & m5.0  & 0.21 $\pm$ 0.11	& 0.19 $\pm$ 0.09   	& 360		\\ 
J12277-032	& A, B		&	1.71 $\pm$ 0.08 	& 10.44& 12.15& M3.5 V	& m3.5  & m5.0  & 0.25 $\pm$ 0.11	& 0.16 $\pm$ 0.08   	& 200 				\\ 
J12332+090	& A, B		&	0.47 $\pm$ 0.10 	& 9.36	& 9.83	& M5.0 V	& m5.0  & m5.5  & 0.14 $\pm$ 0.07	& 0.12 $\pm$ 0.07   	& 16 				\\ 
J13168+170	& A, B		&	2.08 $\pm$ 0.05 	& ...	& 7.562	& ...		& K2	& M0.5	& ...			& 0.46 $\pm$ 0.16   	&... 				\\ 
J13180+022	& A, B		&	0.58 $\pm$ 0.18 	& 10.76& 11.34& M3.5 V	& m3.5  & m4.5  & 0.25 $\pm$ 0.11	& 0.21 $\pm$ 0.09   	& 73 			\\ 
{\it J13317+292}	& A,B		&	1.5 $\pm$ 0.2		& 9.23	& 10.73& M4.0 V	& ...	& ...	& 0.39 $\pm$ 0.05	& 0.17 $\pm$ 0.05	&	8.6		\\	
J13417+582	& A, B		&	0.38 $\pm$ 0.08 	& 10.78& 11.16& M3.5 V	& m3.5  & m4.0  & 0.23 $\pm$ 0.09	& 0.21 $\pm$ 0.08   	& 78			\\ 
J13526+144	& A, B		&	1.62 $\pm$ 0.09 	& 9.42	& 11.04& M2.0 V	& m2.0  & m4.0  & 0.38 $\pm$ 0.17	& 0.23 $\pm$ 0.11   	& 150		\\ 
J14157+594	& A, B		&	0.56 $\pm$ 0.18 	& 10.04& 10.60& ...		& M2.0  & m3.0	& 0.38 $\pm$ 0.17	& 0.31 $\pm$ 0.14   	& 1900			\\ 
J14210+275	& A, B		&	1.77 $\pm$ 0.08		& 10.45& 12.22& M2.5 V	& m2.5	& m4.5	& 0.33 $\pm$ 0.13	& 0.20 $\pm$ 0.09   	& 97			\\ 
J14279-003S	& A, B		&	0.19 $\pm$ 0.03 	& 10.82& 11.01& ...		& M4.5  & m5.0	& 0.18 $\pm$ 0.07	& 0.17 $\pm$ 0.07   	& 3900 				\\ 
J14331+610	& A, B		&	0.62 $\pm$ 0.13 	& 9.99	& 10.61& M2.5 V	& m3.0  & m3.5  & 0.31 $\pm$ 0.12	& 0.25 $\pm$ 0.11   	& 96 			\\ 
J15081+623	& A, B		&	0.44 $\pm$ 0.11 	& 11.37& 11.81& M4.0 V	& m4.0  & m4.5  & 0.21 $\pm$ 0.10	& 0.19 $\pm$ 0.09   	& 140 		\\ 
J15126+457	& A, B		&	0.36 $\pm$ 0.08 	& 11.20& 11.56& M4.0 V	& m4.5  & m5.0  & 0.20 $\pm$ 0.09	& 0.18 $\pm$ 0.09   	& 43 				\\ 
J15191-127	& A, B		&	3.40 $\pm$ 0.15		& 10.03& 13.43& M4.0 V	& m3.0	& m5.5	& 0.33 $\pm$ 0.06	& 0.13 $\pm$ 0.05		& 13		\\
J15400+434N	& A, B		&	1.53 $\pm$ 0.09 	& 9.63	& 11.16& ...		& M3.0  & m4.5	& 0.26 $\pm$ 0.10	& 0.17 $\pm$ 0.08  	& 730 			\\ 
J15496+348	& A,B		&	3.00 $\pm$ 0.10		& 10.30 &13.30& M4.0 V	& m3.5	& m7.0	& 0.26 $\pm$ 0.11	& 0.11 $\pm$ 0.09	&	12	\\
J16487+106	& A, B		&	0.34 $\pm$ 0.09 	& 9.60	& 9.94	& M2.5 V	& m2.5  & m3.0  & 0.33 $\pm$ 0.14	& 0.30 $\pm$ 0.12	& 10		\\ 
J16554-083S	& A, Bab	&	0.43 $\pm$ 0.09 	& 7.15	& 7.58	& M3.0 V	& m3.0	& m3.0	& 0.34 $\pm$ 0.13	& ... 			& 1.4				\\ 
J16554-083S	& Ba, Bb	&	...			& 8.33	& 8.33	& m3.0 V	& m3.5	& m3.5	& 0.23 $\pm$ 0.13	& 0.23 $\pm$ 0.13  	& ...			\\ 
J16578+473	& A, B		&	2.48 $\pm$ 0.02		& ...	& 8.01	& ...		& K0.0 V& M1.5  & ...			& 0.57 $\pm$ 0.18   	&... 			\\ 
J17340+446	& A, B		&	0.50 $\pm$ 0.11 	& 10.74& 11.24& M3.5 V	& m3.5  & m4.0  & 0.26 $\pm$ 0.12	& 0.22 $\pm$ 0.11    	& 60 			\\ 
J17530+169	& A, B		&	0.65 $\pm$ 0.11 	& 10.64& 11.29& M3.0 V	& m3.5	& m4.0  & 0.26 $\pm$ 0.14	& 0.21 $\pm$ 0.12   	& 110 		\\ 
J18180+387E	& A, B		&	1.50 $\pm$ 0.03 	& 9.36	& 10.86& ...		& M3.0  & m4.5	& 0.24 $\pm$ 0.10	& 0.16 $\pm$ 0.07   	& 1700				\\ 
J18264+113 	& A, B		&	5.89 $\pm$ 0.09 	& 10.39& 16.28& ...		& M3.5  & WD	& 0.25 $\pm$ 0.08	& ... 			& ... 				\\ 
J18411+247S	& Aab, B	&	2.25 $\pm$ 0.02 	& 9.18	& 11.43& ...		& M3.5 & m5.5	& ...  			& 0.14 $\pm$ 0.07   	& 460 		\\ 
J18411+247S & Aa, Ab	&	... 			& 9.92	& 9.92	& M3.5		& m4.0 & m4.0	& 0.21 $\pm$ 0.08	& 0.21 $\pm$ 0.08   	& ... 			\\ 
J18427+596N	& A, B		&	0.88 $\pm$ 0.03 	& 6.51	& 7.39	& ...		& M3.0  & m4.0	& 0.28 $\pm$ 0.10	& 0.22 $\pm$ 0.09   	& 380 			\\ 
{\it J18548+109}	& A, B	&	2.33 $\pm$ 0.08 	& 8.29	& 10.62& ...			& M0.0  & m3.0	& 0.56 $\pm$ 0.12	& 0.26 $\pm$ 0.11    &700   \\        
J19463+320	& A, B		&	0.78 $\pm$ 0.04 	& 7.91	& 8.69	& ...		& M0.5  & m2.0	& 0.47 $\pm$ 0.15	& 0.36 $\pm$ 0.12	& 750		\\ 
J19539+444W	& A, B		&	2.34 $\pm$ 0.34 	& 9.94	& 12.28& M5.5 V	& m5.5  & m8.0  & 0.12	$\pm$ 0.07	& 0.07 $\pm$ 0.05  	& 11		\\ 
J19539+444E	& AB, C		&	0.48 $\pm$ 0.25		& ...	& 9.83	& M5.5 V	& ...	& M5.5  & ...			& 0.14 $\pm$ 0.06   	& 260			 \\
J20407+199	& B, C		& 1.4 $\pm$ 0.2		& 9.75	& 11.15	& M2.5	& m2.5	& m4.0	& 0.36 $\pm$ 0.10	& 0.23 $\pm$ 0.09	&	8.4	\\
J20445+089S	& A, Bab	&	1.02 $\pm$ 0.07 	& 9.28	& 10.30& ...		& M1.5  & m3.0	& 0.44 $\pm$ 0.15	& ...  			& 6400			\\ 
J20445+089N	& Ba, Bb	&	... 			& 11.05& 11.05& m3.0 V	& m3.5  & m3.5	& 0.25 $\pm$ 0.10	& 0.25 $\pm$ 0.10   	& ... 			\\ 
J20488+197	& A, B		&	0.32 $\pm$ 0.08 	& 11.48& 11.80& M4.0 V	& m4.5  & m5.0  & 0.27 $\pm$ 0.12	& 0.25 $\pm$ 0.11  	& 22 			\\ 
J21012+332	& Aa, Ab	&	1.68 $\pm$ 0.14 	& 9.77	& 11.45& M3.0 V	& m2.5  & m4.5  & 0.33 $\pm$ 0.14	& 0.18 $\pm$ 0.10  	& 13			\\ 
J21323+245	& A, B		&	0.74 $\pm$ 0.13 	& 10.39& 11.13& M3.5 V	& m3.5  & m4.5  & 0.29 $\pm$ 0.12	& 0.23 $\pm$ 0.11  	& 250 				\\ 
J21518+136	& A, B		&	1.45 $\pm$ 0.08 	& 11.08& 12.53& M4.5 V	& m4.0  & m5.5  & 0.19 $\pm$ 0.08	& 0.13 $\pm$ 0.07   	& 66 				\\ 
J22279+576	& A, B		& 	1.21 $\pm$ 0.03		& 7.35	& 8.56	& M3.0 V	& m3.5	& m5.0	& 0.23 $\pm$ 0.10	& 0.16 $\pm$ 0.09	& 27			\\
J23096-019	& Aab, B	&	1.82 $\pm$ 0.08 	& 10.33& 12.15& M3.5 V	& m3.5  & m5.0  & ...			& 0.17 $\pm$ 0.08   	& 260			\\ 
J23096-019	& Aa, Ab	&	... 			& 11.08& 11.08& m3.0 V	& m3.5  & m3.5  & 0.23 $\pm$ 0.12	& 0.23 $\pm$ 0.12   	& ... 			\\ 
{\it J23293+441N}& A, Bab	&	0.38 $\pm$ 0.02		&  9.39& 9.77	& M3.5 V	& ...	&...	& 0.32 $\pm$ 0.05	& ...		&	4500	\\
{\it J23293+414S}	& Ba, Bb	&	0.73 $\pm$ 0.09 & 9.54	& 10.27& M4.0 V	&  ... & ...  	& 0.34 $\pm$ 0.05	& 0.24 $\pm$ 0.05	&	10	\\
J23318+199E	& Aab, B	&	1.66 $\pm$ 0.02 	& 7.63	& 9.29	& ...		& M3.5 & m5.0	& ...			& ...   		& 230			\\ 
J23318+199E	& Aa, Ab	&	... 			& 8.38	& 8.38	& M3.5		& m3.5  & m3.5	& 0.23 $\pm$ 0.11	& 0.23 $\pm$ 0.11	&... 					\\ 
J23318+199E	& Ba, Bb	&	... 			& 10.04& 10.04& M4.5		& m5.0  & m5.0	& 0.14 $\pm$ 0.09	& 0.14 $\pm$ 0.09	&... 					\\ 
J23517+069	& A, B		&	0.51 $\pm$ 0.11 	& 10.84& 11.35& M3.0 V	& m3.5  & m4.0  & 0.26 $\pm$ 0.13	& 0.22 $\pm$ 0.12	& 450		 		\\ 

	\noalign{\smallskip}
	\hline
	\noalign{\smallskip}

J00169+200	& A, B		&	0.70 $\pm$ 0.12 	& 11.67& 12.37& M3.5 V	& m4.0  & m4.5  & 0.23 $\pm$ 0.14	& 0.19 $\pm$ 0.09	& 210		\\ 
{\it J01221+221}&  A, B	&	0.86 $\pm$ 0.17		& 10.46& 11.32& M4.5 V	& ... & ...	& 0.18 $\pm$ 0.05	& 0.15 $\pm$  0.05	&	8.0		\\
J04352-161	& Aab, B	&	4.60 $\pm$ 0.38		& 12.88& 17.48& M7.0 V	& ...	& ...	& ...			& <0.07			& 1600~\tablefootmark{c}				\\
J04352-161	& Aa, Ab	&	...			& 13.63& 13.63& M7.0 V	& m7.0	& m7.0	& 0.08 $\pm$ 0.04	& 0.08$\pm$ 0.04	&						\\
J06277+093	& A, B		&	1.16 $\pm$ 0.04		& 9.78	& 10.94& M2.0 V	& m2.0	& m3.5	& 0.38 $\pm$ 0.12	& 0.26 $\pm$ 0.11	& 160				\\
J07349+147	& A, B		&	0.76 $\pm$ 0.04 	& 9.19	& 9.95	& M3.0 V	& m3.5  & m4.5  & 0.25 $\pm$ 0.11	& 0.20 $\pm$ 0.09	& 40				\\ 
{\it J10028+484}& A, B	& 	0.31 $\pm$ 0.06 	& 12.61& 12.92& M5.5 V	& ...  & ... &	0.15 $\pm$ 0.08& 0.12 $\pm$ 0.08	&	12		\\

	\noalign{\smallskip}
\end{longtable}
\tablefoot{
	\tablefoottext{a}{Karmn stars in italics are associated to young stellar populations (see Table~\ref{table.mov_groups}).}
	\tablefoottext{b}{The periods given are a lower limit, as we equal our maximum separation measured to the semimajor axis.}
	\tablefoottext{c}{The B component of the system is too faint to estimate its mass with our $M_I$-mass relation and the period was calculated using the stellar mass limit (0.07\,M$_\odot$). Hence, the period of this system should be considered as a lower approximation.}
	}

\end{landscape}

\label{appendix.wide}

\centering
\begin{longtable}{lcccccl}

\label{table.knownwide}\\
\caption{Known binaries at $\rho > 5$\,arcsec.}\\
\hline\hline
\noalign{\smallskip}
	WDS	&	\multicolumn{2}{c}{Primary}	&	\multicolumn{2}{c}{Secondary}	&	$\rho$	&	Notes	\\
	&	Name	&	SpT	&	Name	&	SpT	&	[arcsec]&	\\
	\noalign{\smallskip}
	\hline
	\noalign{\smallskip}

	\endfirsthead
	\caption{continued.}\\
	\hline\hline	
	\noalign{\smallskip}
	WDS	&	\multicolumn{2}{c}{Primary}	&	\multicolumn{2}{c}{Secondary}	&	$\rho$	&	Notes	\\
	&	Name	&	SpT	&	Name	&	SpT	&	[arcsec]&	\\
	\noalign{\smallskip}
	\hline
	\noalign{\smallskip}

	\endhead
	\hline
	\noalign{\smallskip}

	\endfoot

GRB34	&	J00183+440	&	M1.0 V	&	J00184+440&	M3.5 V	&	34.8	&	\tablefootmark{a}	\\
	\noalign{\smallskip}
GIC13	&	J00413+558	& M4.0 V	&	EGGR 245	&	DC	&	10.8	&	\\
	\noalign{\smallskip}
WNO51	&	J01026+623	& M1.5 V	&	J01033+623& M5.0 V	&	293.1	&	\\
	\noalign{\smallskip}		
GIC20	&	J01119+049N	& M3.0 V	&	J01119+049S	& M3.5 V	&	63.6	&	\\
	\noalign{\smallskip}		
GIC27	&	J01518+644	& M2.5 V 	&	GJ3118 B	&	DAs	& 13.4	&	\\
	\noalign{\smallskip}		
PLW32	&	HD 16160 AB	& K3 V + M7.0 V	&	J02362+068	& M4.0 V	& 164.0	&	\\
	\noalign{\smallskip}		
LDS883	&	HD 18143 AB	& G5 V + K7 V	&	J02555+268	& M4.0 V	& 44.0	&	\\
	\noalign{\smallskip}		
LDS5401	&	J02565+554W	& M1.0 V 	& J02565+554E	& M3.0 V 	& 16.7	&	\\
	\noalign{\smallskip}		
KUI11	&	HD 18757	&	G4 V	&	J03047+617	& M3.0 V & 263.2	&	\\
	\noalign{\smallskip}		
LDS884	&	GJ 140	& M0.0 V+	&	J03242+237	& M2.0 V 	&99.5	&	\tablefootmark{b}	\\
	\noalign{\smallskip}		
LDS9158	&	J03396+254E 	& M3.0 V	&	J03396+254W& M3.5 V	&	64.4	&	\\
	\noalign{\smallskip}		
GIC44	&	J03438+166	& M0.0 V	&	J03437+166	& M1.0 V	&	106.50	&\tablefootmark{c} 	\\
	\noalign{\smallskip}		
BU543	&	BD--01 564	& K4 V	&	J03574--011& M2.5 V 	&	11.0	& \tablefootmark{d}\\
	\noalign{\smallskip}		
STF518	& $o$\textsuperscript{02} Eri A	&	K0.5 V	&	J04153--076	& M4.5 V	&	77.9	&	\tablefootmark{e}\\
STF518	& $o$\textsuperscript{02} Eri B	&	DA	&	J04153--076	& M4.5 V	&	8.6	&	\tablefootmark{e}\\
	\noalign{\smallskip}
LDS3584	&	J04252+080S	& M2.5 V+	&	HG 7--207	& M4.0 V	&	73.4	&	\tablefootmark{f}	\\
	\noalign{\smallskip}
STI205	&	J04311+589	& M4.0 V	&	EGGR 180	&	DC	&	10.0	&	\tablefootmark{g}	\\
	\noalign{\smallskip}
LDS6160	&	J05032+213	&	M1.5 V+	&	HD 285190 BC	& M5.0 V + M5.5 V	&	167.0	&	\tablefootmark{h}	\\
	\noalign{\smallskip}
WDK1	&	J05034+531	&	M0.5 V	&	...	&	m7.0	&	5.6	&	\tablefootmark{i}	\\
	\noalign{\smallskip}
DON93	&	J05068--215E&	M1.5 V	&	J05068--215W	&	m3.0 V + m4.0 V	&	8.5	&	\tablefootmark{j}	\\
	\noalign{\smallskip}
LDS6186	&	HD 35956 	&	G0 V+	&	J05289+125	& M4.0 V	&	99.4	&\tablefootmark{k}	\\
	\noalign{\smallskip}
LDS6189	&	J05342+103N  	&	M3.0 V	&	J05342+103S	& M4.5 V	&	5.1	&	\\
	\noalign{\smallskip}
TOK255	&	J05365+113  	&	M0.0 V	&	J05366+112 	& M 4.0 V	&	156.5	&	\\
	\noalign{\smallskip}
GIC61	&	EG Cam  	&	M0.5 V	&	J05599+585	 	& M4.0 V	&	161.2	&	\\
		\noalign{\smallskip}
NAJ1	&	J06105--218 	&	M0.5 V	&	GJ229 B	 	& T7	&	6.8	&	\\
		\noalign{\smallskip}
GIC65	&	J06421+035 	&	M2.0 V 	&	J06422+035	&	M4.0 V	&	49.7	&	\\
		\noalign{\smallskip}
WNO17	&	HD 50281	&	K3 V	&	J06523--051	&	M3.5 V+	&	58.3	&\tablefootmark{l}	\\
		\noalign{\smallskip}
GIC75	&	J07307+481	&	M4.0+	&	EGGR 52	&	DC9+DC9	&	103.4	&	\\
	\noalign{\smallskip}
LDS6206	&	VV Lyn AB	&	M2.5 V+	&	J07319+362N	&	M3.5 V	&	38.20	&	\\
	\noalign{\smallskip}
Pov09	&	V869 Mon	+ GJ 282 B	&	K2 V + K5 V	&	J07361--031	&	M1.0 V	&	3892.0	&	\\
		\noalign{\smallskip}		
LDS3755	&	J07395+334	&	M2.0 V	&	 LP 256--044	&	M6:	&	13.7	&	\\
		\noalign{\smallskip}		
LUY5693	&	EGGR5 54 A	&	DAZ6	&	J07403--174	&	 M6.0 V	&	21.0	&	\tablefootmark{m}\\
		\noalign{\smallskip}		
COU91	&	BD 21+1764A	&	K7 V 	&	J08082+211	&	M3.0 + m3.0 + m5.0	&	10.7	&\tablefootmark{n}	\\
		\noalign{\smallskip}		
LDS204	&	GJ 9255 A	&	F6.5 V	&	J08105--138	&	M2.0 V + M5.5 V	&	97.6	&	\\
		\noalign{\smallskip}		
LUY6218	&	 BD+10 1857 AB	&	M0.0 V+	&	J08428+095	&	M2.5 V	&	114.4	&	\\
		\noalign{\smallskip}		
OSV2	&	J09008+052W	&	M3.0 V	&	Ross 687	&	M3.5 V	&	29.40	&	\\
		\noalign{\smallskip}		
STF1321	&	J09143+526	&	M0.0 V+	&	J09144+526	&	M0.0 V 	&	17.20	&\tablefootmark{o}	\\
		\noalign{\smallskip}		
LDS6226	&	J09187+267	&	M1.5 V	&	LP 313--038	&	M5.0 V	&	 76.3	&	\\
		\noalign{\smallskip}		
GIC87	&	J09288--073	&	M2.5 V	&	GJ 347 B	&	M4.5	&	35.9	&	\\
		\noalign{\smallskip}		
LDS3917	&	J09430+237	&	M1.0 V	&	 LP 370--034	&	M7.0 V	&	131.20	&	\\
		\noalign{\smallskip}		
REB1	&	J10043+503	&	M2.5 V	&	G 196--003 B	&	L2	&	16.0	&	\\
		\noalign{\smallskip}		
LDS3977	&	J10185--117	&	M4.0 V	&	LP 729--055 	&	M5.0 V	&	15.8	&	\\
		\noalign{\smallskip}		
LDS1241	&	J10260+504W	&	M4.0 V	&	J10260+504E 	&	M4.0 V	&	14.4	&	\\
		\noalign{\smallskip}		
LDS3999	&	J10345+465	&	M3.0 V	&	NLTT 24709	&	M4.5 V	&	46.5	&	\\
		\noalign{\smallskip}		
LDS1258	&	 J10448+324	&	M2.0 V + M4.0 V	&	LP 316--605 B	&	M4.5 V	&	35.1	&	\\
		\noalign{\smallskip}		
VBS18	&	BD+44 2051A	&	M1.0 V	&	 J11055+435	&	M5.5 V	&	31.4	&	\\
		\noalign{\smallskip}		
LDS5207	&	J11476+002	&	M4.0 V	&	LP 613--050 B	&	M5.5 V	&	24.8	&	\\
		\noalign{\smallskip}		
LDS4166	&	J12006--138	&	M3.5 V	&	LP 734--010 B	&	M4.5	&	6.8	&	\\
		\noalign{\smallskip}		
		\noalign{\smallskip}		
LDS390	&	LTT 4562	&	M3.0 V	&	J12112--199	&	M3.5 V 	&	85.3	&	\\
		\noalign{\smallskip}		
VYS5	&	J12123+544S	&	M0.0 V	&	J12123+544N	&	M3.0 V	&	14.7	&	\\
		\noalign{\smallskip}		
(Haw96)	&	J12142+006	&	M5.0 V+	&	...	&	...	&	5.0 	&	\tablefootmark{p}\\
		\noalign{\smallskip}		
BU800	&	HD 115404	&	K2 V	&	J13168+170	&	M0.5 V	&	7.5	&	\\
		\noalign{\smallskip}		
LDS448	&	BD--07 3632	&	DA5.0	&	J13300--087	&	M4.0 V	&	503.0	&	\\
		\noalign{\smallskip}		
DEA1	&	J13481--137	&	M4.5 V 	&	LHS 2803B	&	T5.5	&	67.6	&	\\
		\noalign{\smallskip}		
LDS461	&	J13507--216	&	M3.0 V	&	J13503--216	&	M3.5 V	&	374.90	&	\\
		\noalign{\smallskip}		
VVO12	&	BD+46 1951	&	M0.0 V	&	J14173+454	&	M5.0 V	&	59.2	&	\\
		\noalign{\smallskip}		
STT580	&	$\theta$ Boo A	&	F7 V	&	J14251+518	&	M2.5 V	&	69.5	&	\\
		\noalign{\smallskip}		
BUI442	&	 J14257+236W	&	M0.0 V	&	J14257+236E 	&	M0.5 V	&	45.4	&	\tablefootmark{q}\\
		\noalign{\smallskip}		
GIC120	&	J14279--003S	&	M4.5 V	&	J14279--003N	&	M4.5 V	&	13.0	&	\\
		\noalign{\smallskip}		
LDS961	&	J14283+053	&	M3.0 V	&	LP 560--026	&	M3.5 V	&	60.2	&	\\
		\noalign{\smallskip}		
LDS6309	&	Ross 806	&	M2.5 V	&	J15531+347S	&	M3.5 V	&	26.50	&	\\
		\noalign{\smallskip}		
LDS573&	J16554--083S	&	M3.0 V + M4.0V	&	J16554--083N	&	M3.5 V		&	72.2	&\tablefootmark{r}			\\
LDS573&	J16554--083S	&	M3.0 V + M4.0V	&	J16555--083		&	M7.0 V		&	230.8	&\tablefootmark{r}			\\	
		\noalign{\smallskip}		
A184	&	V1090 Her	&	K0 V	&	J16578+473	&	M1.5 V	&	5.08	&	\tablefootmark{e}\\
STFA32	&	V1089 Her	&	K0 V 	&	J16578+473	&	 M1.5 V	&	111.60	&	\tablefootmark{e}\\
		\noalign{\smallskip}		
LDS593	&	J17177--118	&	M3.0 V	&	2MASS J17174454--1148261	&	m4.0	&	30.1	&\tablefootmark{s}	\\
		\noalign{\smallskip}		
BDK9	&	J17578+465	&	M2.5 V	&	G 204--039 B	&	T6.5	&	197.0	&	\\
		\noalign{\smallskip}		
GIC151	&	J18180+387E	&	M3.0 V	&	J18180+387W	&	M4.0 V	&	9.9	&	\\
		\noalign{\smallskip}		
NI38	&	J18264+113	&	M3.5 V	&	2MASS J18262449+1120498	&	``WD''	&	8.1	&	\\
		\noalign{\smallskip}		
LDS6329	&	BD+45 2743	&	M0.5 V	&	J18354+457	&	M2.5 V	&	112.2	&	\\
		\noalign{\smallskip}		
STF2398	&	J18427+596N	&	M3.0 V 	&	J18427+596B	&	M3.5V	&	11.7	&	\\
		\noalign{\smallskip}		
KAM3	&	J19463+320	&	M0.5 V	&	J19464+320	&	M2.5 V	&	5.7	&	\tablefootmark{t}\\
		\noalign{\smallskip}		
GIC159	&	J19539+444W	&	M4.5 V + M8.0 V	&	J19539+444E	&	M5.5 V	&	6.4	&	\\
		\noalign{\smallskip}		
LDS1045	&	GJ 797 A	&	G5 V	&	J20407+199	&	M2.5 V+	&	125.1	&	\tablefootmark{u}\\
		\noalign{\smallskip}		
LDS1046	&	J20445+089S 	&	M1.5 V	&	J20445+089N	&	M3.5 V+	&	15.2	&	\tablefootmark{v}\\
		\noalign{\smallskip}		
LDS6418	&	J20556--140N	&	M4.0 V	&	GJ 810 B	&	M5.0 V	&	107.1	&	\\
		\noalign{\smallskip}		
LDS1049	&	J21012+332	&	m2.5 V + m4.5 V	&	J21013+332	&	M2.0 V + M5.0 V	&	56.9	&	\\
		\noalign{\smallskip}		
LDS1053	&	21160+298E	&	M3.5 V+	&	21160+298W	&	M3.5 V	&	26.1	&	\tablefootmark{w}\\
		\noalign{\smallskip}		
GIC193	&	J23293+414N 	&	M3.5 V	&	J2393+414S	&	M4.0 V	&	17.7	&	\tablefootmark{x}\\
		\noalign{\smallskip}		
WIR1	&	J23318+199E	&	M3.5 V+	&	J23318+199W	&	M4.5 V+	&	5.3	&	\tablefootmark{y}\\
		\noalign{\smallskip}		
LDS830	&	J23573--129E	&	M3.0 V	&	J23573--129W	&	M4.0 V+	&	19.6	&	\tablefootmark{z}\\
	\noalign{\smallskip}
	\hline
	\end{longtable}
\tablefoot{
\tablefoottext{a}{C is background (GRB34).}
\tablefoottext{b}{AB is separated by 2.5\,arcsec (WOR4).}
\tablefoottext{c}{``B-G'' are background (LMP3).}
\tablefoottext{d}{Simbad indicates that J03574--011 is a spectroscopic binary but we did not find any reference.}
\tablefoottext{e}{Triple system.}
\tablefoottext{f}{Primary is SB2 (Llamas 2014).}
\tablefoottext{g}{Primary is an astrometric binary separated by 0.07\,arcseconds (Strand 1977).}
\tablefoottext{h}{The primary is a SB2 (Sch\"ofer 2015).}
\tablefoottext{i}{It was also observed with the CAMELOT low resolution imager at the Observatorio del Teide (Tenerife) in September 2015 with the $BVIgri$ filters in order to obtain more photometric information but we could not avoid the saturation of the primary.}
\tablefoottext{j}{BC is separated by 0.8\,arcsec.}
\tablefoottext{k}{Primary is a SB (Simbad).}
\tablefoottext{l}{Background source at 9.62\,arcsec (TNN6).}
\tablefoottext{m}{WDS VBS41 at 4\,arcsec and 208\,deg was not detected in this work nor in Davison et~al. 2015. It could be an unrelated companion. The WDS "AC" designation refers to the pair in the table.}
\tablefoottext{n}{Hierarchical quadruple.}
\tablefoottext{o}{Primary is a SB1 (Sch\"ofer 2015). Two other WDS entries under the same discoverer code (STF1321) are not physically bound components.}
\tablefoottext{p}{SB2 (Bonfils et~al. 2013). Hawley et~al. (1996) listed in Table~1.(b) a companion at 5.0\,arcsec 1.2\,mag fainter in $V$. 2MASS resolved a source 6.5\,mag fainter in the $J$-band at 5.9\,arcsec and 179\,deg (quality flag: AUU). Neither Law et~al. (2008) nor Dieterich et~al. (2012 -- with NICMOS onboard {\it Hubble}) detected it. We believe that Hawley et~al. (1996) made reference to the background star 2MASS~J12141817+0037297.}
\tablefoottext{q}{Other WDS entries under BU 1442 and STG 6 are unrelated sources.}
\tablefoottext{r}{Quintuple system.}
\tablefoottext{s}{Simbad mixes up the true primary GJ~3999 (``L~845-016''), the true secondary 2MASS J17174454--1148261, and the background star GJ~4000~B (``L~845-015'', see Table~\ref{table.visuals}). We preserve the current (wrong) nomenclature.}
\tablefoottext{t}{C is background (HEL3).}
\tablefoottext{u}{Other WDS entries under RAO 23 are unrelated sources.}
\tablefoottext{v}{Secondary is SB1 (Sch\"ofer 2015).}
\tablefoottext{w}{AB is separated by 0.05\,arcsec (BWL56).}
\tablefoottext{x}{Bab is separated by 0.26\,arcsec (BWL59). AC is background (BWL59).}
\tablefoottext{y}{A and B are SB1 (Delfosse et al. 1999). Other WDS entries under LMP 24 are unrelated sources.}
\tablefoottext{z}{Secondary is SB2 (Sch\"ofer 2015).}
}

\end{appendix}

\end{document}